\documentclass[final,1p,times,number]{elsarticle}

\usepackage{hyperref}
\usepackage{physics}
\usepackage{amsmath}
\usepackage{comment}
\usepackage{float}
\usepackage{units}
\usepackage{amssymb}
\usepackage{xcolor}

\usepackage{subcaption}
\usepackage{caption}

\newcommand{\be}{ \begin{equation} }
\newcommand{\ee}{\end{equation}}
\newcommand{\bea}{ \begin{eqnarray} }
\newcommand{\eea}{\end{eqnarray}}
\newcommand{\mulog}{z}

\journal{Annals of Physics}

\newcommand{\bg}{ \begin{gather} }
\newcommand{\eg}{\end{gather}}

\def\Tr{\mathop{\rm Tr}}

\renewcommand{\Re}{\mathop{\rm Re}}
\renewcommand{\Im}{\mathop{\rm Im}}

\begin{document}

\begin{frontmatter}

\title{From Anderson localization on Random Regular Graphs to Many-Body localization}

%\begin{comment}
%% Group authors per affiliation:

\author{K.\,S.~Tikhonov}
\address{Skolkovo Institute of Science and Technology, Moscow, 121205, Russia}
\address{L.\,D.~Landau Institute for Theoretical Physics RAS, 119334 Moscow, Russia}

\author{A.\,D.~Mirlin}
\address{Institute for Quantum Materials and Technologies, Karlsruhe Institute of Technology, 76021 Karlsruhe, Germany}
\address{Institute for Condensed Matter Theory, Karlsruhe Institute of Technology, 76128 Karlsruhe, Germany}
\address{L.\,D.~Landau Institute for Theoretical Physics RAS, 119334 Moscow, Russia}
\address{Petersburg Nuclear Physics Institute,188300 St.\,Petersburg, Russia.}

\begin{abstract}
The article reviews the physics of Anderson localization on random regular graphs (RRG) and its connections to many-body localization (MBL) in disordered interacting systems. Properties of eigenstate and energy level correlations in delocalized and localized phases, as well at criticality, are discussed. In the many-body part,  models with short-range and power-law interactions are considered, as well as the quantum-dot model representing the limit of the ``most long-range'' interaction. Central themes---which are common to the RRG and MBL problems---include ergodicity of the delocalized phase, localized character of the critical point, strong finite-size effects, and fractal scaling of eigenstate correlations in the localized phase. 

\end{abstract}

\begin{keyword}
Anderson localization, random regular graphs, many-body localization, ergodicity, critical behavior, eigenfunction and energy level statistics
\end{keyword}

\end{frontmatter}

\tableofcontents

\section{Introduction}
\label{sec:introduction}

More than sixty years ago, the celebrated Anderson's paper \cite{anderson58} marked a discovery of Anderson localization, which has greatly influenced the development of the condensed matter physics since then. Transport and localization properties of quantum particles subjected to a random potential or other types of disorder have been systematically explored. It was found that transitions between localized and delocalized phases---known as Anderson transitions---show a remarkably rich physics depending on spatial dimensionality, symmetries, and topologies \cite{evers08}. 

More recently, the physics of the many-body localization (MBL) in disordered interacting systems \cite{gornyi2005interacting,basko2006metal}  attracted a great deal of research attention. The MBL addresses localization or delocalization in highly excited states of interacting many-body systems (i.e., states with a finite energy density). One can thus consider the MBL as a generalization of Anderson localization from single-particle to many-body setting. We refer the reader to Refs.~\cite{nandkishore15,Alet2018a,abanin2019colloquium,gopalakrishnan2020dynamics} for recent reviews on various aspects of the MBL. 

The extension from a non-interacting to an interacting problem strongly complicates the theoretical investigation---analytical as well as computational. 
On the analytical side, the approaches to the MBL transition in Refs.~\cite{gornyi2005interacting,basko2006metal} (and in the later closely related paper \cite{Ros2015420}) were based on the analysis of the corresponding perturbative expansion.  Later, it was shown that matrix elements of Hartree-Fock type (which were discarded in Refs.~\cite{gornyi2005interacting,basko2006metal}) essentially enhance delocalization and parametrically shift the transition point  as given by the perturbative analysis, due to the effect of spectral diffusion \cite{gornyi2017spectral}.
%\cite{gornyi2016manyspectral}. 
The schemes based on the analysis of the perturbation theory do not include, however, effects related to exponentially rare regions of anomalously high or anomalously weak disorder. As was understood in recent years \cite{Agarwal2016a,Thiery2017a}, such regions may likely play an important role for the scaling of the MBL transition and the corresponding critical behavior in the thermodynamic limit of a large system.  Based on these ideas, several phenomenological renormalization-group schemes have been proposed \cite{Dumitrescu2019a,morningstar2020a} that were argued to describe the scaling at the MBL transition. 

On the computational side, exact diagonalization (ED) studies are restricted to systems with $\approx 20$ qubit-like binary degrees of freedom (spins, orbitals of fermions or hard-core bosons, Josephson qubits, etc), with the corresponding Hilbert-space size being $ \sim 2^{20} \sim 10^6$. While numerical simulations on systems of this size do provide a clear evidence of the MBL transition 
\cite{luitz2015many,mace19multifractal}, the corresponding finite-size scaling analysis yields exponents that are inconsistent with the Harris criterion. This is a clear indication of the fact that the system sizes that are accessible by the ED are way too small for the purpose of observing the ultimate large-system critical behavior.
Indeed, it has been estimated that this requires spin chains of the length $L \gtrsim 50$ -- 100 \cite{Khemani2017a,PhysRevX.7.021013,Panda2020a}. Quantum dynamics around the MBL transition in systems of this size can be studied by means of numerical approaches based on matrix product states, and the results are indeed in general agreement with analytical expectations  
\cite{Doggen2018a,Doggen2019a,doggen2020slow,doggen2021many}. These simulations can, however, only probe the dynamics at moderately long time scales. 

Since a controllable analytical treatment of  the MBL problem is notoriously difficult (and still remains a big challenge for future work), simplified models that are amenable to such a treatment are highly useful. The Anderson localization problem on random regular graphs (RRG) serves as such a toy-model of the MBL problem. 

An RRG is a finite-size graph that has locally the structure of a tree with a fixed coordination number $m+1$ but---contrary to a finite tree---does not have boundary. This means that an RRG does have loops but this loops are of large scale. The structure of an RRG mimics that of a graph induced by a Hamiltonian of an interacting many-body system in the corresponding Hilbert space. Specifically, basis states of a many-body system chosen as eigenstates of the non-interacting part of the Hamiltonian (which can be straightforwardly diagonalized) correspond to vertices of the RRG, while interaction-induced couplings between them correspond to links of an RRG. The idea of a connection between a many-body problem and localization on a tree (a Bethe lattice) was put forward in Ref.~\cite{altshuler1997quasiparticle} where decay of a hot quasiparticle in a quantum dot (at zero temperature) was addressed.  As was shown by later work, tree-like graphs can be viewed as approximately modelling the  Hilbert-space structure of a finite many-body system in a much more general context. Also, it has been understood that the appropriate finite graphs are not really trees (i.e. not finite portions of a Bethe lattice) but rather RRG. 
The essential difference between finite Anderson localization on Bethe lattices and RRG has been demonstrated in Refs.~\cite{tikhonov2016fractality,sonner17}; we will only consider RRG in this review, in view of their connection to the MBL problem.

As a first indication of an analogy between the RRG and MBL models, one can inspect the size of the Hilbert space. For a spin chain of length $L$ (a paradigmatic MBL-type system), the Hilbert-space size scales with the length as $2^L$. The RRG model has the same property of an exponential growth of the Hilbert-space volume $N$ with the ``linear size'' $L$ (the distance between the most distant sites on the RRG lattice), $N \sim m^L$. Another key property---which is related to the exponential growth of the lattice---is the suppression of small-scale loops, i.e. the locally tree-like character of the graph.
A similarity between the interaction-induced structure in the Hilbert space of a many-body problem and the RRG structure has recently triggered a surge of interest in Anderson localization on RRG \cite{biroli2012difference,de2014anderson,tikhonov2016anderson,garcia-mata17,metz2017level,Biroli2017,kravtsov2018non,biroli2018,PhysRevB.98.134205,tikhonov19statistics,tikhonov19critical,PhysRevResearch.2.012020,tikhonov2020eigenstate}. It is fair to say that by now we have a rather complete understanding of key properties of this model, including the position of the thermodynamic-limit transition point $W_c$, values of critical exponents and of various observables characterizing properties of eigenstates and energy levels. 

Of course, the RRG model is distinct from a genuine many-body problem.  Specifically, the RRG model discards correlations between matrix elements in the Hilbert space of the many-body problem, thus representing its ``simplified version''. Existence of such correlations is clear from the fact that the number of independent parameters in a many-body Hamiltonian is much smaller than the number of non-zero matrix elements.  Nevertheless, there are remarkable analogies between the localization transitions in the RRG model and in genuine MBL models, which show up in a variety of key physical properties. 
In particular, the most salient qualitative properties of the Anderson-localization transition on RRG include:
\begin{itemize}
\item[(i)]  the critical point of the Anderson transition has a localized character,
\item[(ii)] 
 there are strong finite-size effects which manifest themselves in a drift of the apparent transition point towards stronger disorder with increasing system size, 
 \item[(iii)] 
  the Hilbert-space ``correlation volume'' increases exponentially when the localization transition is approached, 
 \item[(iv)] 
 the delocalized phase is ergodic (which means the Wigner-Dyson (WD) level statistics, the $1/N$ asymptotic scaling of the inverse participation ratio and further associated propertes). 
 \end{itemize}
These properties of the RRG model have been proven analytically \cite{tikhonov19statistics} (see also Refs.~\cite{mirlin1991universality,fyodorov1991localization,fyodorov1992novel} where a related model of sparse random matrices (SRM) was investigated) and also verified numerically \cite{tikhonov2016anderson,garcia-mata17,metz2017level,tikhonov19statistics,biroli2018}. 
Analytical arguments and numerical simulations for the MBL models  lead to analogous conclusions. (The analytical arguments in the case of MBL are, however, less rigorous than for RRG, as was pointed out above.) There are thus strong connections between the RRG and MBL problems. These connections become even closer for models with long-range interaction (decaying sufficiently slowly as a power-law of distance), for which rare regions do not play any essential role, see Ref.~\cite{tikhonov18}.

In this review article, we first overview recent advances on the Anderson model on RRG (Sec.~\ref{sec:RRG}), and then discuss its connections with MBL-type problems (Sec.\ref{sec:MBL}). In the many-body part of the review,  models with short-range and with long-range (power-law) interactions are considered (including the quantum-dot model, which is the limiting case of the ``most long-range'' interaction).

%%%%%%%%  Not needed in the introduction; to be moved (with modification of the wording)  to the MBL section

\vspace{1cm}

%%%%%%%%

%%%%%%%%%%%%%%%%%%%%%%%%%%%%%%%%%%%%%%%%%%%%%%%%%%%%%%%%%%%%%%%%%%%%%%%%%%%%%%%%%%%%%%%%%%%%%%%%%%%%%%%%%%%%%%%%%%%%%%%%%%%%%%%%%%%%%%%%%%%%%%%%%%%%%%%%%%%%%%%%%%%%%%%%%%%%%%%%%%%%%%%%%%%%%%%%%%%%%%%%%%%%%%%%%
%%%%%%%%%%%%%%%%%%%%%%%%%%%%%%%%%%%%%%%%%%%%%%%%%%%%%%%%%%%%%%%%%%%%%%%%%%%%%%%%%%%%%%%%%%%%%%%%%%%%%%%%%%%%%%%%%%%%%%%%%%%%%%%%%%%%%%%%%%%%%%%%%%%%%%%%%%%%%%%%%%%%%%%%%%%%%%%%%%%%%%%%%%%%%%%%%%%%%%%%%%%%%%%%%
%%%%%%%%%%%%%%%%%%%%%%%%%%%%%%%%%%%%%%%%%%%%%%%%%%%%%%%%%%%%%%%%%%%%%%%%%%%%%%%%%%%%%%%%%%%%%%%%%%%%%%%%%%%%%%%%%%%%%%%%%%%%%%%%%%%%%%%%%%%%%%%%%%%%%%%%%%%%%%%%%%%%%%%%%%%%%%%%%%%%%%%%%%%%%%%%%%%%%%%%%%%%%%%%%

\section{Anderson localization on Random Regular Graphs}
\label{sec:RRG}

We study non-interacting spinless fermions hopping over RRG with connectivity $p = m+1$  in a potential disorder,
\begin{equation}
\label{H}
\hat H =t\sum_{\left<i, j\right>}\left(\hat c_i^\dagger \hat c_j + \hat c_j^\dagger \hat c_i\right)+\sum_{i=1} \epsilon_i \hat c_i^\dagger \hat c_i\,,
\end{equation}
where the first sum runs over the nearest-neighbor sites of the RRG. The energies $\epsilon_i$ are independent random variables distributed uniformly on $[-W/2,W/2]$. The hopping $t$ can be set to be $t=1$ without loss of generality, as will be done in the most of the article.

%%%%%%%%%%%%%%%%%%%%%%%%%%%%%%%%%%%%%%%%%%%%%%%%%%%%%%%%%%%%%%%%%%%%%%%%%%%%%%%%%%%%%%%%%%%%%%%%%%%%%%%%%%%%%%%%%%%%%%%%%%%%%%%%%%%%%%%%%%%%%%%%%%%%%%%%%%%%%%%%%%%%%%%%%%%%%%%%%%%%%%%%%%%%%%%%%%%%%%%%%%%%%%%%%
%%%%%%%%%%%%%%%%%%%%%%%%%%%%%%%%%%%%%%%%%%%%%%%%%%%%%%%%%%%%%%%%%%%%%%%%%%%%%%%%%%%%%%%%%%%%%%%%%%%%%%%%%%%%%%%%%%%%%%%%%%%%%%%%%%%%%%%%%%%%%%%%%%%%%%%%%%%%%%%%%%%%%%%%%%%%%%%%%%%%%%%%%%%%%%%%%%%%%%%%%%%%%%%%%

\subsection{Field-theoretical description}
\label{sec:field-theory}
Statistical properties of observables in this model can be expressed in terms of certain functional integrals, either in supersymmetric \cite{mirlin1991localization,mirlin1991universality,tikhonov19statistics} or in the replicated version \cite{PhysRevB.37.3557,PhysRevE.90.052109,PhysRevLett.117.104101}. These approaches are equivalent 
for many purposes (see, for example, their comparison in Ref. \cite{gruzberg96} in a related context). On the other hand, the supersymmetric approach is preferential when properties of individual eigenstates or level statistics at the scale of level spacing is studied.  In view of this, we will use the supersymmetric  formulation in what follows.

The model defined by Eq. (\ref{H}) has two sources of disorder: randomness in the structure of the underlying graph and fluctuations of on-site energies $\epsilon_i$. Various disorder-averaged properties of a disordered system can be derived in terms of averaged products of Green functions.  Such a derivation in the framework of the supersymmetric field theory was performed, with applications to the level statistics and to the scaling of the inverse participation ratio (IPR)  in the SRM model, in Refs.\cite{mirlin1991universality,fyodorov1991localization}. In general, averaged products of retarded and advanced Green functions (with energies $E+\omega/2$ and  $E-\omega/2$, respectively) can be evaluated as  superintegrals of the form\cite{mirlin1991localization}
\be
\label{superint}
 \int\prod_k[d\Phi_k] e^{-\mathcal{L}_H(\Phi)}U(\Phi),
\ee
where the preexponential factor $U(\Phi)$ represents the quantity in question and 
$$[d\Phi_k]=dS_{k,1}^{(1)}dS_{k,1}^{(2)}d\chi_{k,1}^*d\chi_{k,1} dS_{k,2}^{(1)}dS_{k,2}^{(2)}d\chi_{k,2}^*d\chi_{k,2}$$ 
is the supervector integration measure. Here $S$ stay for real commuting and $\chi$ for anticommuting variables. The action $\mathcal{L}_H(\Phi)$ is given by
\be
\label{superaction-rrg}
\mathcal{L}_H(\Phi) =  - \frac{i}{2}\sum_{ij}\Phi_i^\dagger\hat{\Lambda}\left\{ \left [E+ \left (\frac{\omega}{2}+i\eta \right)\hat{\Lambda} \right]\delta_{ij}-H_{ij}\right\}\Phi_j,
\ee
where $\eta >0$ is an infinitesimal imaginary part of frequency. Further, $\hat{\Lambda}$ is a diagonal supermatrix with the first four components (retarded sector) equal to $+1$ and the last four components (advanced sector) equal to $-1$.  

To get the supersymmetric partition function of the RRG model \cite{tikhonov19statistics}, we perform the averaging of the weight $e^{-\mathcal{L}_H(\Phi)}$  over the distribution matrix elements of the Hamiltonian. We consider an ensemble of $N\times N$ Hamiltonians with the following joint distribution of diagonal $H_{ii}$  and off-diagonal $H_{ij} = H_{ji} = A_{ij}t_{ij}$ matrix elements:
\bea
&& {\cal P}(\{H_{ii}\}, \{A_{ij}\}, \{t_{ij}\}) =  \prod_i \gamma (H_{ii}) \nonumber \\
&& \hspace*{1cm} \times  \prod_{i < j} \left [ \left( 1 - \frac{p}{N} \right) \delta(A_{ij}) + \frac{p}{N} \delta(A_{ij} - 1) \right] \nonumber \\
&& \hspace*{1cm} \times   \prod_i  \delta \left ( \sum_{j \ne i} A_{ij} - p \right) \prod_{i<j} h(t_{ij}) .
\label{rrg-distribution}
\eea
Here $A_{ij}$ is the adjacency matrix, and the delta-function in the last line of Eq.~\eqref{rrg-distribution} ensures that the coordination number of each vertex is $p$. For the purpose of generality, we have included in Eq.~(\ref{rrg-distribution}) 
an arbitrary distribution $h(t)$ of non-zero hopping matrix elements. For a RRG model with fixed hoppings $t=1$, we have $h(t) = \delta(t-1)$. Finally, we  decouple the integrations over variables $\Phi_i$ associated with different sites. This is done by means of a functional generalization of the Hubbard-Stratonovich transformation. As a result, we obtain the expression for physical observables in terms of an integral over functions $g(\Phi)$:  
\be
\langle {\cal O} \rangle = \int Dg \: U_{\cal O}(g) e^{-N\mathcal{L}(g)}.
\label{g-funct-integral}
\ee
The integration $\int Dg$ runs over functions of a supervector $g(\Phi)$ with the action $N\mathcal{L}(g)$ where 
\be
\mathcal{L}(g)= \frac{m+1}{2}\int d\Psi d\Psi'g(\Psi)C(\Psi,\Psi')g(\Psi')- \ln\int d\Psi \: F^{(m+1)}_g(\Psi),
\label{Lg}
\ee
with $m = p-1$ and 
\be
F^{(s)}_g(\Psi) = \exp  \left\{ \frac{i}{2} E\Psi^\dagger \hat{\Lambda} \Psi + 
\frac{i}{2}\left(\frac{\omega}{2} + i \eta \right)
\Psi^\dagger\Psi \right\} \tilde{\gamma}(\frac{1}{2}\Psi^\dagger \hat{\Lambda} \Psi ) g^s(\Psi).
\label{Fsg}
\ee
Here the function $\tilde{\gamma}(z)$ is the Fourier transform of the distribution $\gamma(\epsilon)$ of on-site energies, $\tilde{\gamma}(z) = \int d\epsilon\: e^{-i\epsilon z} \gamma(\epsilon)$. 
Finally, $C(\Psi, \Psi')$ is a kernel of an integral operator inverse to that with the kernel $\tilde{h}(\Phi^\dagger \hat{\Lambda}\Psi)$, where $\tilde{h}(z)$ is the Fourier transform of the distribution $h(t)$ of hoppings;  for an RRG model with fixed hoppings $t=1$ we have $\tilde{h}(z) = e^{-iz}$. 

Since the action is proportional to $N$, see Eq. (\ref{g-funct-integral}), in the limit of large $N$ this theory can be treated via the saddle-point approximation. The saddle-point configuration $g_0(\Psi)$ of the action  is determined by varying Eq.~(\ref{Lg}) with respect to $g$, which yields the equation
\be
\label{scvector}
g_0(\Psi)=\frac{\int d\Phi \: \tilde{h}(\Phi^\dagger \hat{\Lambda} \Psi) F^{(m)}_{g_0}(\Phi)}{\int d\Phi \: F^{(m+1)}_{g_0}(\Phi)}.
\ee
This equation is equivalent to the self-consistency equation for the same model but defined on an infinite Bethe lattice. This property is a manifestation of the fact that, with probability unity, RRG has locally (in the vicinity of any of its sites) a structure of a tree with fixed connectivity $p$. [More formally, at large $N$ and $\kappa<1/2$, a $1-o(1)$  portion of RRG nodes have their $\kappa\log_m N$-neighbourhood loopless.]  

Due to supersymmetry, the denominator in Eq.~(\ref{scvector}) is thus equal to unity, so that the saddle-point equation reduces to
\be
\label{sc1}
g_0(\Psi)= \int d\Phi \: \tilde{h}(\Phi^\dagger \hat{\Lambda} \Psi) F^{(m)}_{g_0}(\Phi).
\ee
Equation (\ref{sc1}) is identical to the self-consistency equation describing the model on an infinite Bethe lattice, as derived within the supersymmetry formalism in Ref.~\cite{mirlin1991localization}. For symmetry reasons, the saddle-point solution is a function of two invariants
\be
g_0(\Psi)=g_0(x,y); \quad x=\Psi^\dagger\Psi, \quad y=\Psi^\dagger \hat{\Lambda} \Psi
\label{g0xy}
\ee and the Eq. (\ref{sc1}) reduces to a non-linear integral equation for $g_0(x,y)$ (recall that for the model with non-random hoppings, as defined in Eq.~(\ref{H}), one has $\tilde{h}(z)= e^{-iz}$). For a purely imaginary frequency ($\omega = 0$), the function $g_0(x,y)$ has an important physical interpretation\cite{mirlin1991localization}. Specifically, it equals the Fourier-Laplace transform of the joint probability distribution $f^{(m)}(u', u'')$ of real and imaginary parts of  local Green function,
\be
\label{g0t}
g_0(x,y)=\int du'\int du'' f^{(m)}(u',u'')e^{\frac{i}{2}\left(u'y+iu''x\right)},
\ee
where $G_A^{(m)} (0, 0; E) = \langle 0 | (E- {\cal H} -i\eta)^{-1} | 0 \rangle = u' +i u''$ computed for a slightly modified lattice, with the site $0$ having only $m$ neighbors. When rewritten in terms of $f^{(m)}(u',u'')$,  Eq.~(\ref{sc1}) becomes the self-consistency equation of Abou-Chacra et al., Ref.~\cite{abou1973selfconsistent}.

%%%%%%%%%%%%%%%%%%%%%%%%%%%%%%%%%%%%%%%%%%%%%%%%%%%%%%%%%%%%%%%%%%%%%%%%%%%%%%%%%%%%%%%%%%%%%%%%%%%%%%%%%%%%%%%%%%%%%%%%%%%%%%%%%%%%%%%%%%%%%%%%%%%%%%%%%%%%%%%%%%%%%%%%%%%%%%%%%%%%%%%%%%%%%%%%%%%%%%%%%%%%%%%%%
\begin{figure}[tbp]
\minipage{0.5\textwidth}\includegraphics[width=\textwidth]{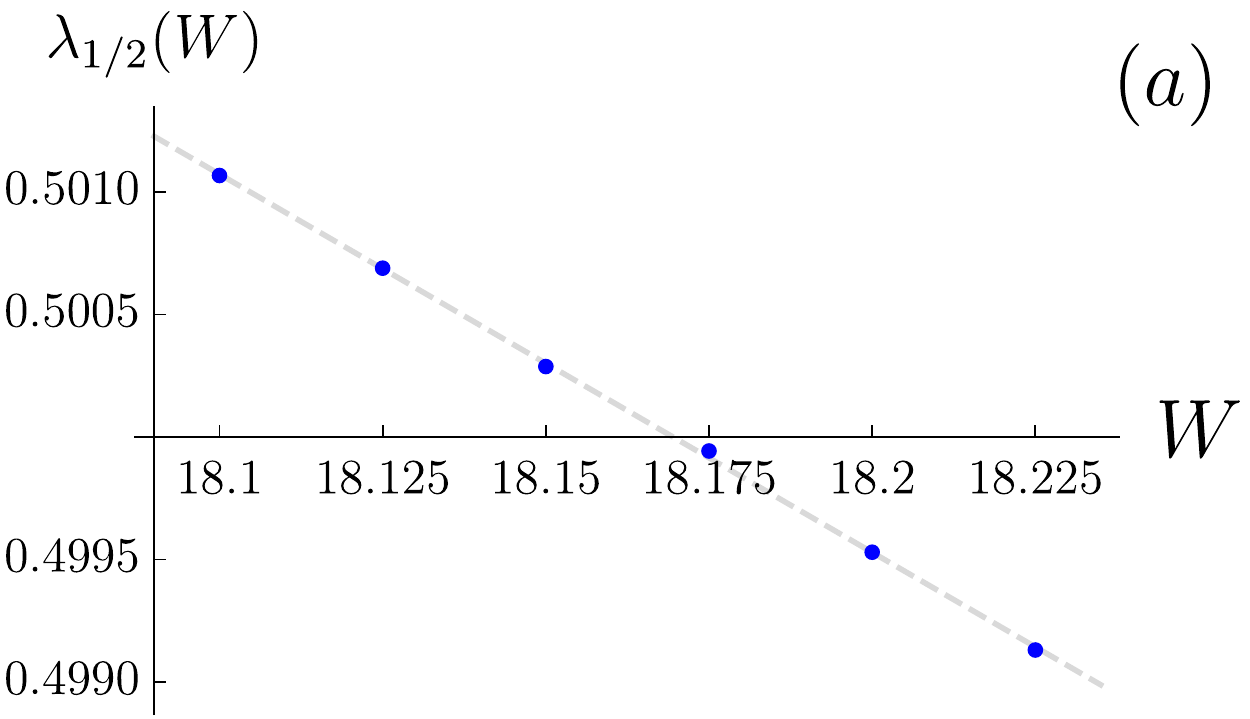}\endminipage
\minipage{0.5\textwidth}\includegraphics[width=\textwidth]{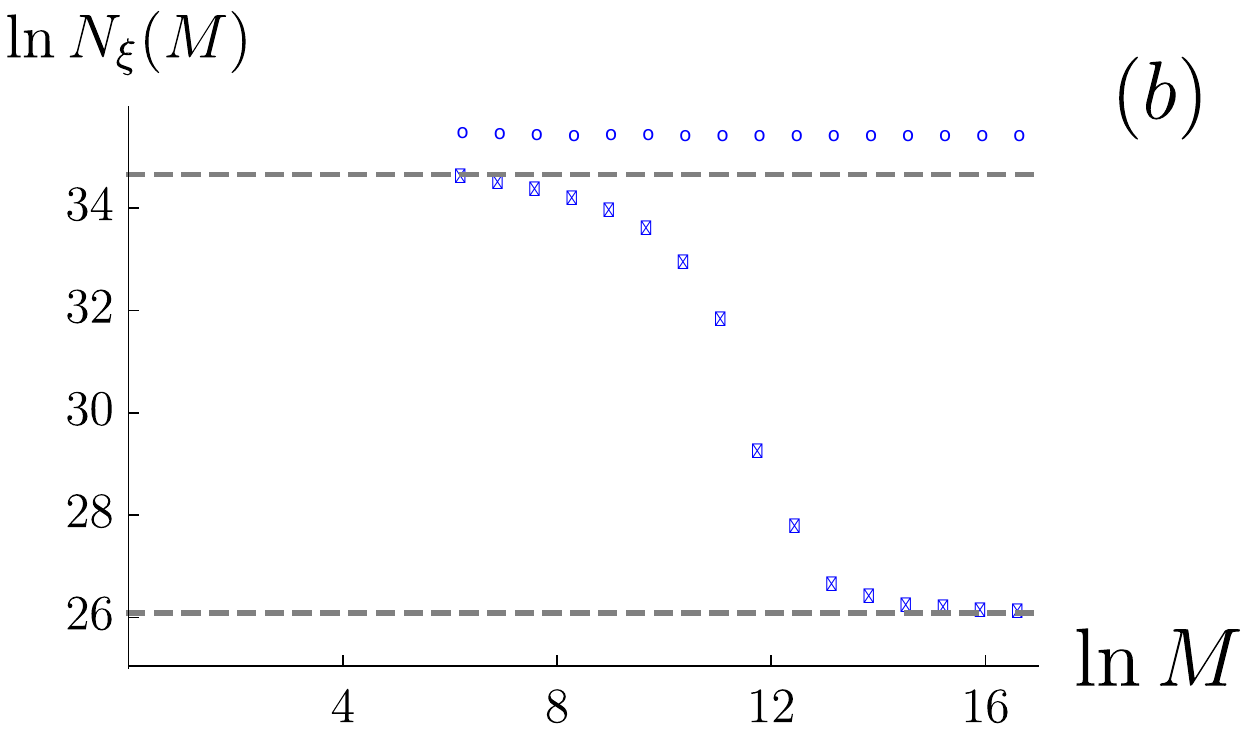}\endminipage
\caption{(a) The disorder-dependence of the largest eigenvalue $\lambda_{1/2}(W)$ of the operator $L_{1/2}$ in Eq. (\ref{eq:linear}) in the vicinity of the transition point. The critical disorder is determined by $\lambda_{1/2} = 1/2$, which gives $W_c = 18.17 \pm 0.01$. (b) $N_{\xi}(M)=\exp\langle - \ln \Im G \rangle_M$ evaluated at pool size $M$ at $\eta=10^{-15}$ for $W=17$ (filled symbols) and $W=19$ (empty symbols).  For $W=17$ (delocalized phase),  $N_{\xi}(M)$ evolves from $\eta^{-1}$ (upper dashed line) to $N_\xi \equiv N_{\xi}(M\to\infty)$ (lower dashed line) which equals the true correlation volume. For $W=19$ (localized phase) $N_{\xi}(M)$ remains constant of order $\eta^{-1}$ for all pool sizes. From Ref.~\cite{tikhonov19critical}.}
\label{fig:pg}
\end{figure}
%%%%%%%%%%%%%%%%%%%%%%%%%%%%%%%%%%%%%%%%%%%%%%%%%%%%%%%%%%%%%%%%%%%%%%%%%%%%%%%%%%%%%%%%%%%%%%%%%%%%%%%%%%%%%%%%%%%%%%%%%%%%%%%%%%%%%%%%%%%%%%%%%%%%%%%%%%%%%%%%%%%%%%%%%%%%%%%%%%%%%%%%%%%%%%%%%%%%%%%%%%%%%%%%%

A closely related object is the function $g_0^{(m+1)}(\Psi)$ which is expressed via $g_0(\Psi)$ as
\be
\label{g0m1}
g_0^{(m+1)}(\Psi)= \int d\Phi \: \tilde{h}(\Phi^\dagger \hat{\Lambda} \Psi) F^{(m+1)}_{g_0}(\Phi).
\ee
The function $g_0^{(m+1)}(x,y)$ is  the Fourier-Laplace transform of the joint probability distribution $f^{(m+1)}(u', u'')$ of real and imaginary parts of  local Green function at any site of the undeformed Bethe lattice. The self-consistency equation, Eq. (\ref{sc1}) can thus be presented in the form
\be
\label{pool_sc}
G^{(m)}\stackrel{d}{=}\frac{1}{E-i\eta-\epsilon-\sum_{i=1}^{m} G_i^{(m)}},
\ee
where the symbol $\stackrel{d}{=}$ denotes the equality in distribution (assuming $G_i^{(m)}$ to be independent copies of $G^{(m)}$). The distribution of the local Green function $G^{(m+1)}$ on an original lattice can be recovered from an auxiliary relation
\be
\label{pool_simple}
G^{(m+1)}\stackrel{d}{=}\frac{1}{E-i\eta-\epsilon-\sum_{i=1}^{m+1} G_i^{(m)}}.
\ee
The distributions of $G^{(m)}$ and $G^{(m+1)}$ are qualitatively
very similar. Below we use a short
notation $G \equiv G^{(m)}$. %%%%%%%%%%%%%%%%%%%%%%%%%%%%%%%%%%%%%%%%%%%%%%%%%%%%%%%%%%%%%%%%%%%%%%%%%%%%%%%%%%%%%%%%%%%%%%%%%%%%%%%%%%%%%%%%%%%%%%%%%%%%%%%%%%%%%%%%%%%%%%%%%%%%%%%%%%%%%%%%%%%%%%%%%%%%%%%%%%%%%%%%%%%%%%%%%%%%%%%%%%%%%%%%%
%%%%%%%%%%%%%%%%%%%%%%%%%%%%%%%%%%%%%%%%%%%%%%%%%%%%%%%%%%%%%%%%%%%%%%%%%%%%%%%%%%%%%%%%%%%%%%%%%%%%%%%%%%%%%%%%%%%%%%%%%%%%%%%%%%%%%%%%%%%%%%%%%%%%%%%%%%%%%%%%%%%%%%%%%%%%%%%%%%%%%%%%%%%%%%%%%%%%%%%%%%%%%%%%%

\subsection{Localization transition and critical behavior}
\label{sec:loctr}
Saddle-point evaluation of the functional integral in Eq. (\ref{g-funct-integral}) provides an interesting perspective on the Anderson localization transition as a spontaneous symmetry breaking phenomenon. In the localized phase, the saddle point solution has the symmetry of the equation, $g_0(x,y)=g_0(y)$. To be more precise, $g_0(x,y)$ depends on the variable $x$ only on the scale $\sim \eta^{-1}$, due to the term with $\eta$ in the action that breaks the symmetry explicitly. Thus, the integral (\ref{g-funct-integral}) in the localized phase (as well as at the critical point) is determined by a contribution of a unique saddle point $g_0$. In the delocalized phase, dependence on the variable $x$ survives even in the limit of $\omega, \eta\to 0$. As a result, a manifold of saddle-points emerges, signifying spontaneous symmetry breaking with the function $g_0(x, y)$ playing the role of an order parameter. In this situation, the  integral (\ref{g-funct-integral})  runs over the manifold of saddle points.

The approach to locating the transition point was first established in Ref.~\cite{abou1973selfconsistent}; equivalent results were later obtained within the supersymmetry formalism in Ref.~\cite{mirlin1991localization}. It amounts to evaluating the stability of the real solution to the self--consistency equation (obtained by by setting $\eta =0$ in Eq. (\ref{pool_sc})) with respect to introducing a small imaginary part. The stability (instability) of the real solution to such a perturbation implies that the system is in the localized (respectively, delocalized) phase. The critical disorder $W_c$ marks the transition between these two types of behavior. The first step in computing $W_c$ is thus to find a real ($\eta=0$) solution $\mathcal{P}_0(G)$ to the distributional Eq. (\ref{pool_sc}). The criterion of its stability can then be expressed in terms of the largest eigenvalue of a certain integral operator, whose kernel can be written in terms of $\mathcal{P}_0(G)$. For $m=2$ (more general form of the operator $L_\beta$ can be found in Refs.~\cite{abou1973selfconsistent,mirlin1991localization}):
\be
L_\beta(x,y)=\frac{|x|^{2\beta}}{y^2}\int d\epsilon\, \gamma(\epsilon)\,  \mathcal{P}_0(y^{-1}-x-\epsilon)
\label{eq:linear}.
\ee
 The real solution is stable if and only if the largest eigenvalue $\lambda_\beta$ of the operator $L_{1/2}$ is smaller than $1/m$.

The largest eigenvalue $\lambda_{1/2}$ as a function of disorder is shown in Fig.~\ref{fig:pg}a. Solving the equation $m\lambda_{1/2}(W)=1$ (with $m=2$), we find \cite{tikhonov19critical}
 \be
 \label{Wc}
 W_c=18.17\pm 0.01.
 \ee
A very close value of $W_c$ was found by a similar method in Ref. \cite{parisi2019anderson}.
%%%%%%%%%%%%%%%%%%%%%%%%%%%%%%%%%%%%%%%%%%%%%%%%%%%%%%%%%%%%%%%%%%%%%%%%%%%%%%%%%%%%%%%%%%%%%%%%%%%%%%%%%%%%%%%%%%%%%%%%%%%%%%%%%%%%%%%%%%%%%%%%%%%%%%%%%%%%%%%%%%%%%%%%%%%%%%%%%%%%%%%%%%%%%%%%%%%%%%%%%%%%%%%%%
\begin{figure}[tbp]
\minipage{0.5\textwidth}\includegraphics[width=\textwidth]{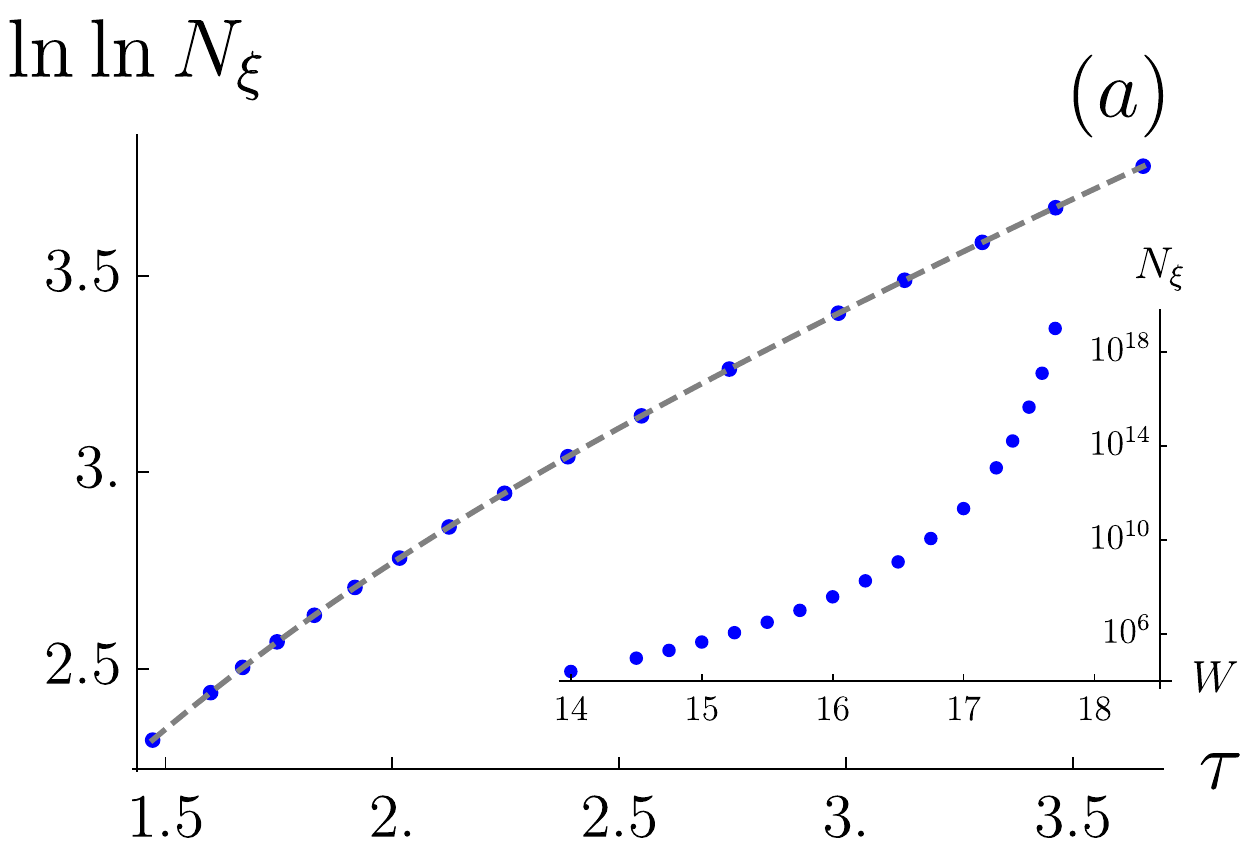}\endminipage
\minipage{0.5\textwidth}\includegraphics[width=\textwidth]{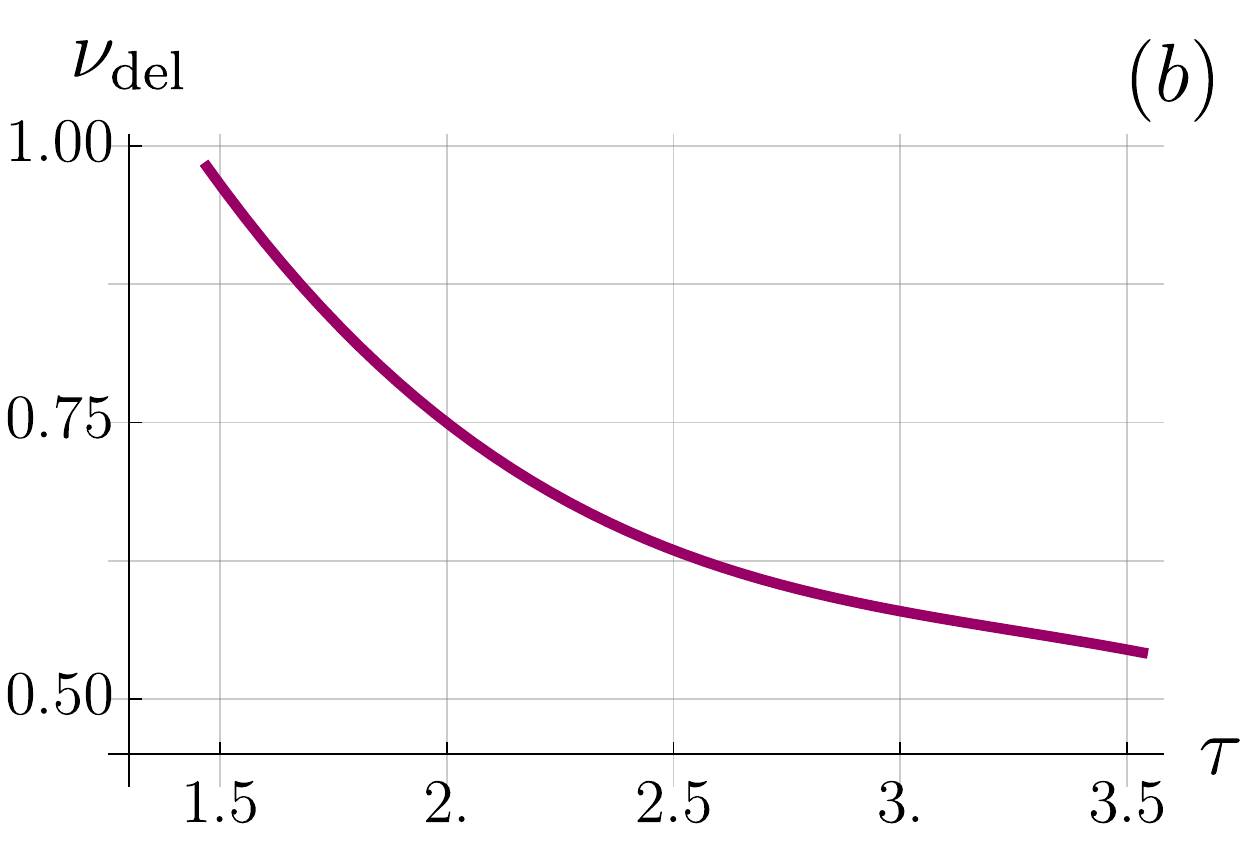}\endminipage
\caption{Critical behavior at the Anderson transition on RRG. (a) Double logarithm of the correlation volume $N_\xi$ as a function of $\tau=-\ln(1-W/W_c)$ (with $W_c=18.17$). Inset: the same data in $(W,\,\ln N_\xi)$ coordinates. (b) Flowing correlation-length exponent $\nu_{\rm del}(\tau)=\partial \ln \ln N_\xi/\partial\tau$. The limiting value, $\nu_{\rm del}(\tau \to \infty)$, gives the critical index of the correlation length $\nu_{\rm del} = 1/2$, Eq.~(\ref{xi}). From Ref.~\cite{tikhonov19critical}.  }
\label{fig:cvol}
\end{figure}
%%%%%%%%%%%%%%%%%%%%%%%%%%%%%%%%%%%%%%%%%%%%%%%%%%%%%%%%%%%%%%%%%%%%%%%%%%%%%%%%%%%%%%%%%%%%%%%%%%%%%%%%%%%%%%%%%%%%%%%%%%%%%%%%%%%%%%%%%%%%%%%%%%%%%%%%%%%%%%%%%%%%%%%%%%%%%%%%%%%%%%%%%%%%%%%%%%%%%%%%%%%%%%%%%

Let us now discuss the critical behavior in the delocalized phase. As has been already discussed above, the purely real solution of the self-consistency equation ($\eta=0$) is unstable to introduction of finite $\eta$ in the delocalized phase where a non-trivial distribution $\mathcal{P}(\Re G, \Im G)$ emerges.  It can be derived from the function $g_0(x,y)$ which acquires dependence on the variable $x$ on a scale $x \sim N_\xi$. This scale diverges exponentially when $W$ approaches the critical value $W_c$ and has the meaning of the correlation volume which can be related to the correlation length $\xi$:
\be 
N_\xi \sim m^\xi \,.
\label{Nxi1}
\ee

In order to evaluate the correlation volume numerically, Eq. (\ref{pool_sc}) was solved 
in Ref.~\cite{tikhonov19critical}
via the pool
method, also known as population dynamics (PD). In this approach, the distribution $\mathcal{P}(G)$ is represented by a large sample of random variables and Eq. (\ref{pool_sc}) is iterated until convergence. The resulting sample of $M$ variables is distributed (at $M\to\infty$) according to a desired distribution. The PD calculation is conducted at a finite (although large) pool size $M$ and a finite (although small) imaginary part of the energy, $\eta$. The resulting distribution can be characterized by ``effective correlation volume'' $N_\xi(\eta,M)$ defined as the PD result for $\exp \langle - \ln \Im G \rangle$, where $\langle .. \rangle$ includes averaging over many iterations after the approximate convergence is reached. The correlation volume $N_\xi$ is given by the double limit
\be
\label{N-xi-true}
N_\xi=\lim_{\eta\to0}\lim_{M\to\infty}N_\xi(\eta,M).
\ee

The role of the pool size, is illustrated by the Fig. \ref{fig:pg}b, where the dependence of $N_\xi(\eta,M)$ on $M$ for $W=17$ and fixed $\eta = 10^{-15}$. This quantity evolves from $\eta^{-1}$ (characteristic for the localized phase) at small pool sizes $N_\xi(\eta,M)$ to the value $N_\xi(\eta,\infty)$ at large $M$. At sufficiently small $\eta$, such that $\eta \ll N_\xi^{-1}$, the behavior of $N_\xi(\eta,M)$ at large $M$ is essentially independent on  $\eta$, illustrating the spontaneous symmetry breaking phenomena. This data should be contrasted to the ones for the localized side of the transition, exemplified by the results for $W=19$, demonstrating the stability of the localized phase with respect to $\eta$.

Apart from studying the specific averages (such as correlation volume $N_{\xi}$), we can characterize the entire function $\mathcal{P}(G)$ and  determine, in particular, the distribution of the LDOS $\rho = (1/\pi) {\rm Im}\, G$. On the delocalized side close to the transition it is expected to be of the following form
\be
\label{Pnu}
\mathcal{P}(\rho) \sim N_\xi^{-1/2} \rho^{-3/2}, \qquad N_\xi^{-1}  < \rho < N_\xi;
\ee
outside of this range the probability is strongly suppressed. This behavior of $\mathcal{P}(\rho)$ is essentially the same as the one found in the $\sigma$ model on the Bethe lattice \cite{mirlin94a,mirlin94b}. Equation (\ref{Pnu}) can be derived via analysis of the symmetry-broken solution near the critical point in both Anderson \cite{mirlin1991localization} and $\sigma$-model \cite{zirnbauer1986localization,efetov1987density}. Main features of this solution are connected to properties of the largest eigenvalue $\lambda_\beta$ of the operator (\ref{eq:linear}) via equation $m\lambda_\beta = 1$. For $W$ below (and close to) $W_c$ the solution is $\beta = 1/2 \pm i\sigma$, with $\sigma \sim (W_c-W)^{-1/2}$. The real part 1/2 of the exponent $\beta$ translates into the exponent 3/2 in the LDOS distribution (\ref{Pnu}), while the imaginary part determines the scale $N_\xi \sim \exp(\pi / \sigma)$ which controls the range of validity of the power-law distribution (\ref{Pnu}). Thus, the critical behavior of the correlation volume  reads
\be
\ln N_\xi \sim (W_c-W)^{-1/2} \,.
\label{N-xi-crit}
\ee
According to Eqs.~\eqref{N-xi-crit} and \eqref{Nxi1}, the critical index of the correlation length on the delocalized side of the transition equals $\frac{1}{2}$:
\be
\xi = \frac{\ln N_\xi}{\ln m} \sim (W_c-W)^{-\nu_{\rm del}} \,; \qquad \nu_{\rm del} = \frac{1}{2} \,.
\label{xi}
\ee

The disorder dependence of the correlation volume $N_\xi$ obtained according to Eq.~(\ref{N-xi-true}) is shown in Fig. \ref{fig:cvol}a.  More specifically, we plot $\ln\ln N_\xi$ as a function of $\tau\equiv-\ln(1-W/W_c)$.  In this representation, we evaluate
\be
\nu_{\rm del} (\tau) = \frac{\partial \ln \ln N_\xi}{\partial \tau},
\ee
and asymptotic value $\nu_{\rm del}(\infty)$ yields the critical index $\nu_{\rm del}$. The $\tau$ dependence of the slope $\nu_{\rm del} (\tau)$ is shown in Fig.~\ref{fig:cvol}b. This figure demonstrates that $\nu_{\rm del}(\tau)$ varies substantially in the range of $\tau$ corresponding to $W = 14$\:--\:$18$. At the same time, it does saturate for $\tau \to \infty$ (i.e., $W\to W_c$) at $\nu_{\rm del}=1/2$, in a perfect agreement with the analytical  prediction Eq.~\eqref{xi}. We can write a more precise equation, which includes a prefactor and subleading correction:
\be
\label{Nxicrit}
\frac{1}{\ln N_\xi} =c_1(W_c-W)^{1/2}+c_2(W_c-W)^{3/2},
\ee
where $c_1=0.0313$ and $c_2= 0.00369$. Equation (\ref{Nxicrit}) is valid with a good accuracy in the range $12 < W < W_c$. It is interesting to compare the numerical values for the critical disorder $W_c$, Eq. (\ref{Wc}), and for the correlation volume, Eq. (\ref{Nxicrit}), with analytical results in the large-$m$ approximation. It turns out that it works quite well already for $m=2$ \cite{tikhonov19critical}, providing the following estimates: $W_c\approx 17.67$ and
$$
\ln N_{\xi} \simeq \pi  \sqrt{\frac{2}{3}}\frac{\ln^{3/2}(W_c/2)}{\ln^{1/2}(W_c/2e)}\sqrt{\frac{W_c}{W_c-W}}, 
$$
which gives $c_1\approx 0.0306$, reproducing a numerically exact value with amazing accuracy.

The relation between the system ``volume'' $N$ and the correlation volume $N_\xi$ plays a crucial role for properties of the system one the delocalized side of the transition, i.e., at $W$ smaller than (but sufficiently close to) $W_c$. For  $N \ll N_\xi$ the system is critical, which means, to the first approximation, that it looks localized.  In the opposite limit, $N \gg N_ \xi$, the system becomes ergodic. We will demonstrate below how this crossover from the  critical regime, $N \ll N_\xi$, to the asymptotic ergodic regime, $N \gg N_ \xi$, manifests itself in key observables.  

It is worth emphasizing once more that the critical point in the RRG model has the localized character. This means that, when the system is on the localized side of the transition, $W > W_c$,   there is no qualitative change in the behavior of many observables of interest with increasing $N$. As an example, the IPR $P_2$ is of order unity for any $N$ for $W>W_c$---in stark contrast to a dramatic change of its behavior for $W<W_c$, from the critical regime $N \ll N_\xi$ to the ergodic regime $N \gg N_\xi$. In view of this, we put more emphasis on the discussion of the delocalized side, $W < W_c$ (including, of course, the critical regime), in this review.  At the same time, there is also a very interesting physics in the localized phase. In particular, Sec.~\ref{sec:eigenfunc_corre_loc} below will address dynamical correlation of eigenstates in the localized phase, which are strongly enhanced when the system approaches the transition point. We also refer the reader to Ref.~\cite{PhysRevResearch.2.012020} where the behavior of other observables in the localized phase was considered.

%%%%%%%%%%%%%%%%%%%%%%%%%%%%%%%%%%%%%%%%%%%%%%%%%%%%%%%%%%%%%%%%%%%%%%%%%%%%%%%%%%%%%%%%%%%%%%%%%%%%%%%%%%%%%%%%%%%%%%%%%%%%%%%%%%%%%%%%%%%%%%%%%%%%%%%%%%%%%%%%%%%%%%%%%%%%%%%%%%%%%%%%%%%%%%%%%%%%%%%%%%%%%%%%%
%%%%%%%%%%%%%%%%%%%%%%%%%%%%%%%%%%%%%%%%%%%%%%%%%%%%%%%%%%%%%%%%%%%%%%%%%%%%%%%%%%%%%%%%%%%%%%%%%%%%%%%%%%%%%%%%%%%%%%%%%%%%%%%%%%%%%%%%%%%%%%%%%%%%%%%%%%%%%%%%%%%%%%%%%%%%%%%%%%%%%%%%%%%%%%%%%%%%%%%%%%%%%%%%%

\subsection{Wavefunction correlations: Ergodic and critical regime}
\label{sec:WF_corr}

\subsubsection{Inverse participation ratio}
\label{sec:IPR}

%%%%%%%%%%%%%%%%%%%%%%%%%%%%%%%%%%%%%%%%%%%%%%%%%%%%%%%%%%%%%%%%%%%%%%%%%%%%%%%%%%%%%%%%%%%%%%%%%%%%%%%%%%%%%%%%%%%%%%%%%%%%%%%%%%%%%%%%%%%%%%%%%%%%%%%%%%%%%%%%%%%%%%%%%%%%%%%%%%%%%%%%%%%%%%%%%%%%%%%%%%%%%%%%%
\begin{figure}[tbp]
\minipage{0.5\textwidth}\includegraphics[width=\textwidth]{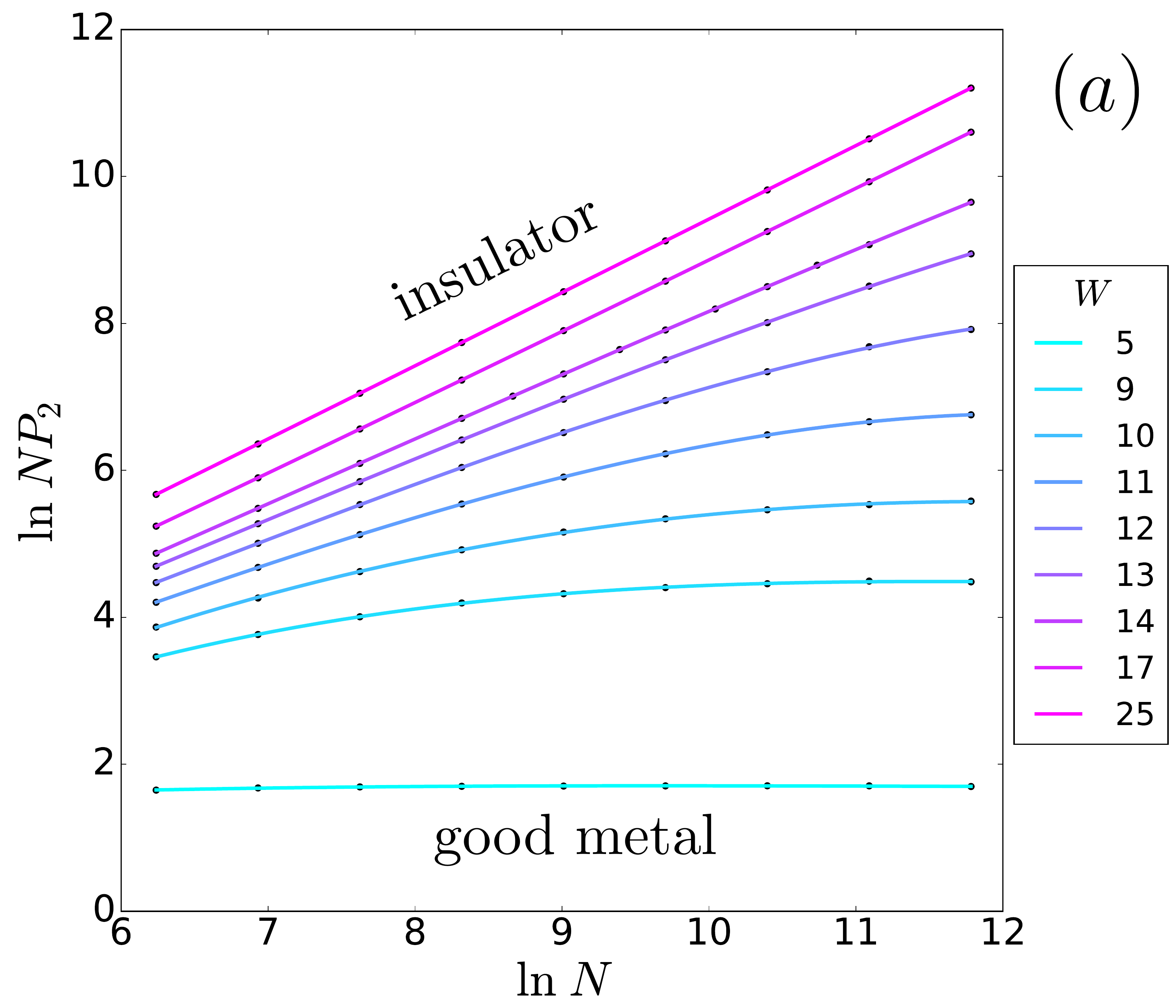}\endminipage
\minipage{0.5\textwidth}\includegraphics[width=\textwidth]{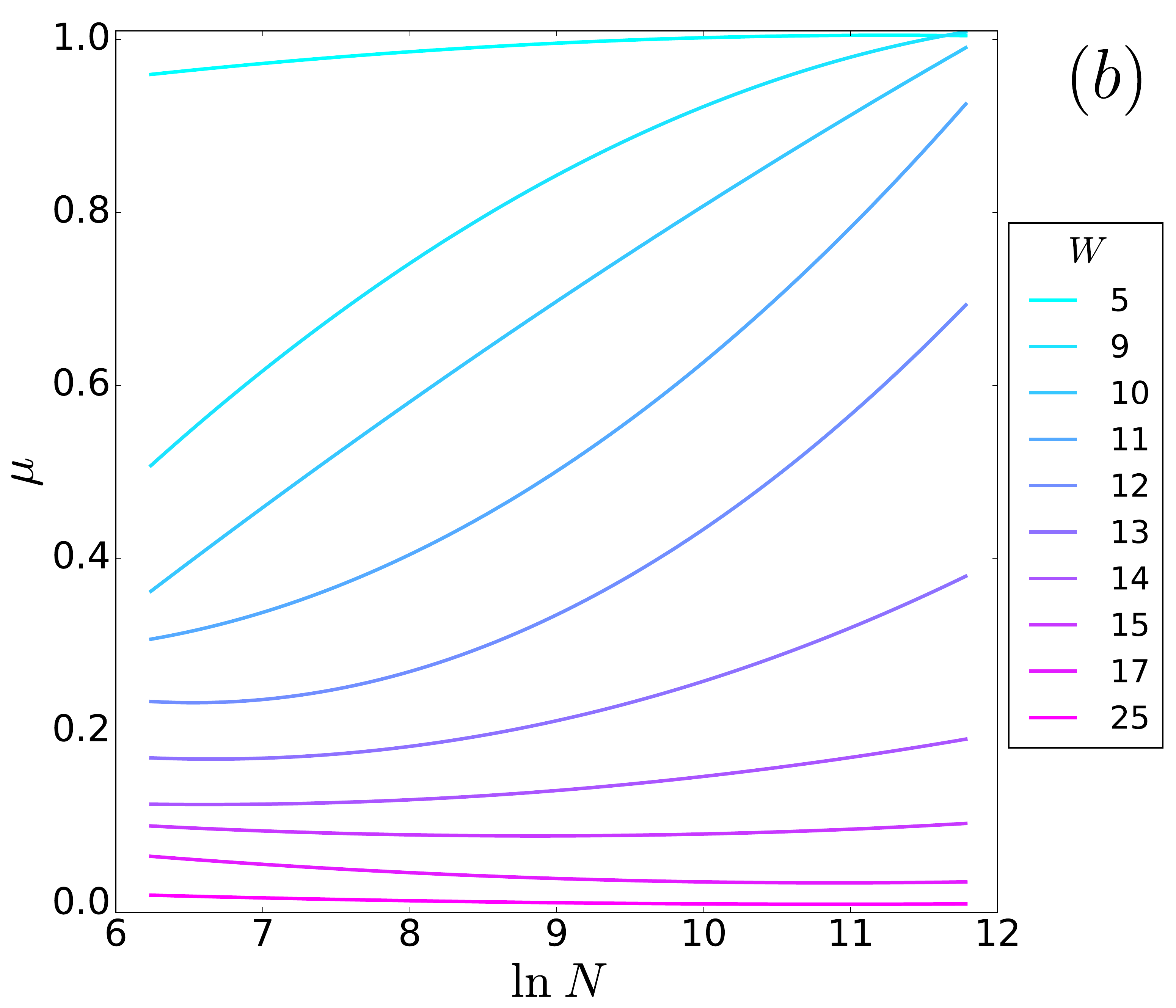}\endminipage
\caption{System size dependence of the inverse participation ratio (IPR) $P_2$ in the Anderson model on RRG as given by ED. (a) $\ln NP_2$ as a function of the system size for various disorder strengths $W$. Dots: simulation, lines: smooth interpolation. (b) System size dependence of the fractal exponent  $\mu$ for various $W$. From Ref.~\cite{tikhonov2016anderson}.}
\label{fig:ipr_rrg}
\end{figure}
%%%%%%%%%%%%%%%%%%%%%%%%%%%%%%%%%%%%%%%%%%%%%%%%%%%%%%%%%%%%%%%%%%%%%%%%%%%%%%%%%%%%%%%%%%%%%%%%%%%%%%%%%%%%%%%%%%%%%%%%%%%%%%%%%%%%%%%%%%%%%%%%%%%%%%%%%%%%%%%%%%%%%%%%%%%%%%%%%%%%%%%%%%%%%%%%%%%%%%%%%%%%%%%%%

The simplest quantity characterizing the wavefunction statistics is ensemble-averaged IPR,
\be
P_2(W,N)= \left \langle\sum_{i=1}^N |\psi_i |^4 \right \rangle.
\ee  
The dependence of $NP_2(W,N)$ on system-size for a set of disorder values as given by ED \cite{tikhonov2016anderson} is shown in Fig.~\ref{fig:ipr_rrg}a. In the localized phase we have $P_2 \sim 1$, so that  $NP_2(W,N)\propto N $ at large $N$, as the data for $W=25$ show. For the delocalized side, analytical calculation in the framework of the theory sketched in Sec.~\ref{sec:field-theory} 
(see also Sec.~\ref{sec:eigenfunc_corr_single} below)
yields, for $N \gg N_\xi$,
\be
P_2 \sim \frac{N_\xi}{N} \,, \qquad N \gg N_\xi \,.
\label{P2-ergodic}
\ee
In the opposite regime of $N \ll N_\xi$, the system is critical, with the result
\be
P_2 \sim 1 \,, \qquad N \ll N_ \xi \,.
\label{P2-critical}
\ee
Thus, in the large-$N$ limit---and more specifically, under the condition $N \gg N_\xi$---the product $NP_2(W,N)$ saturates at an $N$-independent value $C(W)$,
\be
C(W) = \lim_{N\to \infty} NP_2(N) \,,
\label{C_W}
\ee
which is manifestation of ergodicity. This behaviour is indeed supported by the data on the Fig.~\ref{fig:ipr_rrg}a, when the system is far from the transition, $W\lesssim11$. 
The value of $C(W)$ as provided by the theory of Sec.~\ref{sec:field-theory} can be calculated via the PD. As is discussed below, the results are in perfect agreement with those of ED. 
The value of   $C(W)$ increases with $W$ approaching $W_c$ as $C(W) \sim N_\xi$, i.e., exponentially fast according to Eq.~\eqref{N-xi-crit}. Therefore, when $W$ becomes sufficiently close to $W_c$, the condition of ergodic regime, $N \gg N_\xi$, is not fulfilled any more even for the largest $N$ accessible via ED, so that the saturation of  $NP_2(N)$ is not reached.
This is indeed what is observed in Fig.~\ref{fig:ipr_rrg}a for  $11\lesssim{W}< W_c$.

%%%%%%%%%%%%%%%%%%%%%%%%%%%%%%%%%%%%%%%%%%%%%%%%%%%%%%%%%%%%%%%%%%%%%%%%%%%%%%%%%%%%%%%%%%%%%%%%%%%%%%%%%%%%%%%%%%%%%%%%%%%%%%%%%%%%%%%%%%%%%%%%%%%%%%%%%%%%%%%%%%%%%%%%%%%%%%%%%%%%%%%%%%%%%%%%%%%%%%%%%%%%%%%%%
\begin{figure}[tbp]
\centerline{\includegraphics[width=0.5\textwidth]{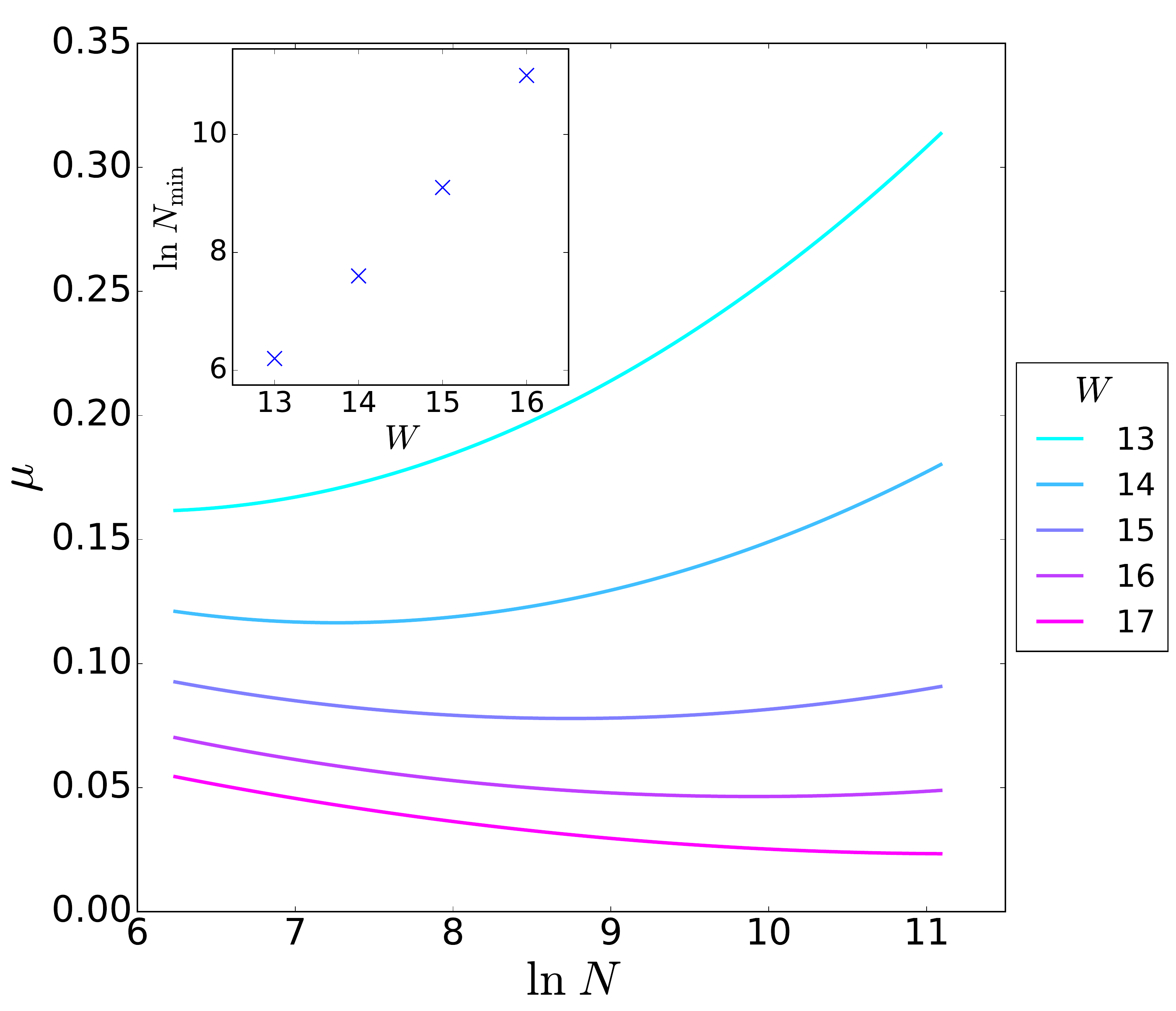}}
\caption{Flowing fractal exponent $\mu$ as a function of the system size for selected disorders in the range from $W=13$ to $W=17$. Inset: disorder-dependence of system size corresponding to the minimum of $\mu(N)$. From Ref.~\cite{tikhonov2016anderson}.}
\label{fig:ipr_rrg_zoomed}
\end{figure}
%%%%%%%%%%%%%%%%%%%%%%%%%%%%%%%%%%%%%%%%%%%%%%%%%%%%%%%%%%%%%%%%%%%%%%%%%%%%%%%%%%%%%%%%%%%%%%%%%%%%%%%%%%%%%%%%%%%%%%%%%%%%%%%%%%%%%%%%%%%%%%%%%%%%%%%%%%%%%%%%%%%%%%%%%%%%%%%%%%%%%%%%%%%%%%%%%%%%%%%%%%%%%%%%%

The data of Fig.~\ref{fig:ipr_rrg}a can be interpreted via the ``flowing fractal exponent'' 
\be
\mu(W, N)=- \frac{\partial\ln{P_2}(W,N)}{\partial\ln{N}},
\ee
 shown in Fig.~\ref{fig:ipr_rrg}b as a function of system size for a set of $W$. For moderate disorder, $W\lesssim11$, the exponent $\mu$  saturates at the ergodic value $\mu=1$ (which is equivalent to the saturation of $NP_2(N)$ in Fig.~\ref{fig:ipr_rrg}a).
 For stronger disorder (see Fig.~\ref{fig:ipr_rrg_zoomed}), $11\lesssim W< W_c$, we observe a non-monotonic behavior: $\mu$ first flows towards its value $\mu_c=0$ at the Anderson-transition critical point. (The IPR is finite at the critical point on RRG, as in the localized phase, thus $\mu_c=0$.) When the size $\log_2{N}$ exceeds  the correlation length $\xi(W)$ (see inset in Fig.~\ref{fig:ipr_rrg_zoomed}), the flow turns towards the delocalized  fixed point with the ergodic value of the exponent, $\mu = 1$. 
 
 Ergodicity of the delocalized phase on RRG (reached at $N \gg N_\xi$) has been also confirmed by numerical simulations in Refs.~\cite{garcia-mata17,biroli2018}. 

\subsubsection{Eigenfunction correlations: From finite $d$ to RRG.}
\label{sec:eigenfunc_corr_finite_d}

Let us now turn to discussion of wavefunction correlations. It is instructive to recall first  the corresponding results for Anderson problem on a cubic $d$-dimensional lattice. Indeed, the RRG  model can be viewed, in a certain sense, as a $d \to \infty$ limit of the $d$-dimensional problem. This limit is, however, very singular. This is clear from comparing the volume as function of the linear size $L$: in $d$ dimensions its a power law $L^d$ and on RRG its an exponential function $m^L$. We discuss below how one can ``guess'' the RRG results on the basis of those for $d$ dimensions. After this, we will present the results derived directly for the RRG model, see Sections \ref{sec:eigenfunc_corr_single} and \ref{sec:eigenfunc_corr_dynamical}. We will see that the ``educated guess'' based on $d$ dimensional results is largely correct but misses some important subleading factors.

Correlation function of the same wavefunction at different spatial points is formally defined as 
\be
\label{alphadef}
\alpha_E \left( r_{1},r_{2}\right) =\Delta \left\langle
\sum_{k}\left\vert \psi _{k}\left( r_{1}\right) \right\vert ^{2}\left\vert
\psi _{k}\left( r_{2}\right) \right\vert ^{2}\delta \left( E-E_{k}\right) \right\rangle.
\ee
Here $\psi_k$ are eigenstates and $E_{k}$ the corresponding energy levels,  $E$ is the energy at which the statistics is studied, $\Delta=1/\nu(E)N$ is the mean level spacing, and  $\nu(E)=N^{-1}\left<\Tr\delta(E-\hat H)\right>$ is the density of states.
For coinciding points, this correlation function reduces to the IPR,  
\be 
P_2 = \int d^dr \, \alpha_E(r,r) = L^d \alpha_E(r,r).
\label{P2-alpha-d}
\ee
 At finite spatial separation $|r_1-r_2|$, a  wavefunction in the delocalized phase near the Anderson transition in a $d$-dimensional system exhibits strong self-correlations up to the correlation length $\xi$,
\be
\label{alphad}
L^{2d}\alpha_E(r_1,r_2)\sim \left(|r_1-r_2|/\min(L,\xi)\right)^{\Delta_2},
\ee
for $|r_1-r_2| < \xi$.
At the transition point (diverging $\xi$), the critical correlations (\ref{alphad}) extend over the whole system. At finite but large $\xi$ (in the vicinity of the transition point) the correlations remain critical as long as $L \ll \xi$.  

The correlation function, characterizing the overlap of different (close in energy) wavefunctions is defined as follows:
\be
\label{sigmadef}
\beta_E \left( r_{1},r_{2},\omega \right) = \Delta
^{2}R^{-1}\left( \omega \right)  \left\langle \sum_{k\neq l}\left\vert \psi
_{k}\left( r_{1}\right) \psi _{l}\left( r_{2}\right) \right\vert ^{2}\delta
\left( E- \frac{\omega}{2}-E_{k}\right) \delta \left( E+\frac{\omega}{2} -E_{l}\right)
\right\rangle,
\ee
where $\omega$ is the energy difference between the states, and the level correlation function
\be
\label{R-omega}
R(\omega)=\frac{1}{\nu^2(E)}\left<\nu(E-\omega/2)\nu(E+\omega/2)\right>
\ee
is introduced. In finite $d$ Anderson problem, the correlation function $\beta_E \left( r_{1},r_{2},\omega \right)$  exhibits the scaling
\be
\label{betad}
L^{2d}\beta_E(r_1,r_2,\omega)\sim \left(|r_1-r_2|/\min(L_\omega,\xi)\right)^{\Delta_2}
\ee
for $|r_1-r_2|<\min(L_\omega,\xi)$. Here $L_{\omega}\sim (\omega\nu)^{-1/d}$ is the length scale associated with the frequency $\omega$ (or, equivalently, with the time $\sim \omega^{-1}$) at criticality. The exponent $\Delta_2$ determines also the scaling of the diffusion propagator at criticality \cite{chalker1990scaling}. 

For the sake of brevity of notations, we will mainly omit below the subscript $E$ indicating dependence of the correlation functions $\alpha_E$ and $\beta_E$ on energy around which the statistics is studied.

Let us now ``translate'' Eq. (\ref{alphad}) to RRG. First, the factor $L^d$ can be interpreted as the system volume and thus replaced on the RRG by the number of sites $N$.  Second, the multifractal exponent can be replaced by its large-$d$ limit $\Delta_2 \to -d$. Finally, the factors of the type $r^d$, should be understood as counting the number of sites in a sphere of a radius $r$ centered at a given site, and replaced by their RRG counterpart $m^r$. As a result, we come to the following conjecture for the scaling of the same-wavefunction-correlator (\ref{alphad}) on RRG:
\be
\label{P2sc2}
\alpha(r_1,r_2)\sim
\left\{
\begin{array}{ll}
 N^{-2} m^{\xi-r_{12}}, & \quad \textrm{metallic}, \ \ r_{12} < \xi;\\
N^{-1}m^{-r_{12}}, & \quad \textrm{critical.}
\end{array}
\right.
\ee
Here $r_{12}$ is the the length of the shortest path between the two points $r_1$ and $r_2$ on RRG. The correlation function of two different eigenfunctions, Eq.~(\ref{betad}), can be obtained in the same manner:
\be
\label{betasc}
\beta(r_1,r_2,\omega)\sim
\left\{
\begin{array}{ll}
 N^{-2} m^{\xi-r_{12}}, & \quad \textrm{metallic}, \ \ r_{12} < \xi;\\
N^{-2} \omega^{-1}m^{-r_{12}}, & \quad \textrm{critical.}
\end{array}
\right.
\ee

We turn now to the analytical derivation of  the correlation functions on RRG as well as  to their numerical investigation. 

\subsubsection{Eigenfunction correlations on RRG: Single wave function}
\label{sec:eigenfunc_corr_single}

The correlation functions introduced above can be evaluated for RRG  using supersymmetric field theory, Sec.~\ref{sec:field-theory} \cite{tikhonov19statistics}. 
Let us start with $\alpha(0)$, which in the delocalized phase (and at sufficiently large $N$) can be expressed in terms of the saddle-point solution:
\be
\label{alpha0}
\alpha(0)=\frac{12}{N^2}\frac{g_{0,xx}^{(m+1)} }{\pi ^2\nu^2}.
\ee
The coefficient $g_{0,xx}^{(m+1)}$ in Eq.~(\ref{alpha0}) has an important physical meaning. Since the function $g_0^{(m+1)}(x,y)$ is the Fourier-Laplace transform of the distribution of local Green functions on an infinite Bethe lattice (see discussion around Eq. (\ref{g0t})), $g_{0,xx}^{(m+1)}$ is proportional to the average square of the local density of states $\nu(j) = - (1/\pi) {\rm Im}\: G(j,j)$:  
\be
g_{0,xx}^{(m+1)} = (\pi^2 / 4) \langle \nu^2 \rangle_{\rm BL} \:.
\label{nu2}
\ee
The subscript ``BL'' here indicates that the average should be computed using the solution of the self-consistency equation that describes  the model on an infinite Bethe lattice. Equation (\ref{alpha0}) can thus be written in the form
\be
\label{alpha0-1}
\alpha(0)=\frac{3}{N^2}\frac{\left<\nu^2\right>_{\rm BL}}{\nu^2}.
\ee
Let us recall a relation between $\alpha(0)$ and the average IPR $P_2$
\be
P_2 = N \alpha(0) \,,
\label{P2-alpha}  
\ee
cf. the analogous formula \eqref{P2-alpha-d} for a $d$-dimensional system.
Equation \eqref{alpha0-1} yields the ergodic ($\sim 1/N$) IPR scaling \eqref{P2-ergodic}. Furthermore, it provides an exact expression for the corresponding prefactor [denoted $C(W)$ in Eq.~\eqref{C_W}] in terms of the Bethe-lattice correlation function (which can be calculated by PD), 
\be
C(W)  = 3 \frac{\left<\nu^2\right>_{\rm BL}}{ \nu^2} \,.
\label{CW-BL}
\ee

Extending this analysis,  one can calculate correlation functions $\alpha(r)$ and $\beta(r,\omega)$ on RRG. The results are expressed, in the limit of large $N$, in terms of averaged products of two Green functions on an infinite Bethe lattice. There are two such correlation functions:
\bea
\label{K1}
K_1(r) &=& \langle G_R(i,i) G_A(j,j) \rangle_{\rm BL}  = 
  \langle \frac{1}{16} (\Psi_{i,1}^\dagger\hat K\Psi_{i,1})(\Psi_{j,2}^\dagger\hat K\Psi_{j,2})\rangle_{\rm BL}; \\
K_2(r) &=& \langle G_R(i,j) G_A(j,i) \rangle_{\rm BL}  = 
 \langle \frac{1}{16} (\Psi_{j,1}^\dagger\hat K\Psi_{i,1})(\Psi_{i,2}^\dagger\hat K\Psi_{j,2})\rangle_{\rm BL}.
 \label{K2}
\eea
 These correlation functions (for $\omega = 0$ and $\eta \to 0$) have been computed in Ref. \cite{mirlin1991localization}. In the localized phase, $W > W_c$, the correlation functions $K_1(r)$ and $K_2(r)$ have $1/\eta$ singularity, are equal to each other, and decay with $r$ as 
\be
\label{K-localized}
K_1(r) = K_2(r) \sim \frac{1}{\eta} m^{-r} e^{-r/\zeta} r^{-3/2},
\ee
where $\zeta$ is the localization length. In the delocalized phase, $W < W_c$, close to the transition point, the result for the function $K_2(r)$ reads
\be
\label{K2-deloc}
K_2(r) \sim N_\xi m^{-r}  r^{-3/2}
\ee
and
\be
\label{K1-deloc}
K_1(r) \simeq K_2(r) + K_{1}^{(d)} \,,
\ee
where $K_{1}^{(d)} = |\langle G_R(j,j) \rangle |^2$ is disconnected part of $K_1(r)$.

The eigenfunction correlations on RRG are expressed \cite{tikhonov19statistics}, using the theory presented in Sec.~\ref{sec:field-theory}, in terms of these correlation functions.
In the localized phase,  $W > W_c$, $U(g)$ in the functional integral \eqref{g-funct-integral}
is simply $\langle \frac{1}{16} (\Psi_{0,1}^\dagger\hat K\Psi_{0,1})(\Psi_{r,2}^\dagger\hat K\Psi_{r,2})\rangle_{\rm BL}$, which is the correlation function $K_1(r)$, so that
\be
\label{P2K1loc}
\alpha(r)=\frac{1}{\pi\nu N}\lim_{\eta\to 0}\eta K_1(r,\eta).
\ee
 Using Eq.~(\ref{K-localized}), we immediately find
\be
\label{alpha-r-loc}
\alpha(r) \sim \frac{1}{N}m^{-r} e^{-r/\zeta} r^{-3/2}.
\ee
This result is extended to the critical point by setting $\zeta = \infty$, which yields
\be
\label{alphaloc}
\alpha(r)\sim \frac{1}{N}\frac{m^{-r}}{r^{3/2}}.
\ee

In the delocalized phase, the functional  integral is reduced to an integral over a manifold of saddle points (see discussion in Sec. \ref{sec:loctr}). 
The result for the correlation of the same wavefunction reads:
\be
\label{alpha-r-deloc}
\alpha(r) = \frac{1}{2\pi^2 N^2} [K_1 (r) + 2 K_2(r)- {\rm Re} \langle G_R(0) \rangle^2 - 2{\rm Re} \langle G_R(r) \rangle^2 ]
\ee
where the last term in square brackets is relatively small for large $N_\xi$. Using Eqs.~(\ref{K1-deloc}) and (\ref{K2-deloc}), we find:
\be
\label{alphadeloc}
\alpha(r)\sim \frac{N_{\xi}}{N^2}\frac{m^{-r}}{r^{3/2}}, \qquad r < \xi.
\ee
For $r > \xi$ the correlation function $\alpha(r)$ is governed by disconnected  parts in Eq.~(\ref{alpha-r-deloc}),  yielding $\alpha(r) \simeq 1$.

The results \eqref{alphaloc} and   \eqref{alphadeloc}  largely confirm the ``educated guess''  \eqref{P2sc2} based on an extrapolation of finite-$d$ results to $d \to \infty$.
They include, however, an additional factor $r^{-3/2}$.

%%%%%%%%%%%%%%%%%%%%%%%%%%%%%%%%%%%%%%%%%%%%%%%%%%%%%%%%%%%%%%%%%%%%%%%%%%%%%%%%%%%%%%%%%%%%%%%%%%%%%%%%%%%%%%%%%%%%%%%%%%%%%%%%%%%%%%%%%%%%%%%%%%%%%%%%%%%%%%%%%%%%%%%%%%%%%%%%%%%%%%%%%%%%%%%%%%%%%%%%%%%%%%%%%
\begin{figure}[tbp]
\minipage{0.5\textwidth}\includegraphics[width=\textwidth]{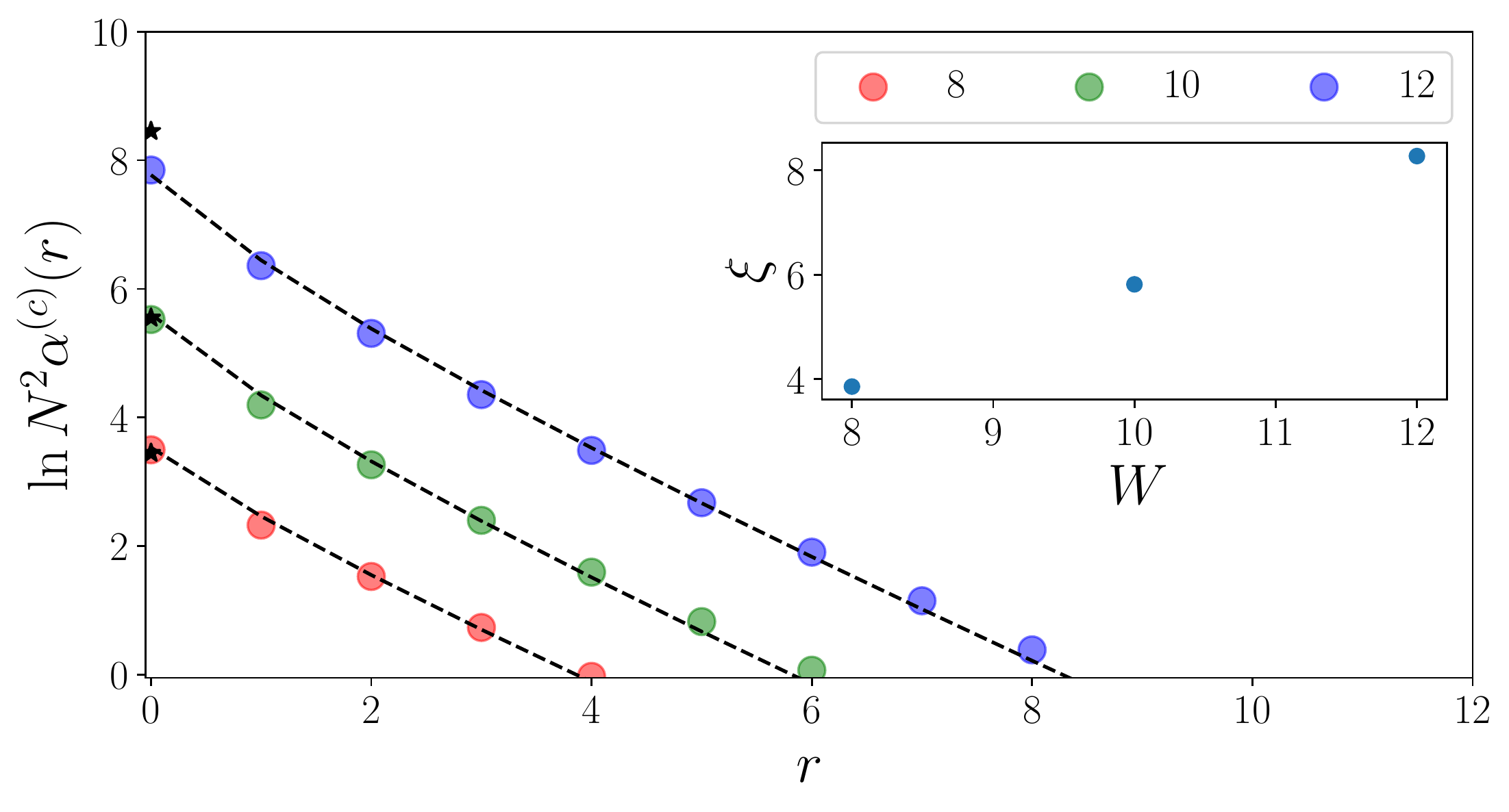}\endminipage
\minipage{0.5\textwidth}\includegraphics[width=\textwidth]{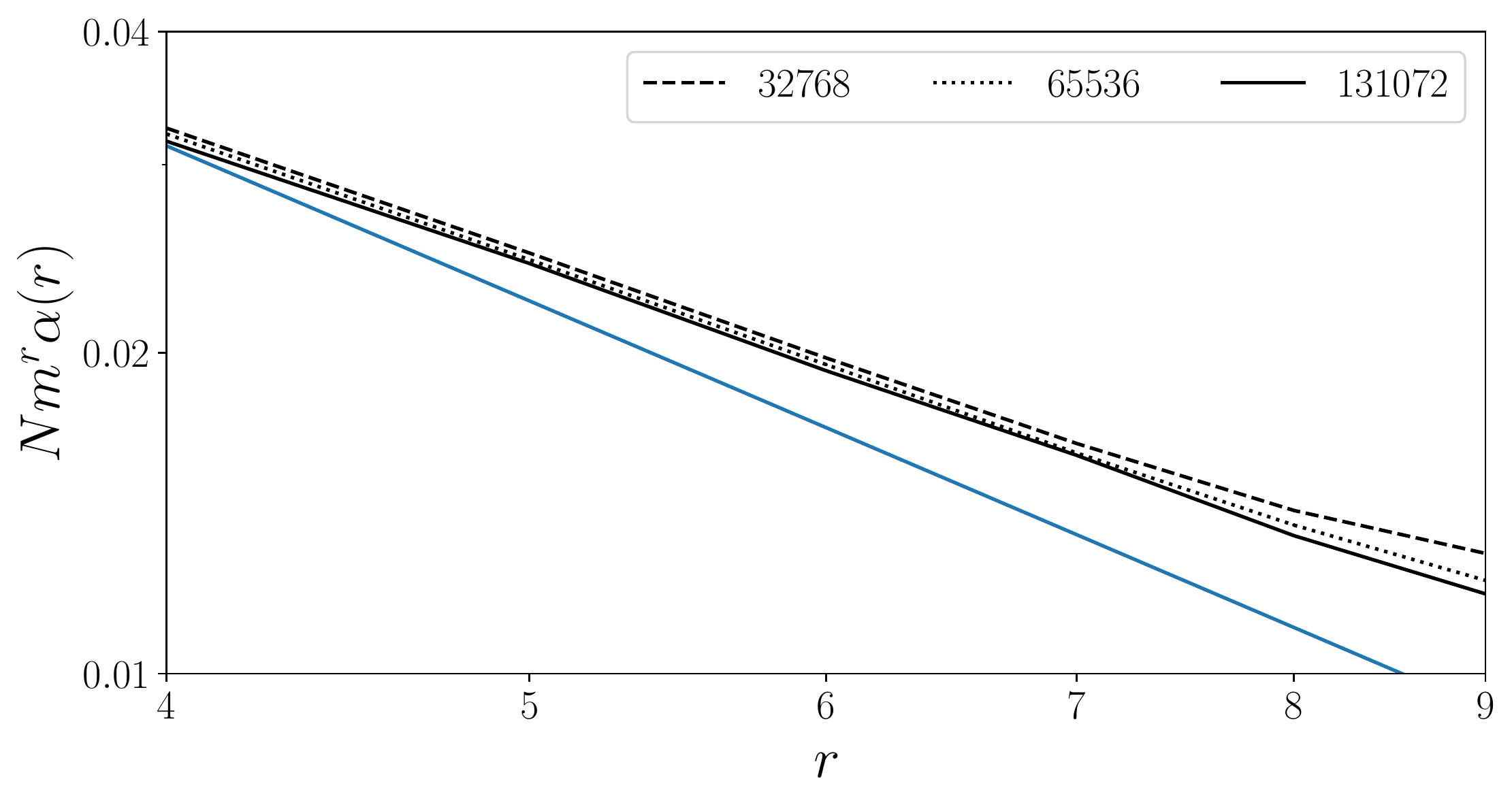}\endminipage
\caption{Eigenfunction self-correlations $\alpha(r)$ on RRG. {\it Left:} connected part $N^2\alpha^{(c)}(r) = N^2\alpha(r) -1$  for $N=2^{17}$ in the delocalized phase, for disorder values $W=8$, 10, and 12. Dashed black lines: fit to $\ln N^2\alpha^{(c)}(r)=-r\ln m-c_1^{(\alpha)}\ln (r+1)-c_2^{(\alpha)}$, see Eq.~(\ref{alphadeloc}). Star symbols: values of $\alpha(0)$ derived from PD, Eq.(\ref{alpha0-1}). Inset: correlation length $\xi(W)$ determined from $\ln N^2\alpha^{(c)}(r)=0$. {\it Right:} $\alpha(r)$  at the critical point ($W=18$) as found by ED: $N m^r\alpha(r)$ as a function of $r$ on double-logarithmic scale.  Blue solid line: $c / r^{3/2}$, see Eq.~(\ref{alphadeloc}). From Ref.~\cite{tikhonov19statistics}. 
}
\label{same}
\end{figure}
%%%%%%%%%%%%%%%%%%%%%%%%%%%%%%%%%%%%%%%%%%%%%%%%%%%%%%%%%%%%%%%%%%%%%%%%%%%%%%%%%%%%%%%%%%%%%%%%%%%%%%%%%%%%%%%%%%%%%%%%%%%%%%%%%%%%%%%%%%%%%%%%%%%%%%%%%%%%%%%%%%%%%%%%%%%%%%%%%%%%%%%%%%%%%%%%%%%%%%%%%%%%%%%%%

These analytical predictions for $\alpha(r)$ on RRG can be compared with results of the ED for the RRG model (\ref{H}) in the middle of the band \cite{tikhonov19statistics}. In Fig.~\ref{same}a we show the connected part $\alpha^{(c)}(r) = \alpha(r) - 1/N^2$ evaluated for RRG ($N= 2^{17}$, $m+1 = 3$) in the delocalized phase. The chosen values of $W$ are sufficiently close to $W_c$, so that the correlation volume is large, $N_\xi \gg 1$ (the system size is still larger, $N\gg N_\xi$).
In this regime, ergodicity of wavefunctions manifests itself via system size independence of $N^2\alpha(r)$, see Eq. (\ref{alphadeloc}). This is fully confirmed by the ED results for three different system sizes, shown in the right panel of Fig. \ref{same}. The finite-size effects can be seen on this plot for $r$ approaching the linear size of the graph $\ln N /\ln m$.

The value of $N^2 \alpha(0) = NP_2 $ is in full agreement with the analytical prediction \eqref{alpha0-1}, \eqref{CW-BL}, where the r.h.s. was evaluated by PD (shown by star symbols). While the agreement is perfect for $W=8$ and $W=10$, a small deviation for $W=12$, which is fully expected since at this value of disorder the ratio $N / N_\xi$ is not so large any more. 

The dependence of $\ln[N^2\alpha^{(c)}(r)]$ on $r$ is approximately linear, in agreement with exponential decay, predicted by Eq.~(\ref{alphadeloc}). The fits by $\ln [N^2\alpha_c(r)] =-r\ln m-c_1^{(\alpha)}\ln (r+1)-c_2^{(\alpha)}$, with $c_1^{(\alpha)}$ as a free parameter, are shown in the left panel of Fig.~\ref{same} by dashed lines. The fitted values of $c_1^{(\alpha)}$ are smaller than $3/2$ but increase upon increase of $\xi$, drifting in the direction of power--law exponent 3/2 (analytical expectation for $\xi\to\infty$). The values of the correlation length $\xi$ defined via condition $\ln[N^2\alpha^{(c)}(r)] = 0$ are shown in the inset of Fig.~\ref{same} (left panel). They match the results for $\xi(W)$ estimated from the analysis of the $N$--dependence of IPR, see inset to Fig.~\ref{fig:ipr_rrg_zoomed}.

The $r$-dependence of $\alpha(r)$  at the critical point is also in a very good agreement with the analytical prediction (\ref{alphaloc}). To demonstrate this, the right panel of Fig.~\ref{same} shows  the product $Nm^r\alpha(r)$ as a function of $r$. At criticality, $\alpha(r)$ is predicted to behave as $1/N$ and the lines indeed almost collapse. With increasing $N$, the finite-size corrections become less relevant and the curves approach the straight line with the slope 3/2,
in agreement with Eq.~(\ref{alphaloc}).

\subsubsection{Dynamical correlations of eigenfunctions on RRG}
\label{sec:eigenfunc_corr_dynamical}

We now turn to the analysis of the  correlation function  $\beta_E(r,\omega)$, defined by Eq.~(\ref{sigmadef}). 
This correlation function depends on the frequency $\omega$, thus describing dynamical correlations.

To evaluate  $\beta_E(r, \omega)$ by using supersymmetric field theory, Sec.~\ref{sec:field-theory} \cite{tikhonov19statistics}, one first expresses it in terms of Green functions on RRG: 
%The result reads,
% in the delocalized phase at $N \gg N_\xi$, 
%\be
%\label{betamain}
%\beta(r,\omega)=\frac{1}{2\pi^2 N^2}\Re\left[K_1(r,\omega)-\left<G_R(0)\right>^2\right] ,
%\ee
%where
%\bea
%\label{K1w}
%K_1(r,\omega) = \langle G_R(i,i,E+\frac{\omega}{2}) G_A(j,j,E-\frac{\omega}{2}) \rangle_{\rm BL}.
%\eea
%
\be
\label{alphasigmaQ}
\alpha_{ij} \left(E\right) \delta \left(
\omega/\Delta\right) +\beta_{ij} \left(E,\omega \right) \bar{R
}\left( \omega \right) =\Delta^2 B_{ij}(E,\omega),
\ee
where $i$ and $j$ are two cites on RRG separated by a distance $r$,
\be
\bar{R}\left( \omega \right) =R\left( \omega \right) -\delta \left(\omega /\Delta \right)
\label{R-bar}
\ee
 is the non-singular part of the two-level correlation function (\ref{R-omega}) and LDOS correlation function is introduced as follows
\bea
\label{defB}
B_{ij}(E,\omega) &=& \left<\nu_{i}(E-\omega/2)\nu_{j}(E+\omega/2)\right> \nonumber \\
 &=& \frac{1}{2\pi^2}\Re\left[\left<G_R(i,i,E+\frac{\omega}{2})G_A(j,j,E-\frac{\omega}{2})-G_R(i,i,E+\frac{\omega}{2})G_R(j,j,E-\frac{\omega}{2})\right>\right]. \nonumber\\
\label{Bij}
\eea
In the delocalized phase at $N \gg N_\xi$, one finds \cite{tikhonov19statistics}
\be 
B_{ij}(E,\omega) = \bar{R}_{\rm WD}(\omega) \frac{1}{2\pi^2}
\Re\left[K_1(r,\omega)-\left<G_R(0)\right>^2\right],
\label{Bij-result}
\ee
where $\bar{R}_{\rm WD}(\omega)$ is the WD level correlation function and
\bea
\label{K1w}
K_1(r,\omega) = \langle G_R(i,i,E+\frac{\omega}{2}) G_A(j,j,E-\frac{\omega}{2}) \rangle_{\rm BL},
\eea
which (for $W<W_c$ and $N \gg N_\xi$) is given by its RMT form, $\bar{R}_{\rm WD}(\omega)$. Thus, one finally gets in this regime
\be
\label{betamain}
\beta(r,\omega)=\frac{1}{2\pi^2 N^2}\Re\left[K_1(r,\omega)-\left<G_R(0)\right>^2\right].
\ee
Similarly to Eq. (\ref{alpha-r-deloc}), this equation expresses a correlation function of eigenfunctions on RRG with large $N$ in terms of a correlation function defined on an infinite Bethe lattice (or, equivalently, via a self-consistency equation). In this review, we will focus on the  correlation function $\beta(r,\omega)$ at $r=0$, introducing the short notation $\beta(0,\omega)\equiv \beta(\omega)$. For analytical and numerical study of $r$--dependence of $\beta(r,\omega)$, see Ref. \cite{tikhonov19statistics}.

To calculate $\beta(\omega)$,  as given in terms of infinite-Bethe lattice correlation function by Eqs.~(\ref{betamain}), (\ref{K1w}),
one can use self-consistency equations for the joint distribution function of two Green functions on different energies, $u=G_R(i,i,E+\omega/2)$ and $v=G_A(i,i,E-\omega/2)$, which provide a generalization of Eqs. (\ref{pool_sc}) and (\ref{pool_simple}) \cite{PhysRevLett.117.104101,metz2017level,tikhonov19statistics}:
\begin{eqnarray}
f^{(m)}(u,v) &=& \int d\epsilon \: \gamma(\epsilon)\int \left( \prod_{r=1}^{m} du_r \,  dv_r \, f^{(m)}(u_r,v_r)  \right) \nonumber \\
& \times & \delta\left[ u - \frac{1}{E+\frac{\omega}{2}+i\eta -\epsilon - \sum_{r=1}^{m} {u_r}} \right]    
\delta\left[ v - \frac{1}{E-\frac{\omega}{2}-i\eta -\epsilon  - \sum_{r=1}^{m}{v_r}} \right]  ;
\label{2tra}  \\
f^{(m+1)}(u,v) &=& \int d\epsilon \: \gamma(\epsilon) \int \left( \prod_{r=1}^{m+1} du_r \,  dv_r \, f^{(m)}(u_r,v_r) \right)  \nonumber  \\
& \times &  \delta\left[ u - \frac{1}{E+\frac{\omega}{2}+i\eta -\epsilon - \sum_{r=1}^{m+1} {u_r}} \right]    
\delta\left[ v - \frac{1}{E- \frac{\omega}{2}-i\eta -\epsilon  - \sum_{r=1}^{m+1}{v_r}} \right].
\label{1tra}
\end{eqnarray}

Now we discuss application of the general formula (\ref{betamain}) to specific regimes. At criticality ($W=W_c$, or, more generally, $N_\xi \gg N$), performing in Eq.~(\ref{K-localized}) an analytical continuation to real frequency, $\eta\to - i\omega/2$, and setting $r=0$, one gets
\be
\label{K1loc}
K_1(r=0, \omega)\sim \frac{1}{-i\omega}.
\ee
According to Eq.~(\ref{betamain}), the term  (\ref{K1loc}) does not contribute to $\beta(\omega)$. Thus, we need to evaluate corrections at criticality which are expected to be governed by inverse powers of $\ln1/\eta$:
\be
\label{K1-eta-with-subleading}
K_1(r=0) \simeq \frac{c_1^{(K)}}{\eta}+\frac{c_2^{(K)}}{\eta\ln^{\mulog}1/\eta},
\ee
The subleading factor in a form of a power--law of the logarithm of frequency is a natural counterpart of the subleading $r^{-3/2}$ factor in the $r$-dependence, see Eq.~(\ref{K2-deloc}). Equation (\ref{K1-eta-with-subleading}) was verified in Ref.\cite{tikhonov19statistics} by numerical solution of the self-consistency equation. The numerical value of the exponent $\mulog$ is $\mulog \simeq 1/2$. It seems likely that $\mulog=1/2$ is in fact an exact value for this model; it remains to be seen how this can be derived analytically.

When an analytical continuation to real frequency, $\eta\to - i\omega/2$, is performed in Eq.~(\ref{K1-eta-with-subleading}) and the result is substituted into Eq.~(\ref{betamain}), the first term in Eq.~(\ref{K1-eta-with-subleading}) drops and the following scaling of $\beta(\omega)$ at criticality is found:
\be
\label{betacrit}
\beta(\omega)\sim \frac{1}{N^2 \omega\ln^{\mulog+1} 1/\omega}.
\ee
The applicability of the critical scaling (\ref{betacrit}) is limited, on the side of small $\omega$, by the finite size $N$ of the system. In the limit $\omega \to 0$ the correlation function is given by
\be
\label{beta-crit-0}
\beta(\omega \to 0)\sim 1/N.
\ee 
This can be shown by using the fact that $\beta(\omega \to 0) / \alpha(0) = 1/3$ in the delocalized phase at $N \gg N_\xi$, see below. It follows that $\beta(\omega \to 0) / \alpha(0) \sim 1$ by continuity also for $N \ll N_\xi$. Using Eq.~(\ref{alphaloc}), we find Eq.~(\ref{beta-crit-0}). The critical scaling (\ref{betacrit}) matches the low-frequency value (\ref{beta-crit-0}) at the scale
\be 
\label{omegaN}
\omega_N\sim \frac{1}{N \ln^{\mulog+1}N},
\ee
which is parametrically (by a logarithmic factor) smaller than the mean level spacing $\Delta\sim 1/N$.

We consider now the behavior of $\beta(\omega)$ in the delocalized phase, $W<W_c$ and $N\gg N_\xi$. In the small-frequency limit, a
comparison of Eqs.~(\ref{betamain}) and (\ref{alpha-r-deloc}) yields
\be
\label{beta-alpha}
\beta(\omega \to 0) = \frac{1}{3} \alpha(0) \sim \frac{N_\xi}{N^2}.
\ee
The factor $1/3$ in Eq.~(\ref{beta-alpha}) is the same as in the Gaussian ensemble of RMT.  Its emergence here is one more manifestation of the ergodicity of the delocalized phase on RRG.  The correlation function $K_1(0,\omega)$ and hence $\beta(\omega)$ remain nearly frequency independent for not too high frequencies before the system enters the critical regime, see Eq.~(\ref{betacrit}). The crossover frequency $\omega_\xi$ is determined by matching Eqs.~(\ref{betacrit}) and (\ref{beta-alpha}); it is found as $\omega_\xi\sim N_{\xi}^{-1}$ (up to a logarithmic factor).

Summarizing, the results for the dynamical correlation function $\beta(\omega)$ in the delocalized phase read
\be
\label{betascres}
\beta(\omega)\sim
\left\{
\begin{array}{cc} \displaystyle
 \frac{N_\xi}{N^2}, & \quad \omega<\omega_{\xi},\\[0.4cm]
 \displaystyle 
\frac{1}{N^2 \omega \ln^{\mulog+1}1/\omega}, & \quad \omega>\omega_{\xi}.
\end{array}
\right.
\ee
These results largely confirm the expectation, Eq. (\ref{betasc}), based on the $d \to \infty$ extrapolation (up to an additional logarithmic frequency--dependent factor).

%%%%%%%%%%%%%%%%%%%%%%%%%%%%%%%%%%%%%%%%%%%%%%%%%%%%%%%%%%%%%%%%%%%%%%%%%%%%%%%%%%%%%%%%%%%%%%%%%%%%%%%%%%%%%%%%%%%%%%%%%%%%%%%%%%%%%%%%%%%%%%%%%%%%%%%%%%%%%%%%%%%%%%%%%%%%%%%%%%%%%%%%%%%%%%%%%%%%%%%%%%%%%%%%%
\begin{figure}[tbp]
\minipage{0.5\textwidth}\includegraphics[width=\textwidth]{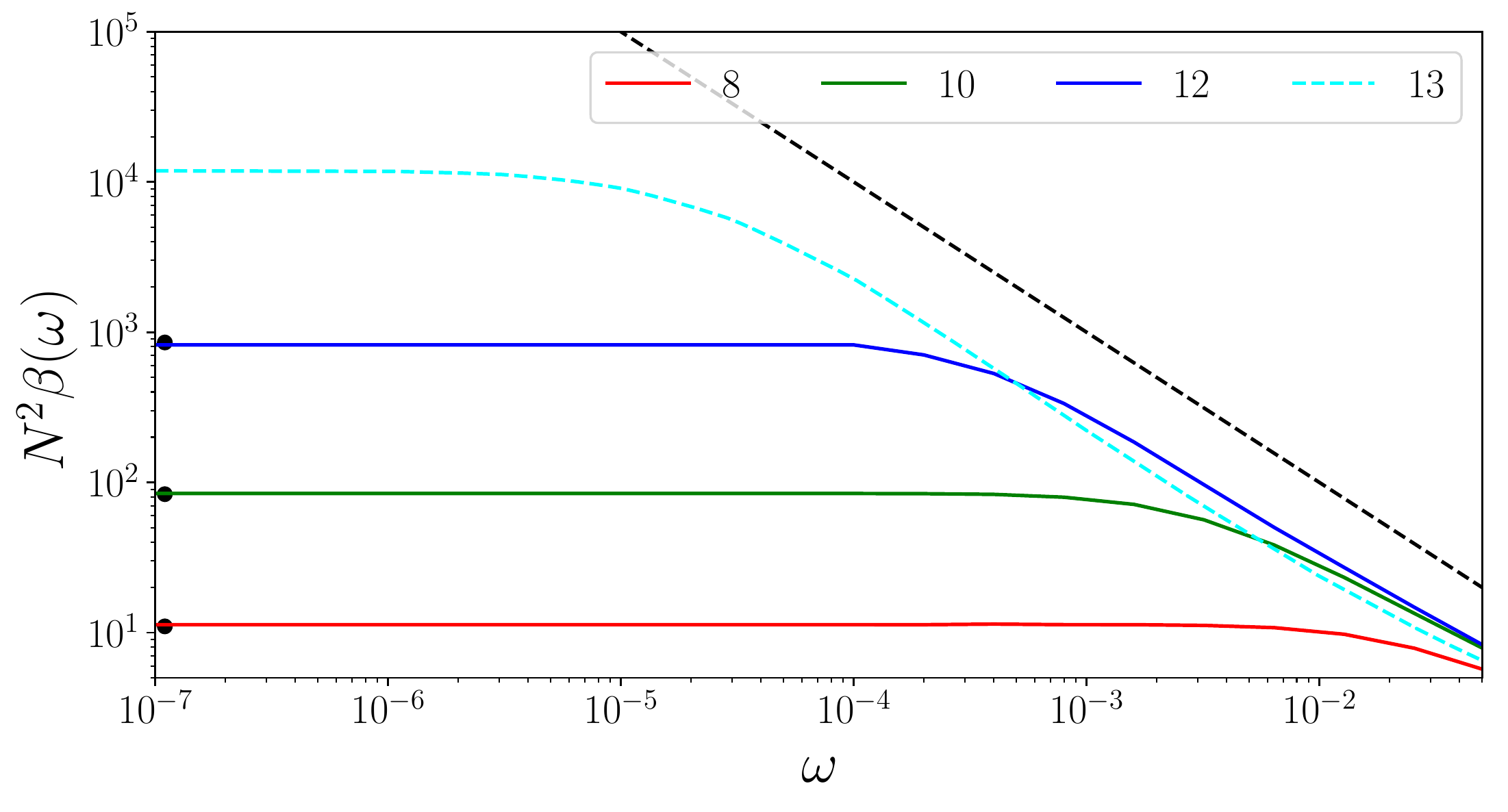}\endminipage
\minipage{0.5\textwidth}\includegraphics[width=\textwidth]{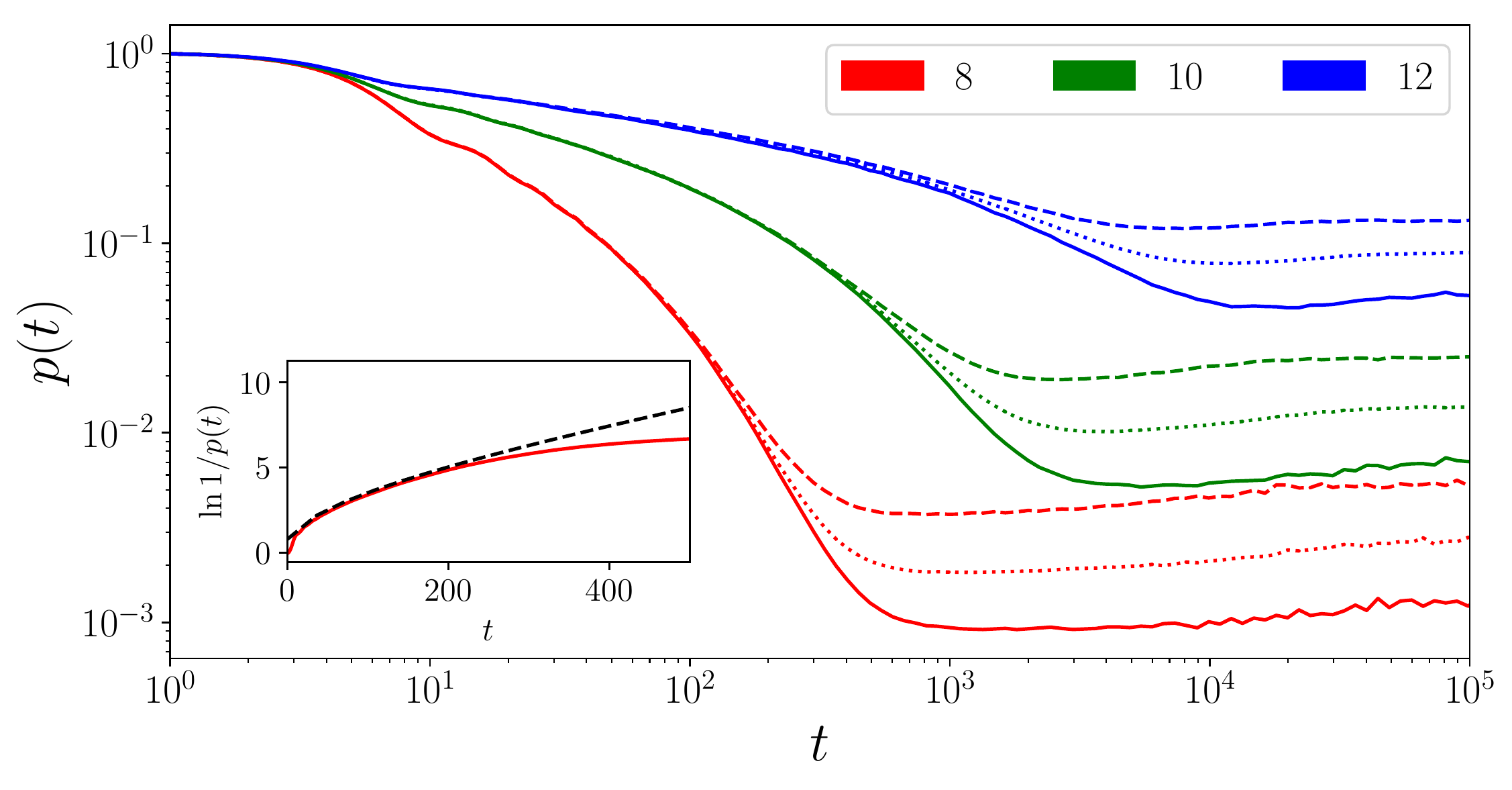}\endminipage
\caption{Correlation function $\beta(\omega)$ of different eigenfunctions at the same spatial point and return probability $p(t)$ on RRG. {\it Left:} 
$N^2  \beta(\omega)$ for $W<W_c$. Solid lines: ED results ($N=2^{17}$), cyan dashed line: result of Eq. (\ref{betamain}) with $K_1(0, \omega)$ determined from numerical solution of Eqs. (\ref{1tra}), (\ref{2tra}), black dashed line: $1/\omega$ scaling. Black dots: $\alpha(0)/3$. {\it Right:} 
return probability $p(t)$ (ED result) for $W=8$ (red), $10$ (green), and $12$ (blue). The fast decay of $p(t)$ at initial time crosses over to finite-size saturation at $p_\infty = N \alpha(0) \sim N_{\xi}/N$ (the dashed, dotted, and solid lines correspond to various system sizes ($N=2^{15}$, $2^{16}$, and $2^{17}$), respectively). Inset: comparison of $p(t)$ (solid, $W=8$) to, classical diffusion over the infinite Bethe lattice, $p(t) = a p_0(Dt)$ (dashed).
From Ref.~\cite{tikhonov19statistics}.  }
\label{different}
\end{figure}
%%%%%%%%%%%%%%%%%%%%%%%%%%%%%%%%%%%%%%%%%%%%%%%%%%%%%%%%%%%%%%%%%%%%%%%%%%%%%%%%%%%%%%%%%%%%%%%%%%%%%%%%%%%%%%%%%%%%%%%%%%%%%%%%%%%%%%%%%%%%%%%%%%%%%%%%%%%%%%%%%%%%%%%%%%%%%%%%%%%%%%%%%%%%%%%%%%%%%%%%%%%%%%%%%

The numerical results for the correlation function $\beta(\omega)$ are shown in Fig.~\ref{different}.  In the left panel, the results for the delocalized phase are presented. Solid lines ($W=8,\;10,\;12$) are obtained by ED. 
This plot also shows (by dashed line) results for 
$W=13$ as obtained  from Eq.~(\ref{betamain}) and (\ref{K1w}), with finite-$\omega$  correlations on an infinite Bethe lattice derived from the self-consistency equations (\ref{1tra}), (\ref{2tra}). These two types of numerical results are in a very good agreement between each other as well as with the analytical prediction Eq. (\ref{betascres}). Both the critical behavior ($1/\omega$, up to corrections that are difficult to observe in this plot) and the low-frequency saturation are evident. As an additional check, this figure also includes  the numerically obtained values of $\alpha(0)/3$ (dots); it is seen that 
$\beta(\omega\to 0)/\alpha(0)=1/3$, see Eq.~(\ref{beta-alpha}), is perfectly fulfilled. It is possible to extract the correlation length $\xi$ from either the value of $\beta(\omega\to 0)$ or from the crossover scale in the frequency--dependence. Both ways give values of $\xi(W)$ close to those shown in the inset of Fig.~\ref{same}. 

\subsubsection{Return probability}
\label{sec:return_prob}

In this section, we consider spreading of a state, localized at $t=0$ at a given site $j$ and evaluate probability $p(t)$ to find it at the same site at a later time $t>0$. Formally, it is defined as follows:
\be
p(t)=\left<\frac{1}{N}\sum_{j} |\bra{j}e^{-i\hat H t}\ket{j}|^2\right>,
\label{pt}
\ee
where averaging over the initial site $j$ is performed. It is straightforward to express the Fourier transform of return probability $p(\omega)$ in terms of eigenstates correlations functions,  defined in Eqs. (\ref{alphadef}) and (\ref{sigmadef}):
\be
\label{pomega}
p(\omega) = N\delta(\omega)\int dE\: \nu(E)\alpha_E(0) + N^2 \int dE\: \nu^2(E) R_E(\omega) \beta_E(\omega).
\ee
Thus, return probability encodes both $\alpha_E(0)$ (or, equivalently, IPR) via its $t\to\infty$ limit, and $\beta_E(\omega)$ via its time--dependence. Below we assume that the sum in Eq. (\ref{pt}) is projected to the states in the vicinity of a certain energy $E$ and omit the subscript $E$ in the notation for $p_E(t)$.

In the critical regime, we find by a straightforward calculation \cite{tikhonov19statistics}, from Eq. (\ref{pomega}):
\be
\label{pcrit}
p(t) \simeq p_{\infty} + \frac{c^{(p)}}{\ln^{\mulog}t}, \qquad t \to \infty,
\ee
with a numerical constant $c^{(p)}$.

In the metallic regime ($W<W_c$ and $N \gg N_\xi$), the return probability $p(t)$ can be described by a classical random walk over the tree. Such a random walk is described by
\be
\label{pdiff}
p(t) = p_0(Dt) \sim \frac{1}{(Dt)^{3/2}}e^{-Dt},
\ee
which gives the probability for a particle to be found at the starting point after time $t$ \cite{helfand1983statistics,monthus1996random}.

The diffusion coefficient can be expressed in terms of a certain integral involving the solution of the self-consistency equation \cite{mirlin1991localization} (see also a similar computation for $\sigma$ model  \cite{efetov1987density}); the corresponding asymptotics at $N_\xi\gg 1$ reads
\be
\label{diff}
D \sim N_\xi^{-1} \ln^3 N_ \xi \ ,
\ee
where $N_\xi$ scales according to Eq.~(\ref{xi}). In a finite system the decay, described by Eq. (\ref{pdiff}) saturates at a value, given by the first term in Eq. (\ref{pomega}):
\be
p_\infty \sim N \alpha(0) \sim \frac{N_\xi}{N},
\label{p-infty-deloc}
\ee
with $1/N$ scaling of $p_\infty$ being another manifestation of ergodicity of the delocalized phase on the RRG.

The right panel of Fig.\ref{different} presents results for return probability $p(t)$ as obtained by ED of the RRG model (projected to the middle of the band). At long times, this decay is limited by the system size $N$, and $p(t)$ saturates at the value $p_\infty$. While for $W=8$, the diffusive exponential decay is apparent from the plot, the curve for disorder $W=12$ shows a different type of behavior: $p(t)$ exhibits a nearly flat part up to $t \sim 10^{-3}$. This is sign of a critical behavior which, for longer times, crosses over to an exponentially fast decay representative for the delocalized regime. However, it does not have much time to develop, since the saturation dictated by the system size sets in. 

In the inset to the right panel of Fig. ~\ref{different}, the return probability is compared to the solution of a classical diffusion problem on the Bethe lattice \cite{helfand1983statistics,monthus1996random}. The correspondence is very good, until finite-size effects become important at $t \gtrsim 200$.

\subsubsection{Adjacent eigenstate correlations}
\label{sec:RRG-adjacent-state-corr}

%%%%%%%%%%%%%%%%%%%%%%%%%%%%%%%%%%%%%%%%%%%%%%%%%%%%%%%%%%%%%%%%%%%%%%%%%%%%%%%%%%%%%%%%%%%%%%%%%%%%%%%%%%%%%%%%%%%%%%%%%%%%%%%%%%%%%%%%%%%%%%%%%%%%%%%%%%%%%%%%%%%%%%%%%%%%%%%%%%%%%%%%%%%%%%%%%%%%%%%%%%%%%%%%%
\begin{figure}[tbp]
\minipage{0.5\textwidth}\includegraphics[width=\textwidth]{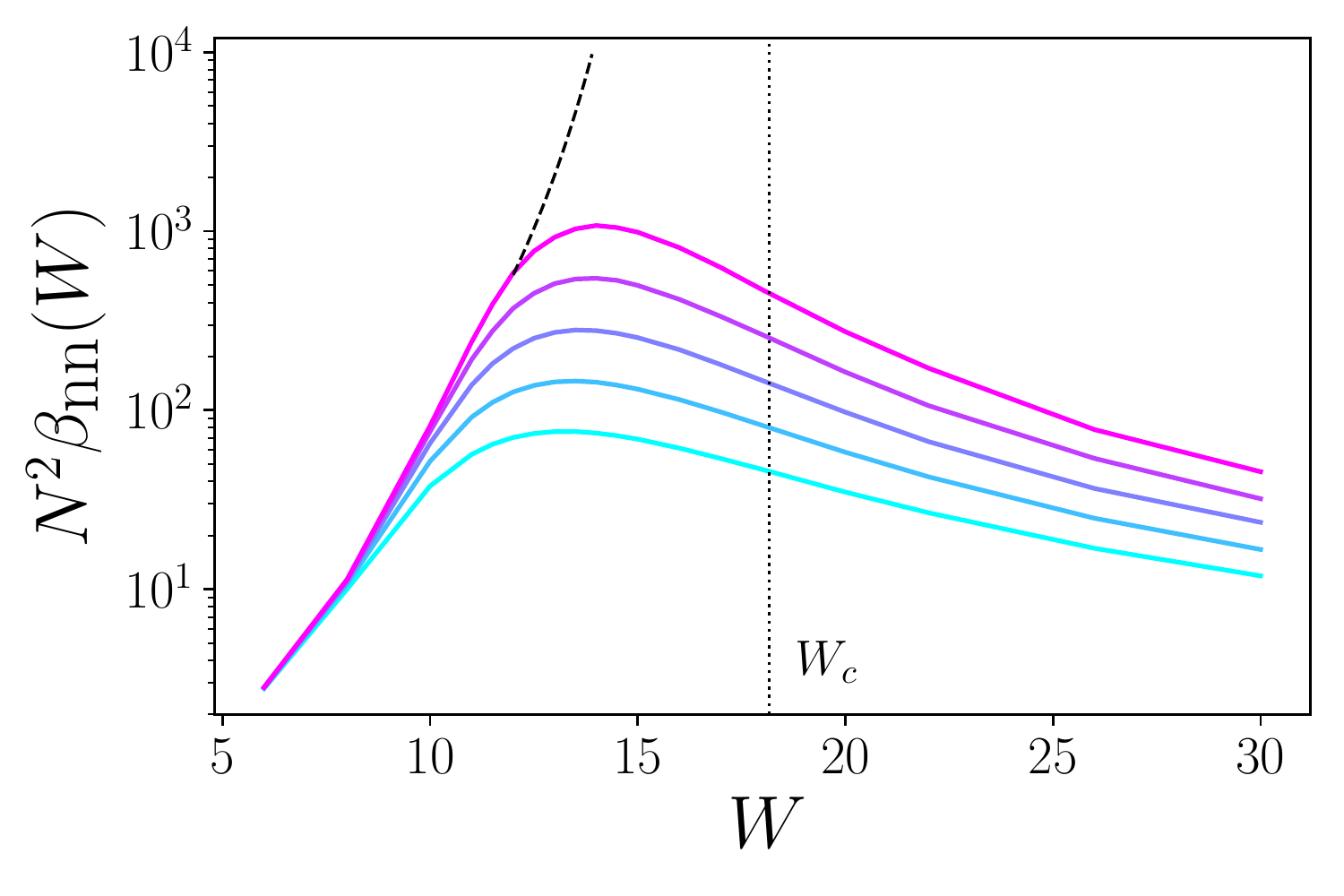}\endminipage
\minipage{0.5\textwidth}\includegraphics[width=\textwidth]{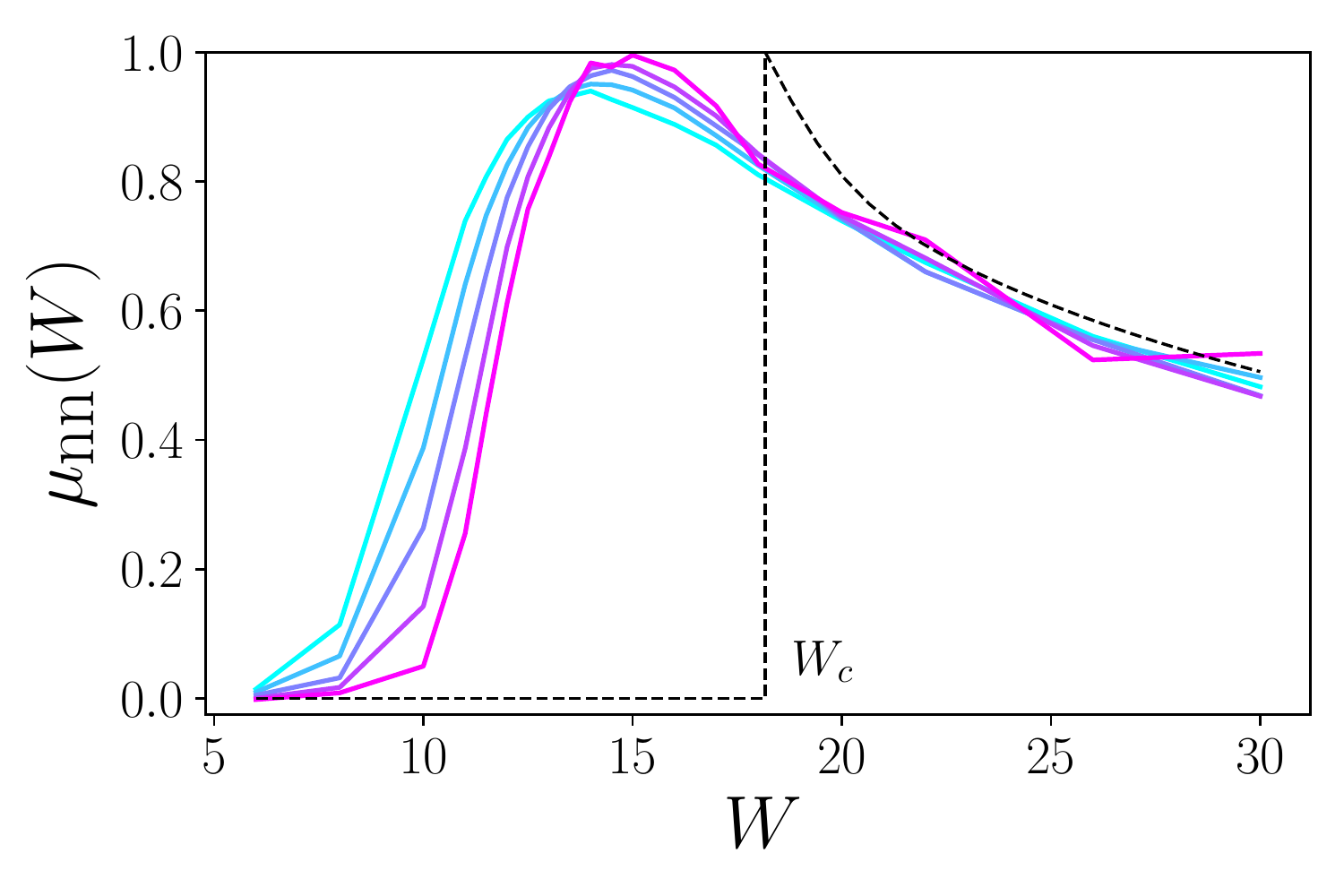}\endminipage
\caption{Correlation of adjacent wavefunctions on RRG (ED results) for $N=2^{12}, 2^{13}, 2^{14}, 2^{15}, 2^{16}$ (from cyan to magenta).  {\it Left:} Correlation function $\beta_{\textrm{nn}}(W)$. Dashed line: the asymptotic behaviour of $N^2\beta_{\textrm{nn}}$ for $W<W_c$ according to Eqs. (\ref{betann}) and (\ref{Nxicrit}). Vertical dotted corresponds to $W=W_c$.  {\it Right:} Exponent $\mu_{\textrm{nn}}$ characterizing the $N$ scaling of adjacent-state correlations, see Eq. (\ref{alphafit}). Dashed line shows the analytically expected $N\to\infty$ behaviour, see Eq.~(\ref{betann-del}) for the delocalized phase and Eq.~(\ref{betann}) for the localized phase (this part of the dashed line is a guide for an eye.) From Ref.~\cite{tikhonov2020eigenstate}. }
\label{alpha_rrg}
\end{figure}
%%%%%%%%%%%%%%%%%%%%%%%%%%%%%%%%%%%%%%%%%%%%%%%%%%%%%%%%%%%%%%%%%%%%%%%%%%%%%%%%%%%%%%%%%%%%%%%%%%%%%%%%%%%%%%%%%%%%%%%%%%%%%%%%%%%%%%%%%%%%%%%%%%%%%%%%%%%%%%%%%%%%%%%%%%%%%%%%%%%%%%%%%%%%%%%%%%%%%%%%%%%%%%%%%

Let us now turn to another correlator that is closely related to $\beta(\omega)$---a correlation function of adjacent-in-energy eigenstates:
\be
\beta_{\textrm{nn}} = \Delta\left<\sum_k\delta(E_k-E)\left\vert \psi
_{k}\left( j\right) \psi _{k+1}\left( j\right) \right\vert ^{2}\right>.
\label{beta-nn}
\ee
Here the subscript ``nn'' stands for ``nearest neighbor'' (in energy space).  We follow the analysis of $\beta_{\textrm{nn}}$ on RRG that was performed in Ref.~\cite{tikhonov2020eigenstate}.

Clearly, $\beta_{\textrm{nn}} \simeq \beta(\omega \sim \Delta)$, where $\Delta$ is the level spacing.
Thus, in the delocalized phase and in the large-$N$  limit (the condition is $N\gg N_\xi$) one has
\be
\label{betann-del}
N^2\beta_{\textrm{nn}} = N_\xi/3.
\ee
The coefficient $1/3$ in this equation holds if normalization of correlation volume is fixed by the condition $P_2 \simeq N_\xi / N$ at $N \gg N_\xi$. The behavior in the localized phase is discussed in detail in Sec.~\ref{sec:eigenfunc_corre_loc} below; the result is, according to Eq.~(\ref{betafit}),
\be
\label{betann}
N^2\beta_{\textrm{nn}} \sim N^{\mu(W)}.
\ee

To characterize the evolution of $\beta_\textrm{nn}$ with the system size $N$, it is useful to define  a disorder-  and size-dependent exponent
(cf. a similar procedure in Sec.~\ref{sec:IPR} where ED data for IPR are analyzed in a similar way):
\be
\label{alphafit}
\mu_{\textrm{nn}}(W, N) = \frac{\partial\ln\left(N^2\beta_{\textrm{nn}}\right)}{\partial\ln N}.
\ee
On the delocalized side, $ N^2\beta_{\textrm{nn}}$ is independent on $N$ at large $N$, which implies that $\mu_{\textrm{nn}}(W<W_c,N) \to 0$  at $N\gg N_\xi(W)$. At the critical point, $W=W_c$, we have $\mu_{\textrm{nn}}(W_c) \to 1$  at $N\to \infty$. On the localized side,   Eq.~(\ref{betann}) yields $\mu_{\textrm{nn}}(W>W_c, N) \to \mu(W)$ in the large-$N$ limit. 

In Fig. \ref{alpha_rrg}, ED data for $N^2\beta_{\textrm{nn}}(W, N)$ (left panel) and the corresponding results for $\mu_{\textrm{nn}}(W, N)$ (right panel) are shown. As expected, for $W<W_c$  the $\mu_{\textrm{nn}}(N)$ curves gradually drift downwards, towards zero, with increasing $N$. Closer to $W_c$, this drift is in fact non-monotonic (first upward, then downward); the reason for this was discussed in Sec.~\ref{sec:IPR}.  Still closer to the critical point, for $15 \lesssim W < W_c$ only upward drift is observed. This is related to finite--size limitation and the upward is expected to be superseded by a downward drift at sufficiently large $N$ to give $\mu_{\textrm{nn}}(W) \to 0$ in the $N\to\infty$ limit. On the localized side, $W > W_c$, numerical simulations yield a nearly $N$-independent $\mu_{\textrm{nn}}(W, N)$, in consistency with the expected limiting behavior  $\mu_{\textrm{nn}}(W>W_c, N) \to \mu(W)$.

%%%%%%%%%%%%%%%%%%%%%%%%%%%%%%%%%%%%%%%%%%%%%%%%%%%%%%%%%%%%%%%%%%%%%%%%%%%%%%%%%%%%%%%%%%%%%%%%%%%%%%%%%%%%%%%%%%%%%%%%%%%%%%%%%%%%%%%%%%%%%%%%%%%%%%%%%%%%%%%%%%%%%%%%%%%%%%%%%%%%%%%%%%%%%%%%%%%%%%%%%%%%%%%%%
%%%%%%%%%%%%%%%%%%%%%%%%%%%%%%%%%%%%%%%%%%%%%%%%%%%%%%%%%%%%%%%%%%%%%%%%%%%%%%%%%%%%%%%%%%%%%%%%%%%%%%%%%%%%%%%%%%%%%%%%%%%%%%%%%%%%%%%%%%%%%%%%%%%%%%%%%%%%%%%%%%%%%%%%%%%%%%%%%%%%%%%%%%%%%%%%%%%%%%%%%%%%%%%%%

\subsection{Level statistics}

%%%%%%%%%%%%%%%%%%%%%%%%%%%%%%%%%%%%%%%%%%%%%%%%%%%%%%%%%%%%%%%%%%%%%%%%%%%%%%%%%%%%%%%%%%%%%%%%%%%%%%%%%%%%%%%%%%%%%%%%%%%%%%%%%%%%%%%%%%%%%%%%%%%%%%%%%%%%%%%%%%%%%%%%%%%%%%%%%%%%%%%%%%%%%%%%%%%%%%%%%%%%%%%%%
\begin{figure}[tbp]
\minipage{0.50\textwidth}\includegraphics[width=\textwidth]{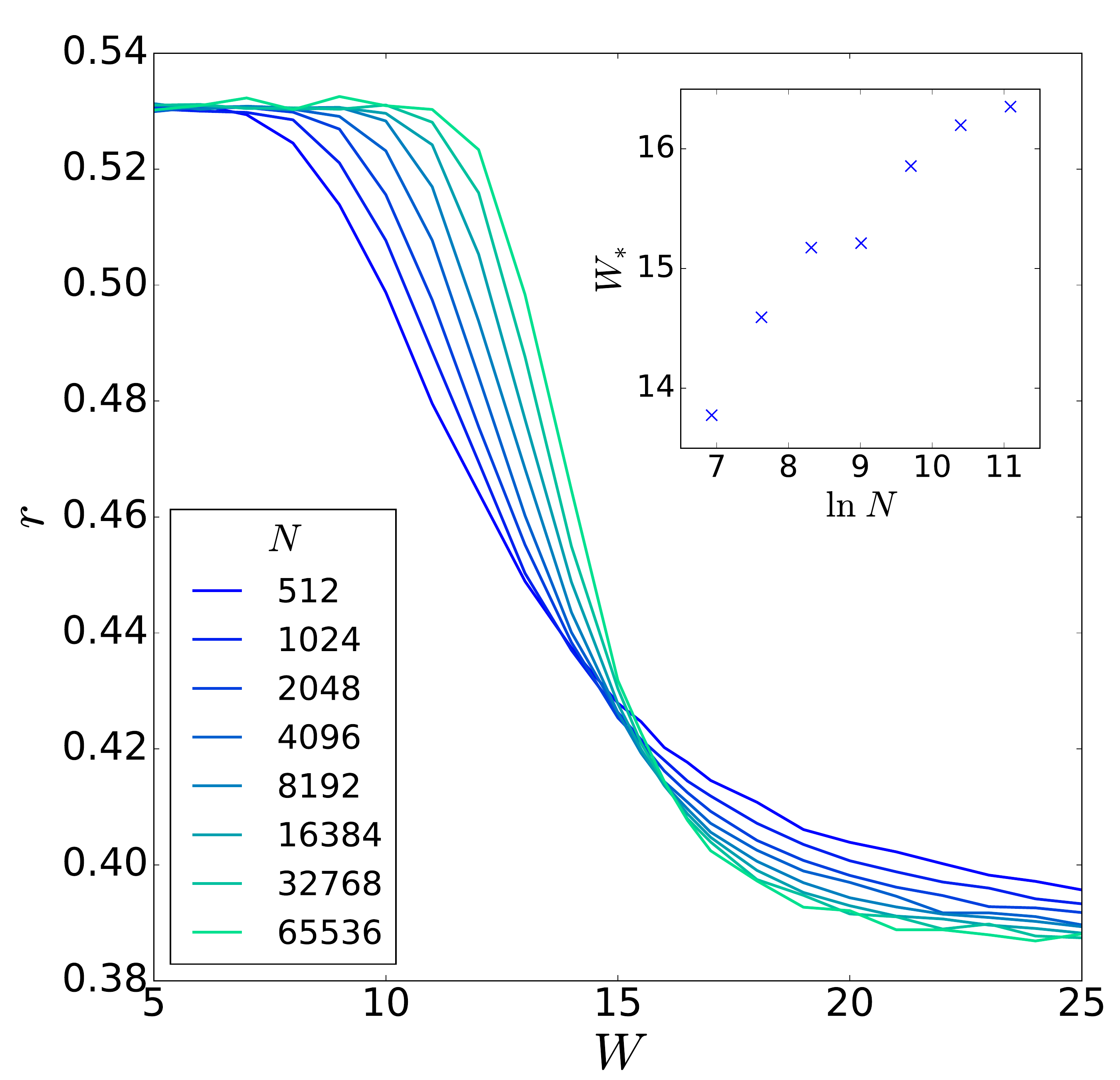}\endminipage
\minipage{0.50\textwidth}\includegraphics[width=\textwidth]{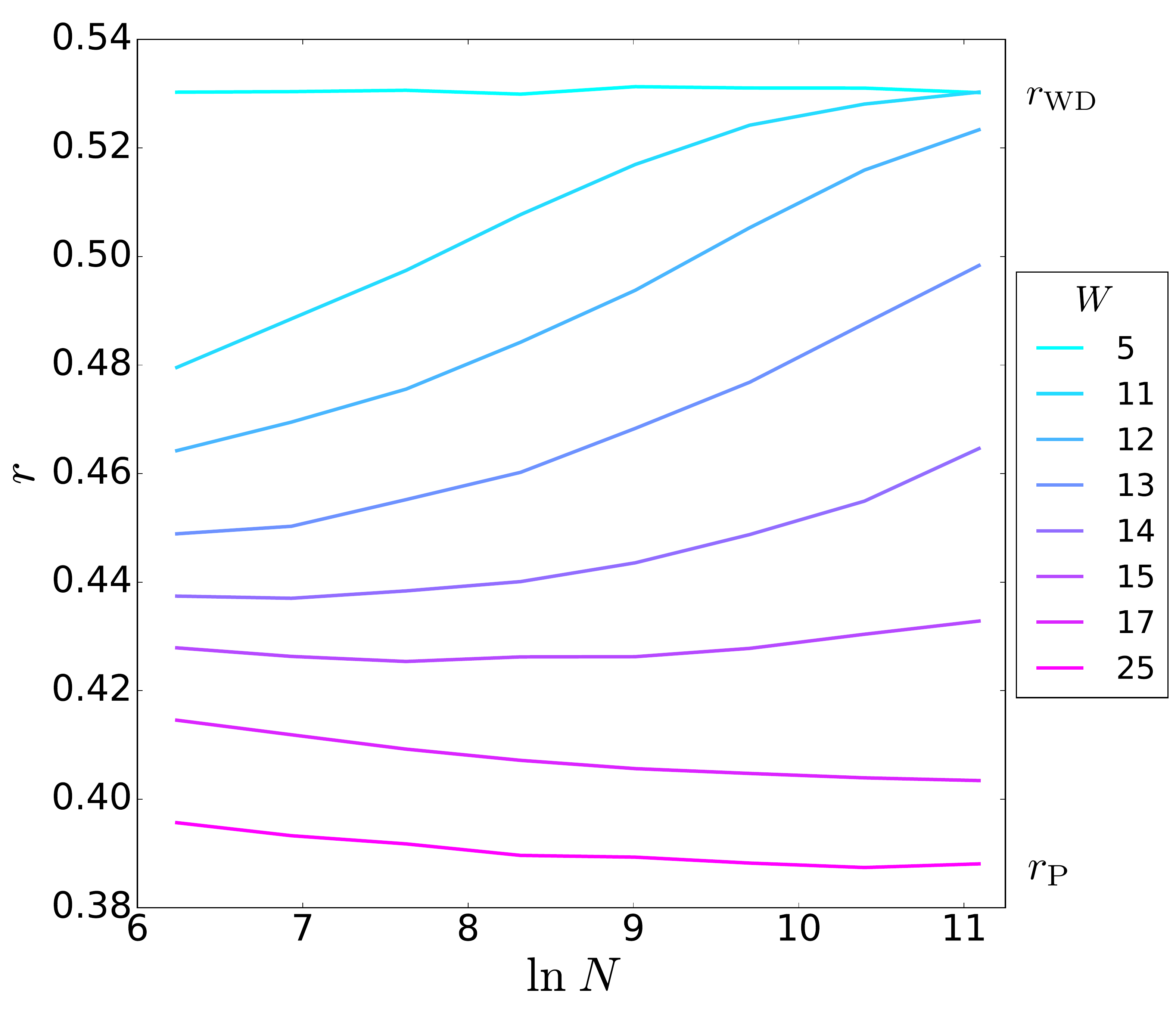}\endminipage
\caption{Mean adjacent gap ratio $r$. {\it Left:} $r(W)$ at various $N$. Inset: drift of the crossing point $W_*$ with linear system size $\ln N$. {\it Right:} $r(\ln N)$, at various $W$. From Ref.~\cite{tikhonov2016anderson}.}
\label{rrg:adjgap}
\end{figure}
%%%%%%%%%%%%%%%%%%%%%%%%%%%%%%%%%%%%%%%%%%%%%%%%%%%%%%%%%%%%%%%%%%%%%%%%%%%%%%%%%%%%%%%%%%%%%%%%%%%%%%%%%%%%%%%%%%%%%%%%%%%%%%%%%%%%%%%%%%%%%%%%%%%%%%%%%%%%%%%%%%%%%%%%%%%%%%%%%%%%%%%%%%%%%%%%%%%%%%%%%%%%%%%%%

Eigenenergies $E_k$ of disordered tight-binding models  are correlated random variables whose statistics on $d$-dimensional lattices has been investigated intensively for several decades. Two most popular means to characterize the multivariate distribution function of energy levels $\mathcal{P}(\{E_i\})$ are the statistics of $P(\omega)$ of spacings between adjacent levels  and the two-level correlation function $R(\omega)$. 
\subsubsection{Gap ratio}
\label{sec:gr}
The energy levels have qualitatively distinct statistical properties in the localized and delocalized phase of the model. This transition in the level statistics, which becomes a crossover for a finite system size, has been studied in detail in finite-$d$ models \cite{hofstetter1993statistical,zharekeshev1995scaling,zharekeshev1997asymptotics,varga1995shape,kaneko1999three,milde2000energy}. Following Refs.~\cite{oganesyan2007localization,biroli2012difference}, we use the ensemble-averaged ratio $r=\langle{r_n}\rangle$ of two consecutive spacings,  
\be
\label{agr}
r_n = \min(\delta_n,\delta_{n+1})/\max(\delta_n,\delta_{n+1}),
\ee
which changes from $r_{\rm{P}}=0.386$ to $r_{\rm{WD}}=0.530$, with limiting values corresponding to the Poisson and the WD Gaussian orthogonal ensemble (GOE) limits.

The disorder--dependence of $r$ for a set of $N$ is shown in Fig.~\ref{rrg:adjgap}, left panel. As expected, the increase of $W$ produces a crossover from the GOE to the Poisson value. This crossover becomes sharper for larger $N$, remaining rather broad even for $N=2^{16}$. This implies that the critical regime is broad up to the largest $N$ studied.

The curves in Fig.~\ref{rrg:adjgap}, left panel apparently demonstrate a crossing point near $W=15$. A closer inspection shows that with $N$ increasing from $2^{9}$ to $2^{16}$, the apparent crossing point drifts from $W_*\simeq{14}$ to $W_*\simeq{16}$. At the same time, the value of $r$ at the ``moving crossing point'' gradually decreases towards the Poisson value.  This is a manifestation of the localized nature of the Anderson transition critical point on tree-like graphs. Close to the lower critical dimension $d=2$, the critical point corresponds to weak disorder, and the critical level statistics is close to the WD one (multifractality is weak). With increasing $d$ the critical point moves towards strong disorder, so that the level statistics approaches the Poisson form (multifractality takes its strongest possible form) in the limit $d\to\infty$. As we discussed in Sec. \ref{sec:WF_corr}, the latter limit corresponds to tree-like models. Therefore, in contrast to finite--$d$ models, no intermediate true crossing point for curves $r(W)$ is expected: the crossing point should necessarily drift towards the Poisson limit with increasing $N$. This is exactly what we observe in Fig.~\ref{rrg:adjgap}, left panel. 

A complementary view on the same data is provided on  Fig.~\ref{rrg:adjgap}, right panel, where a set of curves $r(N)$ corresponding to different $W$ is shown. For moderate disorder ($W<W_c$), dependence $r(N)$ is non-monotonic. This behavior is a simple consequence of the gradual drift of the apparent crossing point which is explained above. Exactly at critical disorder, $W=W_c$, the system develops (with increasing $N$) the properties, specific to the critical point and, $r$ decreases down to $r_{\rm P}$. On the delocalized side ($W<W_c$) in the vicinity of the transition, it behaves as a critical system as long as its linear size $\log_2 N$ is smaller than the correlation length $\xi(W)$. Thus, for $N<N_\xi(W)\sim{2^{\xi(W)}}$, spectral statistics (as well as other observables) develops as at criticality ($r$ decreases with growing $N$). As soon as $N$ reaches $N_\xi(W)$, the system ``feels'' that it is actually in the delocalized phase, and $r$ drifts towards its large-$N$ limit $r_{\rm WD}$. Thus, $N\sim N_\xi$ marks the point of the minimum of $r(N)$ curve. 

%%%%%%%%%%%%%%%%%%%%%%%%%%%%%%%%%%%%%%%%%%%%%%%%%%%%%%%%%%%%%%%%%%%%%%%%%%%%%%%%%%%%%%%%%%%%%%%%%%%%%%%%%%%%%%%%%%%%%%%%%%%%%%%%%%%%%%%%%%%%%%%%%%%%%%%%%%%%%%%%%%%%%%%%%%%%%%%%%%%%%%%%%%%%%%%%%%%%%%%%%%%%%%%%%
\begin{figure}[tbp]
\minipage{0.50\textwidth}\includegraphics[width=\textwidth]{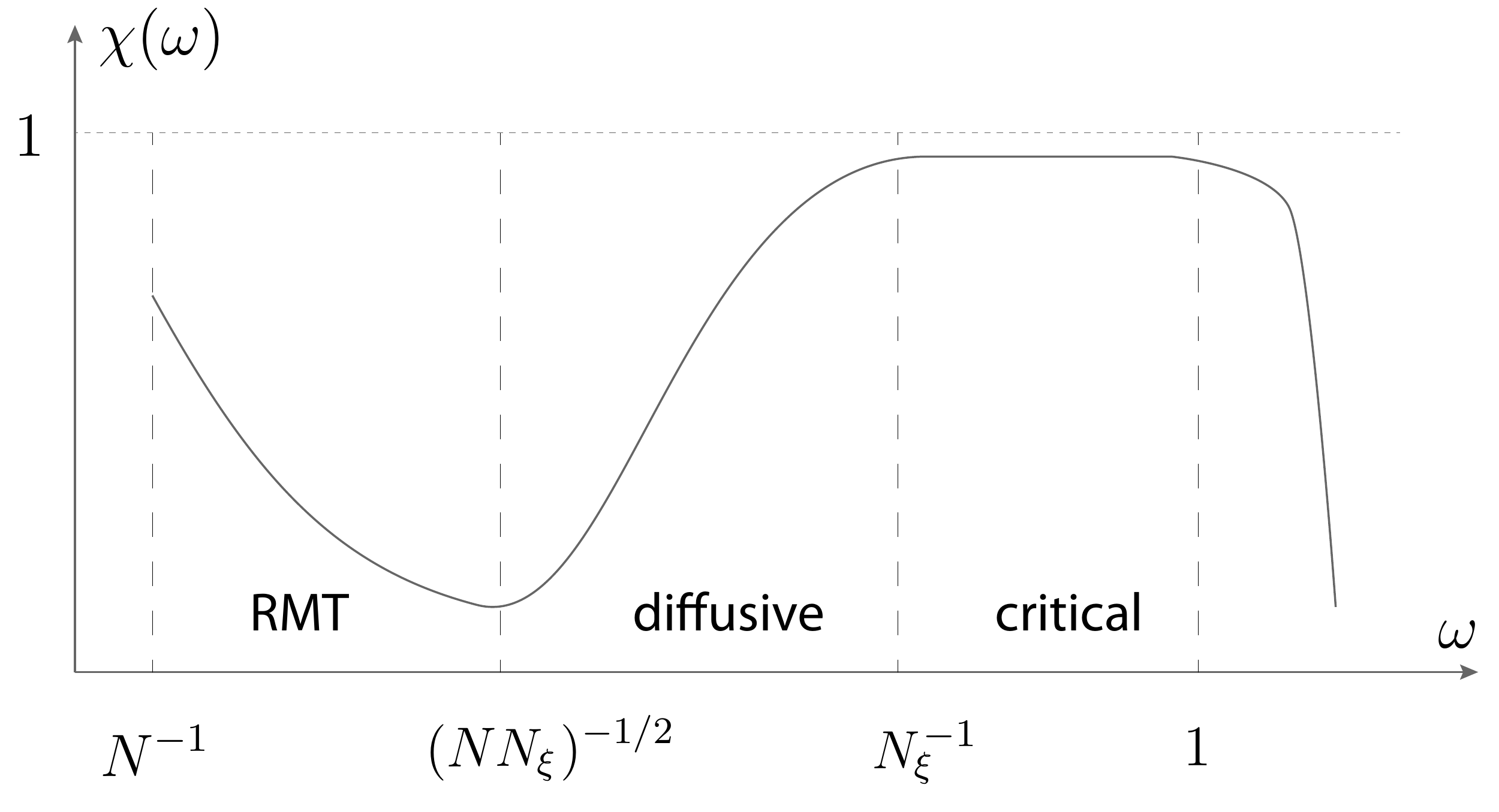}\endminipage
\minipage{0.50\textwidth}\includegraphics[width=\textwidth]{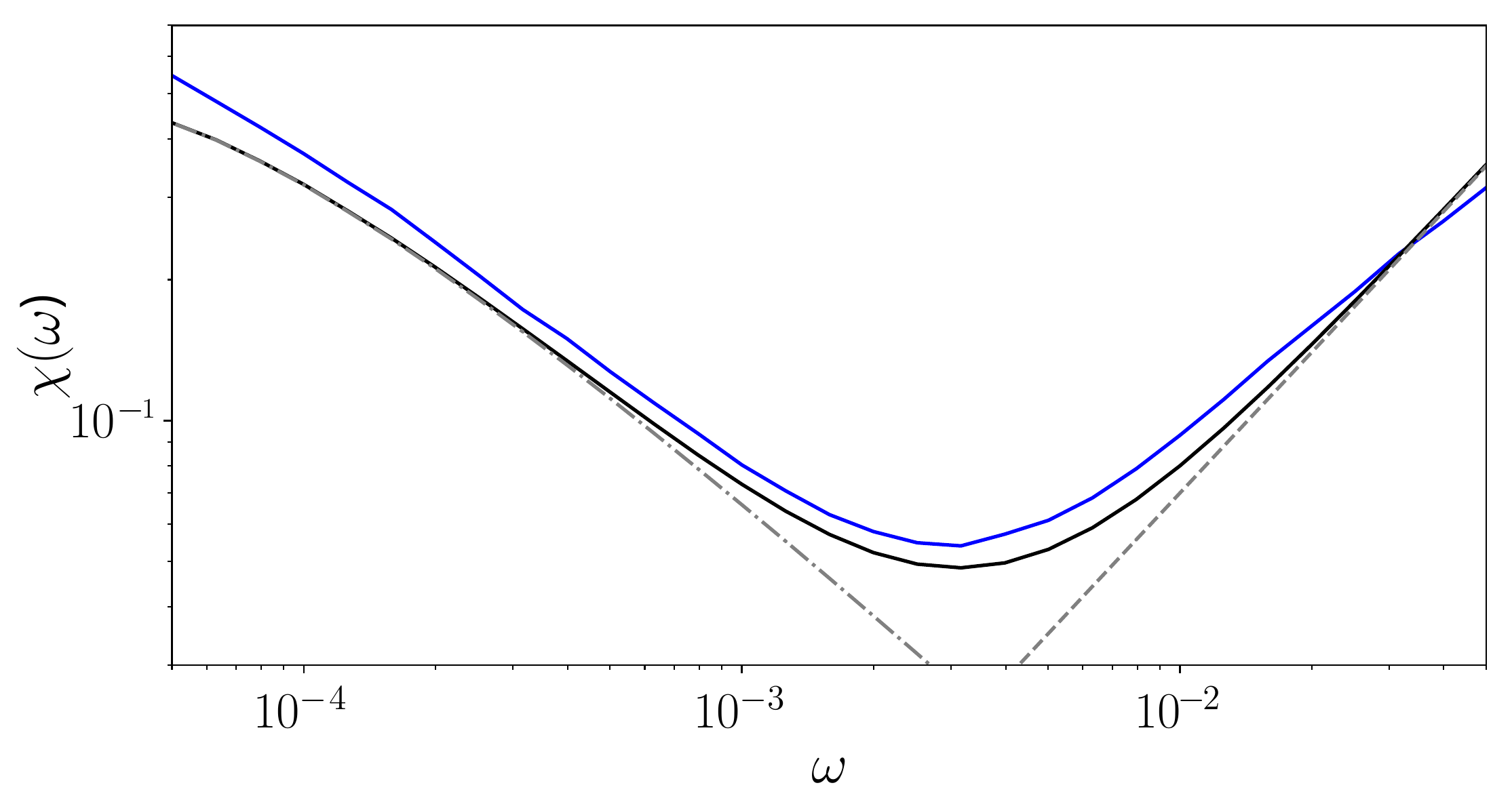}\endminipage
\caption{ Level statistics on RRG: Relative level number variance $\chi(\omega)$.  {\it Left:} Schematic representation of analytical predictions for $\chi(\omega)$ for $W < W_c$ in the vicinity of the transition, $N \gg N_\xi \gg 1$, see Eqs.~(\ref{chi-WD}), (\ref{chi-RRG-diff}) and Eq.~(\ref{chi-star}) for RMT, diffusive and critical results, correspondingly. {\it Right:} Blue line: ED results for $\chi(\omega)$ ($N=2^{17}$, $W=8$). Black solid line: Eq.~(\ref{RWDDiff}), dash-dotted line:  RMT contribution, Eq.~(\ref{chi-WD}), dashed line: diffusive contribution (\ref{chi-RRG-diff}).  From Ref.~\cite{tikhonov19statistics}. }
\label{rrg:spec}
\end{figure}
%%%%%%%%%%%%%%%%%%%%%%%%%%%%%%%%%%%%%%%%%%%%%%%%%%%%%%%%%%%%%%%%%%%%%%%%%%%%%%%%%%%%%%%%%%%%%%%%%%%%%%%%%%%%%%%%%%%%%%%%%%%%%%%%%%%%%%%%%%%%%%%%%%%%%%%%%%%%%%%%%%%%%%%%%%%%%%%%%%%%%%%%%%%%%%%%%%%%%%%%%%%%%%%%%

\subsubsection{Level correlations}
There exists a close correspondence between the two-level correlation function $R(\omega)$ and the variance $\Sigma_2(\omega)$ of the number of levels $I(\omega)$ within a band of the width $\omega$,
\be
\label{variance}
\Sigma_2(\omega)=\left< I^{2}(\omega )\right>-\left<I(\omega )\right>^{2},
\ee
which is a convenient characteristics of the rigidity of the spectrum at $\omega \gg \Delta$. 
In particular, for the Poisson and GOE statistics the level number variance reads 
\begin{eqnarray}
& \Sigma_2(\omega)  = \omega/\Delta, & \qquad \text{Poisson}, \\[0.1cm]
& \displaystyle \Sigma_2(\omega) \simeq \frac{2}{\pi^2}\ln\frac{2\pi\omega}{\Delta}, & \qquad \text{GOE}; \ \omega \gg \Delta.
\end{eqnarray}
Numerically, it is advantageous to evaluate
\be
\label{chi-omega}
\chi(\omega)=\Sigma_2(\omega)/\left< I(\omega )\right>.
\ee
For the Poisson statistics it is equal to unity; in the WD case it becomes
\be
\displaystyle \chi(\omega) \simeq \frac{2}{\pi^2}\frac{\Delta}{\omega} \ln\frac{2\pi\omega}{\Delta},  \qquad \text{GOE}; \ \omega \gg \Delta. 
\label{chi-WD}
\ee
A review of the behavior of $R(\omega)$ and $\Sigma_2(\omega)$ in a metallic sample of spatial dimensionality $d < 4$ can be found in Refs.~\cite{aronov1995fluctuations,mirlin00}.

By definition, the RRG two-level correlation function can be written as
\be
\label{R-omega-B}
R(\omega)= \Delta^2 \sum_{ij} B_{ij}(E, \omega),
\ee
where $B_{ij}(E,\omega)$ is the correlation function of local densities of states defined by Eq.~(\ref{defB}). Using the result Eq. (\ref{Bij-result}), the connected part $R^{(c)}(\omega)$ of the level correlation function can be presented in the form
\be
\label{RWDDiff}
R^{(c)}(\omega) = R^{(c)}_{\rm WD}(\omega)+R^{(c)}_{\textrm{diff}}(\omega),
\ee
where 
\be
\label{Rdiff}
R_{\textrm{diff}}^{(c)}(\omega) = \frac{\Delta}{2\pi^2 \nu} \sum_r (m+1)m^{r-1}\Re K_{1}^{(c)}(r,\omega)
\ee
and $K_{1}^{(c)}(r,\omega) = K_1(r,\omega)-\left|\left<G_R(0)\right>\right|^2$ is the connected part of the Bethe-lattice correlation function (\ref{K1w}).  For $\omega \ll N_\xi^{-1}$ the correlation function $K_{1}^{(c)}(r,\omega)$ is essentially independent of $\omega$ and thus can be replaced by $K_{1}^{(c)}(r,0) \equiv K_{1}^{(c)}(r) \simeq K_2(r)$. Using Eq.~(\ref{K2-deloc}), we see that the sum over $r$ in Eq.~(\ref{Rdiff}) converges at $r \sim 1$,  yielding
\be
\label{RRes}
R_{\textrm{diff}}^{(c)}(\omega)\sim\frac{N_\xi}{N}, \qquad   \omega<N_{\xi}^{-1}.
\ee
The corresponding behavior of the relative number variance $\chi(\omega)$ is
\be
\label{chi-RRG-diff}
\chi(\omega) \sim  N_\xi\, \omega.
\ee

The Eq. (\ref{RWDDiff}) suggests that the RMT level statistics describes the level correlations in a rather wide frequency range when the system is sufficiently large, $N \gg N_\xi$. The frequency scale at which the level correlation function loses its universal character is determined by comparison of the two contributions in Eq.~(\ref{RWDDiff}). The crossover scale reads
\be
\omega_c\sim \frac{1}{ (N N_\xi)^{1/2}}.
\label{omega-c}
\ee
In $d<4$ dimensions the universal (RMT) behavior ceases to be valid at the Thouless energy $E_{\rm Th}$. In the case of RRG, the situations is different. Indeed, the Thouless energy, is defined as inverse of the time $t_{\rm Th}$ required for particle to diffusely reach all points of the system. This time scales only logarithmically with the volume $N$ of RRG:
\be
\label{thouless-time}
 t_{\rm Th} \sim \frac{D}{\ln N /\ln  m},
\ee
where $D$ is the diffusion coefficient, see Eq.~(\ref{diff}). 

Let us now discuss the spectral statistics right at the transition. As Eq.~(\ref{omegaN}) suggests, the critical eigenstates are fully correlated only up to the scale $\omega_N$, which is  smaller that the level spacing $\Delta$ by a logarithmically large factor $\ln^{3/2}N \gg 1$. This implies that the level repulsion is significant only for small frequencies $\omega \lesssim \omega_N$. Therefore, the level statistics of the RRG model at criticality approaches the Poisson statistics with increasing $N$ and the critical level compressibility is Poissonian
\be
\chi_* = 1,
\label{chi-star}
\ee
in contrast to a finite-$d$ system, for which the critical statistics is intermediate between WD and Poisson forms \cite{shklovskii1993statistics} and $0 < \chi_* < 1$. The approach of $\chi$ to the critical value (\ref{chi-star}) is expected to be logarithmically slow in frequency due to logarithmic dependence of $\omega_N / \Delta$ on $N$.

The analytical results for the relative level number variance  $\chi(\omega)$ for a delocalized RRG system close to the metal-insulator transition, $N\gg N_{\xi}\gg 1$, are summarized in the left panel of Fig.~\ref{rrg:spec}.

%%%%%%%%%%%%%%%%%%%%%%%%%%%%%%%%%%%%%%%%%%%%%%%%%%%%%%%%%%%%%%%%%%%%%%%%%%%%%%%%%%%%%%%%%%%%%%%%%%%%%%%%%%%%%%%%%%%%%%%%%%%%%%%%%%%%%%%%%%%%%%%%%%%%%%%%%%%%%%%%%%%%%%%%%%%%%%%%%%%%%%%%%%%%%%%%%%%%%%%%%%%%%%%%%
\begin{figure}[tbp]
\minipage{0.50\textwidth}\includegraphics[width=\textwidth]{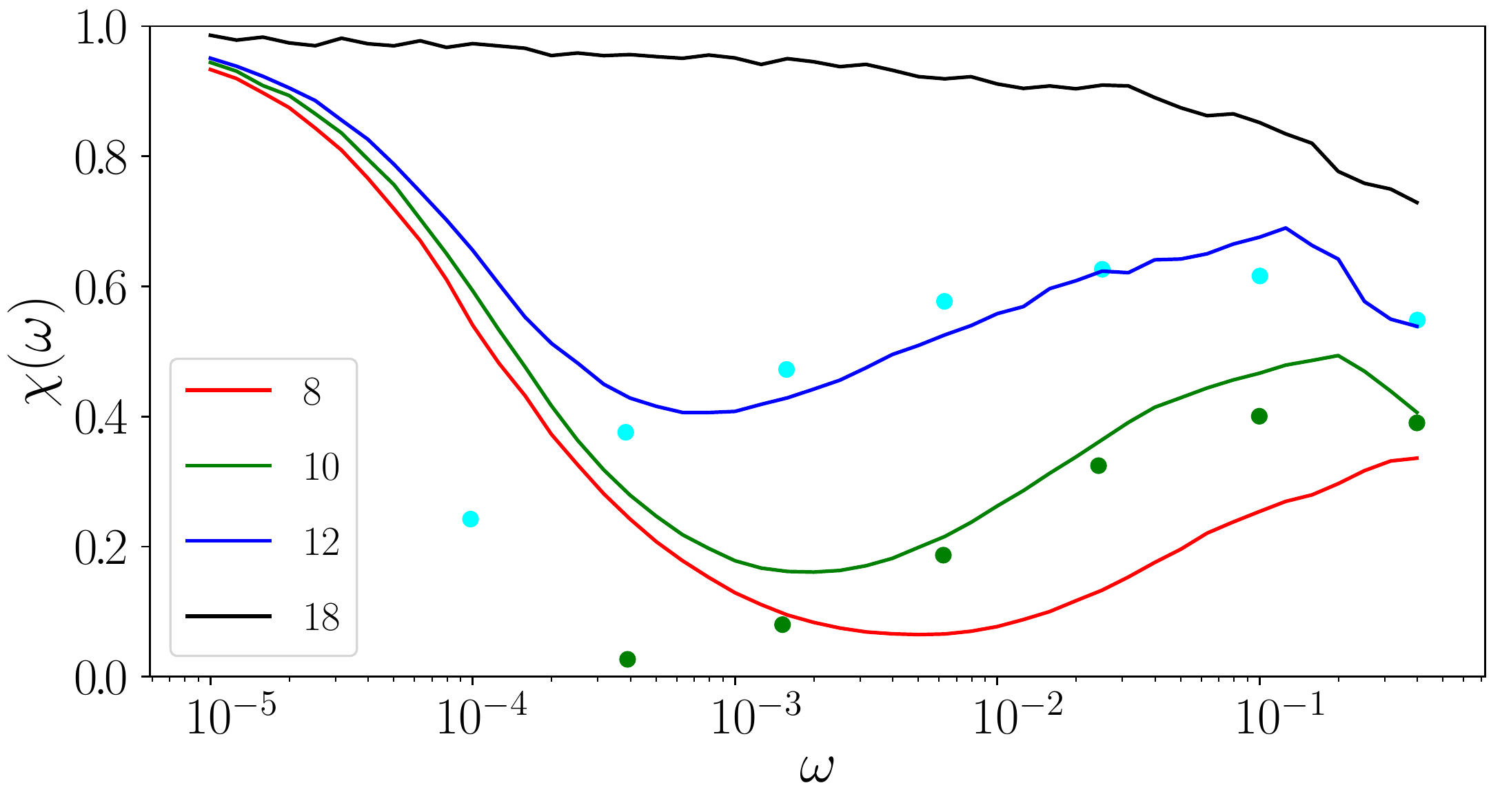}\endminipage
\minipage{0.50\textwidth}\includegraphics[width=\textwidth]{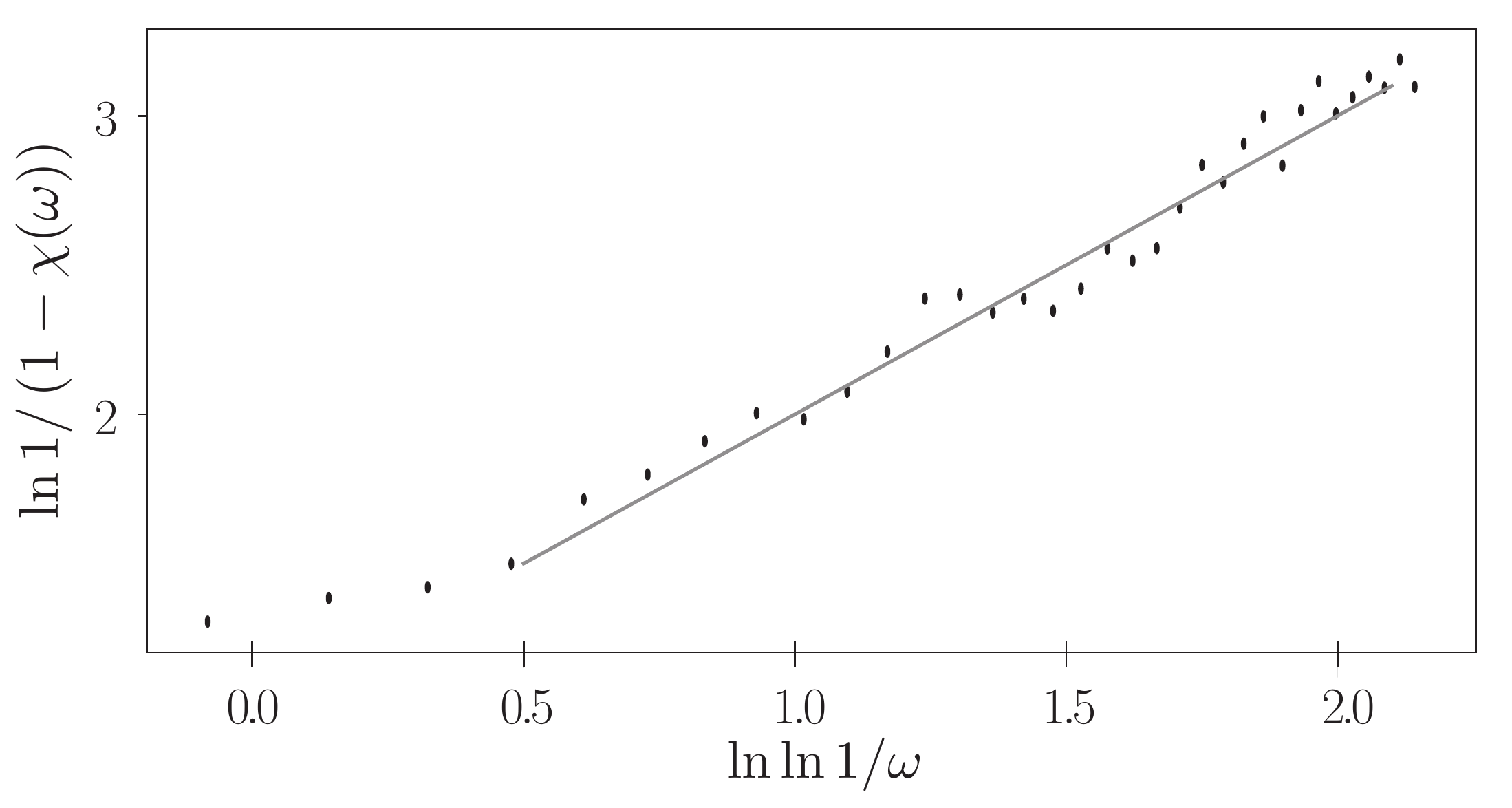}\endminipage
\caption{{\it Left:} Evolution of level number variance $\chi(\omega)$ on the disorder. Solid lines: ED result ($N=2^{16}$) for delocalized phase ($W=8$, 10, and 12) and for the critical point ($W=18$). Dots: the results of Ref.~\cite{metz2017level} obtained by solution of the self-consistency equation, corresponding to the limit of $N \to \infty$ at fixed $\omega$, note that RMT contribution is discarded in this limit. Green dots: $W=10$, cyan dots: $W=12.5$.  {\it Right:}  $\ln (1-\chi(\omega))^{-1}$ vs $ \ln \ln (1/\omega)$ at the critical point. The straight line: Eq.~(\ref{chi-critical-fit}) with $\mu'=1$. From Ref.~\cite{tikhonov19statistics}. }
\label{rrg:spec2}
\end{figure}
%%%%%%%%%%%%%%%%%%%%%%%%%%%%%%%%%%%%%%%%%%%%%%%%%%%%%%%%%%%%%%%%%%%%%%%%%%%%%%%%%%%%%%%%%%%%%%%%%%%%%%%%%%%%%%%%%%%%%%%%%%%%%%%%%%%%%%%%%%%%%%%%%%%%%%%%%%%%%%%%%%%%%%%%%%%%%%%%%%%%%%%%%%%%%%%%%%%%%%%%%%%%%%%%%

ED numerical results for the spectral correlations are shown in the Figs.~\ref{rrg:spec}, right panel and in Fig.~\ref{rrg:spec2}. On the first plot, $\chi(\omega)$ is shown on the log-log scale. The universal RMT and diffusive regimes are clearly seen, cf. the left panel of this figure. On the same plot, we show the analytical prediction (\ref{RWDDiff}) which matches numerical result nicely. 

On the Fig.~\ref{rrg:spec2}, left panel we show the evolution of $\chi(\omega)$ curve with disorder increasing towards $W_c$. In the metallic domain ($W<W_c$), both RMT and diffusive regions are observed. The point of the minimum of $\chi$ marks the crossover scale, which drifts towards smaller $\omega$ with increasing disorder (i.e., increasing $N_\xi$). The largest value of $\chi(\omega)$, reached at the right border of the diffusive regime, drifts towards the Poisson value, Eq. (\ref{chi-star}) and the critical regime starts to develop. In order to numerically observe the critical regime, the simulations were performed also directly  at the critical disorder  ($W=18$, black curve). 
The gradual approach of $\chi(\omega)$ to its critical value Eq. (\ref{chi-star}) can indeed be described by 
\be
\label{chi-critical-fit}
\chi(\omega) = 1 - \frac{c^{(\chi)}}{\ln^{\mu'} (1/\omega)},
\ee
with $\mu' \simeq 1$, as shown in the Fig.~\ref{rrg:spec2}, right panel (it is probable that $\mu' = 1$ is an exact value of this index).

In Ref.~\cite{metz2017level}, the relative level number variance on RRG was studied by means of the saddle-point method that yields the self-consistency equations (\ref{2tra}) and (\ref{1tra}). The results of Ref.~\cite{metz2017level}  are shown in the Fig.~\ref{rrg:spec2} by green ($W=10$) and cyan $W=12.5$ dots. A good agreement between these data and ED results of Ref.~\cite{tikhonov19statistics} is observed for not too low frequency, $\omega > \omega_c$. For lower frequencies, the results of ED cross over to the RMT behavior, while the data of Ref.~\cite{metz2017level} continue to follow the diffusive behavior, Eq. (\ref{chi-RRG-diff}) (the analysis of Ref.~\cite{metz2017level} was performed in the limit $N\to \infty$ at fixed $\omega$ which does not allow to include the RMT part).

%%%%%%%%%%%%%%%%%%%%%%%%%%%%%%%%%%%%%%%%%%%%%%%%%%%%%%%%%%%%%%%%%%%%%%%%%%%%%%%%%%%%%%%%%%%%%%%%%%%%%%%%%%%%%%%%%%%%%%%%%%%%%%%%%%%%%%%%%%%%%%%%%%%%%%%%%%%%%%%%%%%%%%%%%%%%%%%%%%%%%%%%%%%%%%%%%%%%%%%%%%%%%%%%%
%%%%%%%%%%%%%%%%%%%%%%%%%%%%%%%%%%%%%%%%%%%%%%%%%%%%%%%%%%%%%%%%%%%%%%%%%%%%%%%%%%%%%%%%%%%%%%%%%%%%%%%%%%%%%%%%%%%%%%%%%%%%%%%%%%%%%%%%%%%%%%%%%%%%%%%%%%%%%%%%%%%%%%%%%%%%%%%%%%%%%%%%%%%%%%%%%%%%%%%%%%%%%%%%%

\vspace{1cm}

\subsection{Wavefunction correlations: Localized phase}
\label{sec:eigenfunc_corre_loc}

In Sec.~\ref{sec:eigenfunc_corr_dynamical}, dynamical correlations of eigenstates $\beta(\omega)$ on RRG were studied in the ergodic phase and in the critical regime.
In this subsection, we consider, following Ref.~\cite{tikhonov19statistics}, the  RRG correlation function $\beta(\omega)$ in the localized phase. In combination with Sec.~\ref{sec:eigenfunc_corr_dynamical}, this yields   a full description of eigenstates correlations around the localization transition on RRG. 

Let us start with a qualitative discussion. For $W> W_c$, individual eigenstates are localized on different sites with exponentially decaying wave functions and typically overlap very weakly. However, with certain probability two such states form a resonance, which strongly enhances the overlap.  The probability of a resonance is enhanced for small energy difference $\omega$, hence $N^2 \beta(\omega)$ in the localized phase should decay with increase of $\omega$. For Anderson localization problem in $d$ dimensions, this decay was characterized in Ref.~\cite{cuevas07multifrac}, with the result
\be
\label{finite-d-loc}
N^2 \beta(\omega) \sim \xi^{d-d_2} \ln^{d-1} (\delta_\xi / \omega)\,, \qquad \omega < \delta_\xi \,,
\ee
where $\xi$ is the localization length, $\delta_\xi \sim \xi^{-d}$ is the level spacing in the localization volume, and $d_2$ the multifractal exponent. It was pointed out in Ref.~\cite{cuevas07multifrac} that the logarithmic enhancement of correlations with lowering $\omega$ in Eq.~(\ref{finite-d-loc}) is closely related to the Mott's law for the ac conductivity. It is not straightforward to translate the result in Eq. (\ref{finite-d-loc}) to the the RRG model.  Equation (\ref{finite-d-loc}) suggests that the enhancement of correlations for small $\omega$ on RRG should be faster than a power of $\ln \omega$. As we show below, in the localized phase of the RRG model the function $N^2 \beta(\omega)$ has a power--law dependence on $\omega$, with an exponent that is a slowly decaying function of disorder.

Let us first consider the limit of strong disorder $W$. In this case, almost every single-particle state is localized within a small localization length $\zeta$ around a certain lattice site.  Typically, two localized states are separated by a distance of the order of system size $L = \ln N / \ln m$ and have an overlap $\propto e^{-L / \zeta}$. However, via rare resonant events, two states located far apart may form a resonant pair and strongly hybridise. Such a pair gives a maximal possible contribution to the correlation function $\beta(\omega)$. Even though such events are rare, they determine the average value $\beta(\omega)$ in the case of $d$-dimensional system with $d>1$, see Ref.~\cite{cuevas07multifrac}.
This resonant enhancement is reflected by the factor $\ln^{d-1} (\delta_\xi / \omega)$ in Eq.~(\ref{finite-d-loc}). Its role is clearly increasing with increasing $d$. As discussed below, the power-law scaling of $\beta(\omega)$ on RRG is a direct consequence of this resonance mechanism.

Localized eigenstates on a tree-like graph decay as follows \cite{zirnbauer1986localization,mirlin1991localization,tikhonov19statistics,PhysRevResearch.2.012020}, see Eq.~(\ref{alpha-r-loc}):
\be
\langle |\psi^2(r)| \rangle \sim m^{-r} \exp\{-r/\zeta(W) \} \,,
 \label{eigenstate-exp-decay}
 \ee
where $r$ is the distance from the ``localization center'' of the state and $\zeta(W)$ is the localization length. It behaves as $\zeta(W) \sim (W-W_c)^{-1}$ in the vicinity of the transition and $\zeta(W) \sim 1/ \ln (W / W_c)$ for $W\gg W_c$. While Eq.~(\ref{eigenstate-exp-decay}) is valid on average, let us assume  for a moment that all eigenstates decay in this way; this will be sufficient to understand the $W$-dependent power-law in eigenstate correlations.

%%%%%%%%%%%%%%%%%%%%%%%%%%%%%%%%%%%%%%%%%%%%%%%%%%%%%%%%%%%%%%%%%%%%%%%%%%%%%%%%%%%%%%%%%%%%%%%%%%%%%%%%%%%%%%%%%%%%%%%%%%%%%%%%%%%%%%%%%%%%%%%%%%%%%%%%%%%%%%%%%%%%%%%%%%%%%%%%%%%%%%%%%%%%%%%%%%%%%%%%%%%%%%%%%
\begin{figure}[tbp]
\centerline{\includegraphics[width=0.5\textwidth]{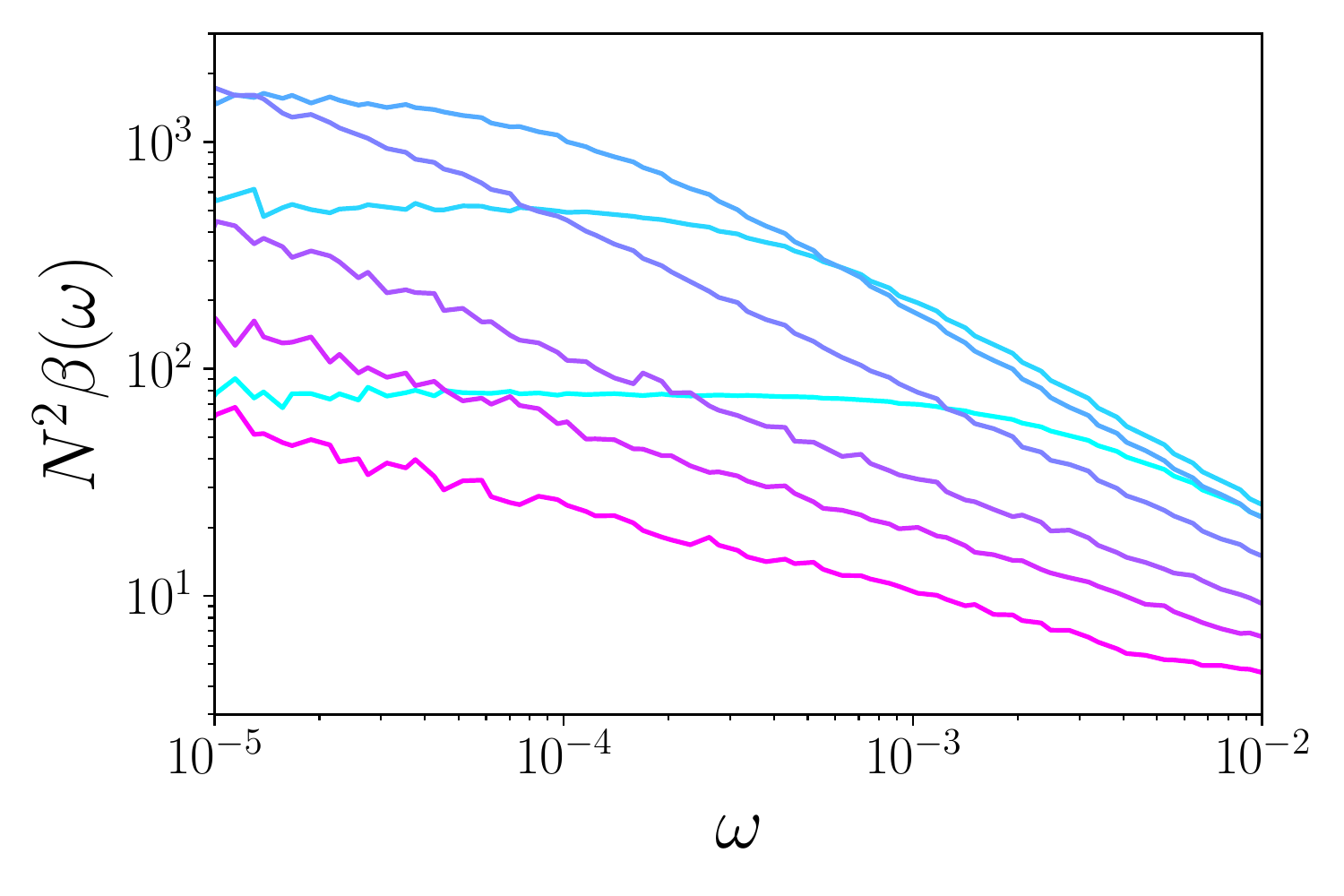}}
\caption{Correlation of different eigenstates $\beta(\omega)$ for RRG (ED, $N=2^{15}$) and, from cyan to magenta: $W= 10, 12, 14$ on the delocalized side, $W=18$ for the critical point, and $W=24, 30, 42$ on the localized side. From Ref.~\cite{tikhonov2020eigenstate}. }
\label{beta_rrg_full}
\end{figure}
%%%%%%%%%%%%%%%%%%%%%%%%%%%%%%%%%%%%%%%%%%%%%%%%%%%%%%%%%%%%%%%%%%%%%%%%%%%%%%%%%%%%%%%%%%%%%%%%%%%%%%%%%%%%%%%%%%%%%%%%%%%%%%%%%%%%%%%%%%%%%%%%%%%%%%%%%%%%%%%%%%%%%%%%%%%%%%%%%%%%%%%%%%%%%%%%%%%%%%%%%%%%%%%%%

Consider two eigenstates $\psi_k$ and $\psi_l$ localized at sites separated by the distance $R$. The corresponding overlap matrix element is $M \sim m^{-R} \exp\{-R/\zeta(W)\}$. The optimal condition of the Mott-like resonance for two eigenstates with the energy difference $\omega$ is $M \sim \omega$. Under this condition, $\psi_k$ and $\psi_l$ get strongly hybridized:
\be 
\sum_j |\psi_k(j) \psi_l(j)|^2 \sim 1.
\label{resonance-overlap}
\ee  
The resonant condition can be written as follows
\be
R(\omega)  \simeq \frac{\ln (1/\omega)}{\ln m + \zeta^{-1}(W)} \,.
\label{R-omega-res}
\ee
The total number of states in a sphere of radius $R$ centered at the state $\psi_k$ equals
\be
N_{R(\omega)} \sim m^{R(\omega)} \sim  \omega^{-\mu(W)} \,,
\ee
where
\be
\mu(W) = \frac{\zeta(W) \ln m}{\zeta(W) \ln m+1} \,.
\label{mu-omega}
\ee
The resonant condition implies that energy of one of these states is separated by $\sim \omega$ from energy of the state $\psi_k$. Thus, the probability $p_\omega$ of the resonance in the frequency interval $[\omega, 2 \omega]$ involving the given state $\psi_k$  is equal to
\be
p_\omega \sim \omega N_{R(\omega)} \sim \omega^{1- \mu(W)} \,.
\label{p-omega}
\ee
Using the definition (\ref{sigmadef}), we find
\be
N^2 \omega \beta(\omega) \sim N^2 \int_\omega^{2\omega} d\omega' \beta(\omega') =   \sum_{l \,:\:   \omega < | E_k - E_l | < 2\omega}
\left\langle \sum_j |\psi_k(j) \psi_l(j)|^2 \right\rangle \sim  \omega^{1- \mu(W)} \,.
\label{beta-omega-res}
\ee
In this consideration,  the state $k$ is fixed; we have used Eq.~(\ref{p-omega}) for the probability of a resonance in this interval and Eq.~(\ref{resonance-overlap}) for the resonant overlap.  As a result, we arrive at the following result:
\be
\label{betafit}
N^2\beta(\omega) \sim \omega^{-\mu(W)},
\ee
 where the exponent $\mu(W)$ is given by Eq.~(\ref{mu-omega}). 

Let us evaluate asymptotic behavior of the exponent $\mu(W)$. In the vicinity of the critical point (on the localized side), Eq.~(\ref{mu-omega}) gives 
\be 
\mu(W) \to 1 \,,  \qquad W \to W_c +0 \,.
\label{mu-critical}
\ee
This matches  (up to a logarithmic correction) the critical behavior $\beta(\omega) \propto 1/\omega$, see second line of Eq.~\eqref{betascres}. In the opposite limit of $W\gg W_c$,  we find
\be
\mu(W) \sim \frac{1}{ \ln(W/W_c)}\,, \qquad W\gg W_c \,.
\label{mu-strong-disorder}
\ee
Thus, $\mu(W)$ decays to zero at $W \to \infty$ but this decay is logarithmically slow.

In our reasoning above, we relied on Eq. (\ref{eigenstate-exp-decay}) which describes the average decay of a wavefunction. At the same time, wavefunctions fluctuate strongly; in particular, decay of the typical wavefunction amplitude is described by a different localization length \cite{PhysRevResearch.2.012020}. An account of strong fluctuations of eigenstates around the average does not change the main conclusion about the power-law scaling, Eq.  (\ref{betafit}) and yields qualitatively the same results for $\mu(W)$ \cite{tikhonov2020eigenstate}. 

%%%%%%%%%%%%%%%%%%%%%%%%%%%%%%%%%%%%%%%%%%%%%%%%%%%%%%%%%%%%%%%%%%%%%%%%%%%%%%%%%%%%%%%%%%%%%%%%%%%%%%%%%%%%%%%%%%%%%%%%%%%%%%%%%%%%%%%%%%%%%%%%%%%%%%%%%%%%%%%%%%%%%%%%%%%%%%%%%%%%%%%%%%%%%%%%%%%%%%%%%%%%%%%%%
\begin{figure}[tbp]
\minipage{0.5\textwidth}\includegraphics[width=\textwidth]{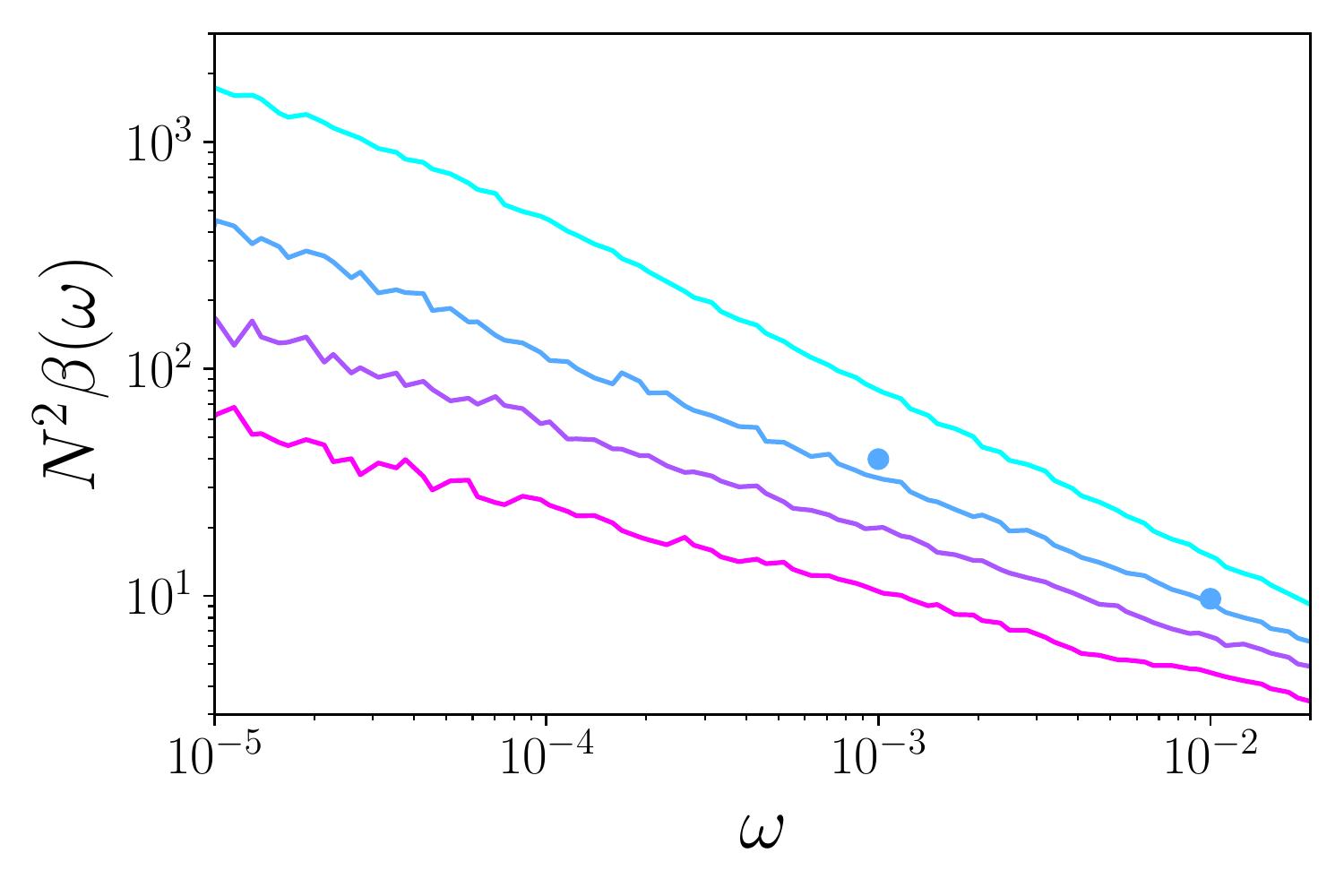}\endminipage
\minipage{0.5\textwidth}\includegraphics[width=\textwidth]{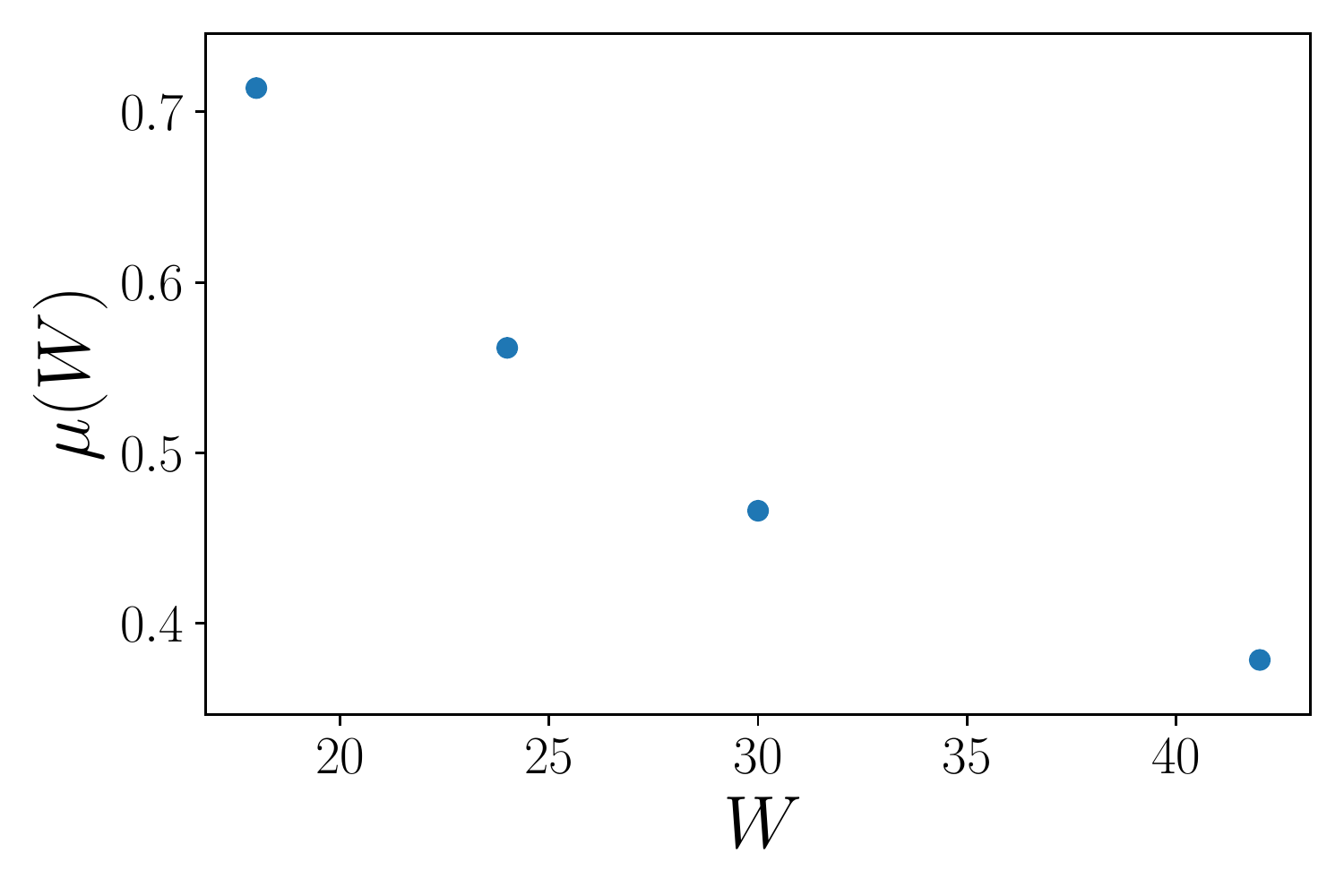}\endminipage
\caption{Correlation of different eigenstates in the localized phase of RRG (ED, $N=2^{15}$). {\it Left:} Disorder $W = 18$ (essentially the critical point), 24, 30, and 42 (from cyan to magenta), illustrating a power-law form of $\beta(\omega)$ in the localized phase, with exponent $\mu(W)$ depending on disorder. Dots: PD results for $W=24$. {\it Right:} Exponent $\mu(W)$ characterizing the frequency dependence of $\beta(\omega)$, see Eq. (\ref{betafit}).  From Ref.~\cite{tikhonov2020eigenstate}.}
\label{beta_rrg}
\end{figure}
%%%%%%%%%%%%%%%%%%%%%%%%%%%%%%%%%%%%%%%%%%%%%%%%%%%%%%%%%%%%%%%%%%%%%%%%%%%%%%%%%%%%%%%%%%%%%%%%%%%%%%%%%%%%%%%%%%%%%%%%%%%%%%%%%%%%%%%%%%%%%%%%%%%%%%%%%%%%%%%%%%%%%%%%%%%%%%%%%%%%%%%%%%%%%%%%%%%%%%%%%%%%%%%%%

Figures \ref{beta_rrg_full} and  \ref{beta_rrg} show results of numerical evaluation of $\beta(\omega)$ by ED of the RRG model with the connectivity $p=3$, in the vicinity of the band center, $E=0$, and for system sizes $N$ in the range from $2^{12}$ to $2^{16}$. Figure \ref{beta_rrg_full} displays the results for $\beta(\omega)$ for $N=2^{15}$ and several disorder values covering the while phase diagram. In the delocalized phase, $W=10$, $12$, and $14$, the power-law ($1/\omega$) behavior at high frequencies and a saturation at lower frequencies are observed, agreement with Eq. (\ref{betascres}); see also the left panel of Fig.~\ref{different}. The saturation frequency $\omega_\xi$ decreases with approach to the critical point, so that at criticality, $W=18$, the power-law (approximately $1/\omega$) behavior is observed in the whole range of frequencies. Remarkably, the power-law behavior of $\beta(\omega)$  survives in the localized phase $W=24$, 30, and 42, where it is characterized by a disorder--dependent exponent $\mu(W)$. These numerical results support the power-law scaling, Eq.  \eqref{betafit}. The exponent $\mu(W)$ satisfies Eq.~\eqref{mu-critical} at criticality and gradually decreases towards zero as $W$ grows, in consistency with Eq.~\eqref{mu-strong-disorder}. 

The function $\beta(\omega)$ for the localized phase is highlighted in the left panel of Fig. \ref{beta_rrg}. The right panel shows the numerically determined exponent $\mu(W)$, as obtained from the data presented in the left panel. These values were derived by the fit of numerical data by a pure power-law dependence (\ref{betafit}). At the critical point ($W=18$) the presence of an additional subleading logarithmic factor, together with finite-size effects, lead to reduction of the obtained value of $\mu$ in comparison with its exact value, Eq. \eqref{mu-critical}. 

%%%%%%%%%%%%%%%%%%%%%%%%%%%%%%%%%%%%%%%%%%%%%%%%%%%%%%%%%%%%%%%%%%%%%%%%%%%%%%%%%%%%%%%%%%%%%%%%%%%%%%%%%%%%%%%%%%%%%%%%%%%%%%%%%%%%%%%%%%%%%%%%%%%%%%%%%%%%%%%%%%%%%%%%%%%%%%%%%%%%%%%%%%%%%%%%%%%%%%%%%%%%%%%%%
%%%%%%%%%%%%%%%%%%%%%%%%%%%%%%%%%%%%%%%%%%%%%%%%%%%%%%%%%%%%%%%%%%%%%%%%%%%%%%%%%%%%%%%%%%%%%%%%%%%%%%%%%%%%%%%%%%%%%%%%%%%%%%%%%%%%%%%%%%%%%%%%%%%%%%%%%%%%%%%%%%%%%%%%%%%%%%%%%%%%%%%%%%%%%%%%%%%%%%%%%%%%%%%%%
%%%%%%%%%%%%%%%%%%%%%%%%%%%%%%%%%%%%%%%%%%%%%%%%%%%%%%%%%%%%%%%%%%%%%%%%%%%%%%%%%%%%%%%%%%%%%%%%%%%%%%%%%%%%%%%%%%%%%%%%%%%%%%%%%%%%%%%%%%%%%%%%%%%%%%%%%%%%%%%%%%%%%%%%%%%%%%%%%%%%%%%%%%%%%%%%%%%%%%%%%%%%%%%%%

\section{Many-body localization and its connections with localization on RRG}
\label{sec:MBL}

After the discussion of the main properties of the Anderson-localization problem on RRG, we now turn to the problem of many-body localization (MBL). In this section, we will review basic features of the MBL transition in system with various interaction range: from quantum dots (unbounded interaction) to short--range interacting spin chains. As an intermediate case, we will consider systems with power-law interactions, which interpolate between short and infinite interaction ranges. Our main focus will be on those features of the MBL transitions which can be compared to the corresponding properties of the RRG model and, in some cases, can be inferred from an (approximate) mapping between the RRG and MBL problems. This includes, in particular, the  localization criterion (i.e., the scaling of the localization transition) and the statistical properties of many-body eigenstates and energy levels.

%%%%%%%%%%%%%%%%%%%%%%%%%%%%%%%%%%%%%%%%%%%%%%%%%%%%%%%%%%%%%%%%%%%%%%%%%%%%%%%%%%%%%%%%%%%%%%%%%%%%%%%%%%%%%%%%%%%%%%%%%%%%%%%%%%%%%%%%%%%%%%%%%%%%%%%%%%%%%%%%%%%%%%%%%%%%%%%%%%%%%%%%%%%%%%%%%%%%%%%%%%%%%%%%%

\subsection{Quantum dot or modified SYK model}
\label{sec:qdot}

%%%%%%%%%%%%%%%%%%%%%%%%%%%%%%%%%%%%%%%%%%%%%%%%%%%%%%%%%%%%%%%%%%%%%%%%%%%%%%%%%%%%%%%%%%%%%%%%%%%%%%%%%%%%%%%%%%%%%%%%%%%%%%%%%%%%%%%%%%%%%%%%%%%%%%%%%%%%%%%%%%%%%%%%%%%%%%%%%%%%%%%%%%%%%%%%%%%%%%%%%%%%%%%%%
%%%%%%%%%%%%%%%%%%%%%%%%%%%%%%%%%%%%%%%%%%%%%%%%%%%%%%%%%%%%%%%%%%%%%%%%%%%%%%%%%%%%%%%%%%%%%%%%%%%%%%%%%%%%%%%%%%%%%%%%%%%%%%%%%%%%%%%%%%%%%%%%%%%%%%%%%%%%%%%%%%%%%%%%%%%%%%%%%%%%%%%%%%%%%%%%%%%%%%%%%%%%%%%%%

A model of disordered quantum dot can be described by the following Hamiltonian written in the basis of exact eigenstates of the non-interacting problem:
\be
\label{qdot}
\hat H = \sum_i \epsilon_i \hat c_i^\dag \hat c_i + \sum_{ijkl} V_{ijkl}\hat c_i^\dag \hat c_j^\dag \hat c_k \hat c_l,
\ee
where single-particle orbital energies $\epsilon_i$ and interaction matrix elements $V_{ijkl}$ are random quantities, and we consider spinless fermions for simplicity. The single-particle spectrum $\epsilon_i$ is characterized by mean level spacing
$\Delta$,  while the interaction matrix elements (which are zero in average) by their root--mean--square value $V$. The ratio of these two energy scales yields a dimensionless interaction strength:
\be
\label{gdef}
 \frac{V}{\Delta}  = g^{-1} \ll 1 \,.
\ee
The parameter $g \gg 1$  has a meaning of the dimensionless conductance for the non-interacting model.

\subsubsection{Analytical considerations}

In Ref.~\cite{altshuler1997quasiparticle}, a problem of a line width of a hot quasiparticle in a quantum dot characterized by the Hamiltonian \eqref{qdot} was addressed. The authors pointed out inapplicability of the Fermi golden-rule calculation for decay rate at sufficiently low energies. They proposed an hierarchical Fock-space model for this problem and approximately reduced it to a tight-binding Anderson model on a Bethe lattice with a large coordination number. In this framework, they concluded that there is an Anderson localization transition in Fock space as a function of the quasi-particle energy. This basic physical picture was corroborated and further developed in subsequent works \cite{jacquod1997emergence,mirlin1997localization,silvestrov1997decay,silvestrov1998chaos,gornyi2016many,gornyi2017spectral}. Related physics was also considered in earlier numerical simulations of a similar model in the nuclear-physics context \cite{aaberg1990onset, aaberg1992quantum}. 

We use now this framework to study properties of typical many-body states at energy $E \gg \Delta$ above the ground state (filled Fermi see). Such a state contains typically $ \sim n = \sqrt{E/\Delta} \gg 1$ excited particles and holes, each one  with energy $\epsilon_i \sim n \Delta$. The problem is essentially equivalent to a model with $n/2$ electrons populating $n$ orbitals around the energy corresponding to the middle of the corresponding many-body spectrum. Eigenstates of the non-interacting problem ($V=0$) form the basis of the Fock space. We define now a tight-binding model on a graph representing the many-body problem. Each Fock-space basis state serves as a vertex of the graph, the corresponding eigenvalues of the non-interacting part of the Hamiltonian serve as as diagonal matrix elements, and the interaction matrix elements translate into hopping. The emergent graph is similar to RRG, and thus the original problem in this representation bears a similarity with the problem of Anderson localization on RRG. 
Although there are essential differences with the RRG model (that will be discussed below), let us first discard them and estimate parameters of the effective RRG model.

The ``volume'' (number of sites) $\mathcal{N}$ of the random graph is determined by the Hilbert-space volume of the quantum dot model, $\mathcal{N} \sim 2^n$. More accurately, there will be subleading prefactors of power-law type with respect to $n$ in this relation (in particular, since the number of electrons is fixed). They do not affect, however, the leading behavior of $\ln \mathcal{N}$:
\be
\ln \mathcal{N} \simeq n \ln 2 \,.
\ee
The coordination number $m+1$ is determined by the number of basis states that are connected by an interaction operator with a given basis states. There are $n/2$ occupied electronic orbitals and $n/2$ empty orbitals in a basis state, and the interaction operator can move two electrons from occupied to empty orbitals. Simple combinatorics shows that the number of states that are obtained in this way in $\simeq n^4/64$. Discarding numerical factors, we thus have
\be
m \sim n^4 \,.
\label{qdot-rrg-m}
\ee
Since the energy of each particle in the quantum dot model is $\sim n \Delta$, and since the interaction moves only two particles, the spread of energies of states connected to a given basis state is
$\sim n\Delta$. This is associated with disorder of the RRG model:
\be
W \sim n \Delta \,.
\label{qdot-rrg-W}
\ee
Finally, the hopping matrix element $V$ of the RRG model is nothing but the interaction matrix element of the quantum-dot model:
\be
V \sim \frac{\Delta}{g} \,.
\label{qdot-rrg-V}
\ee

With the effective RRG model at hand, we can readily determine the corresponding localization transition point. Specifically, for  RRG with hopping $V$ and a large connectivity $m+1$, the critical disorder is given by
\be
\label{rrg-large-m}
W_c \sim V m\ln m\,.
\ee
Substituting here Eqs.~\eqref{qdot-rrg-m}, \eqref{qdot-rrg-W}, and \eqref{qdot-rrg-V}, we find the critical value of the parameter $g$ for given $n$:
\be
\label{qdot-rrg-gc}
g_c \sim n^3\ln n \,.
\ee
Equivalently, this can be written as an equation for critical $n$ at given $g$: 
\be
\label{qdot-rrg-nc}
n_c \sim g^{1/3} \ln^{-1/3}g \,.
\ee
To clarify the physics beyond Eqs.~\eqref{rrg-large-m} and \eqref{qdot-rrg-gc}, it is worth recalling that, in the above identification, $W/m$ is the level spacing of states directly coupled by interaction to a given basis state. The condition $V \gtrsim W/m$ is thus the condition of strong hybridization (i.e. ``delocalization'') on the first step of the hierarchy. The additional factor $\ln m$ comes from enhancement of delocalization in higher-order processes via virtual non-resonant states. 

This mapping on RRG is, however, only approximate. Indeed the effective graph in the Fock space of a quantum dot contains short-scale loops, at variance with RRG. For example, consider a second-order process, in which we first move two electrons $i,j \mapsto k,l$ and then another pair of electrons $i', j' \mapsto k',l'$. There is another second order process, in which the same is done in the opposite order. Clearly, we come to the same state, at variance with the tree-like RRG structure on short scales. Putting it differently, there are correlations between amplitudes of higher-order processes in the quantum-dot model that are discarded within the RRG approximation. The presence of such correlations is not surprising if one recalls that the number of independent random parameters grows exponentially with $n$ in the RRG model but only as a power law of $n$ in the quantum-dot model. This questions the direct applicability of Eq.~\eqref{qdot-rrg-gc} to the quantum-dot model. 
However, it turns out \cite{gornyi2017spectral} that there is a mechanism in the quantum dot model that greatly reduces the effect of the above correlations, thereby enhancing many-body delocalization and making the quantum dot problem much more similar to the RRG problem than one might think. This mechanism is the spectral diffusion: when an electron moves, energies of other electrons get modified, which leads to reshuffling of energies of many-body states, thus making the problem more ``RRG-like''.
 The corresponding detailed analysis was performed in Ref.~\cite{gornyi2017spectral}, with the result that the equation \eqref{qdot-rrg-gc} obtained from the mapping to RRG holds with respect to the leading (power-law) factor but the exponent of the logarithmic factor is possibly modified:
\be
\label{qdot-modified}
g_c \sim n^3\ln^\mu n \,,
\ee
with $-\frac{3}{4} \le  \mu \le 1$. The upper bound on $\mu$ corresponds to the RRG result  \eqref{qdot-rrg-gc}. The corresponding modification of Eq.~\eqref{qdot-rrg-nc} is 
\be
\label{qdot-rrg-nc-modified}
n_c \sim g^{1/3} \ln^{\tilde \mu}g \,, \qquad \qquad \tilde \mu = - \frac{\mu}{3} \,,
\ee
with $-1/3\le \tilde \mu\le 1/4$. 
We will discuss the spectral diffusion as a mechanism of ergodization of the system in more detail in Sec. \ref{sec:long_range} in the context of systems with power-law interaction (for which the analysis turns out to be somewhat simpler). 

Recently, a closely related problem is considered in Refs.~\cite{micklitz2019nonergodic, monteiro2020minimal} where a modified SYK model of $2n$ Majorana fermions $\chi_i$ was studied. The model is described by following Hamiltonian:
\be
\label{syk}
\hat H = \frac{1}{2}\sum_{ij}J_{ij}\hat \chi_i\hat \chi_j + \frac{1}{4!} \sum_{ijkl} J_{ijkl}\hat \chi_i\hat \chi_j\hat \chi_k\hat \chi_l,
\ee
where two-fermion and four-fermion couplings are random, with zero averages and variances
\be
\left<J_{ij}^2\right> = \frac{\delta^2}{2n} \,, \qquad \qquad  \left<J_{ijkl}^2\right> = \frac{6J^2}{(2n)^3}\,.
\ee
The system of $2n$ Majorana fermions can be also represented as a system of conventional (complex) fermions, with $n$ single-particle fermionic states. It is thus expected that the model \eqref{syk} is 
essentially equivalent, in what concerns the MBL physics, to the quantum-dot problem \eqref{qdot} discussed above. The number of electronic orbitals was denoted by $n$ in both models.
Let us establish a correspondence between the remaining parameters of the SYK model and those of the above quantum-dot model ($\Delta$ and $V$, yielding the ratio $g= \Delta/V$). The single-particle bandwidth of the SYK model is  given by $\left<J_{ij}^2\right>^{1/2}n^{1/2}\sim \delta$, so that the single-particle level spacing is
\be
\Delta\sim \frac{\delta}{n} \,.
\ee 
Further, the typical matrix element is of the order of 
\be
V\sim \frac{J}{n^{3/2}}\,.
\ee
 We thus find the following expression for the parameter $g$ characterizing the interaction strength in terms of the parameters of the model \eqref{syk}:
 \be 
 g\sim \frac{\Delta}{V} \sim  \frac{\delta \: n^{1/2}  }{J}\,.    
\ee
Therefore, the RRG transition-point condition \eqref{qdot-rrg-gc}, when expressed in terms of parameters of the modified SYK model, reads
\be
\delta_c \sim J n^{5/2} \ln n \,.
\label{delta-c-SYK-RRG}
\ee
The authors of Ref.~\cite{monteiro2020minimal} choose at some point the normalization $J\sim n^{-1/2}$. Then Eq.~\eqref{delta-c-SYK-RRG} becomes $\delta_c \sim n^2 \ln n$, which is exactly the result they find. In other words, the conclusion of  Ref.~\cite{monteiro2020minimal} is that the RRG equation \eqref{qdot-rrg-gc}  [or, equivalently, \eqref{delta-c-SYK-RRG} ]  for the critical point holds precisely, including the power of the logarithm. It should be mentioned, however, that the derivation in Ref.~\cite{monteiro2020minimal} employs an approximation of effective-medium type that essentially approximates the structure of the graph by a locally tree-like structure. The status of this approximation is not fully clear to us at this stage. As was pointed out above, the analysis in Ref.~\cite{gornyi2017spectral} leaves a window for the power of the logarithm, \eqref{qdot-modified}, which translates in the corresponding modification of Eq.~\eqref{delta-c-SYK-RRG}:
$\delta_c \sim J n^{5/2} \ln^\mu n$ with $-\frac{3}{4} \le  \mu \le 1$. This question remains to be fully settled in future work. 

Despite this uncertainty in the power of $\ln n$ in the formula for the critical disorder strength $g_c(n)$ [or, equivalently $\delta_c(n)$ in the terminology of the perturbed SYK model], it is clear that the MBL transition in the quantum-dot problem is at least closely related to the localization transition in the RRG model. It is thus strongly expected that key properties of both phases (localized and delocalized) in the quantum-dot model are the same as on RRG. Specifically, we expect the delocalized phase to be ergodic, implying, in particular, that asymptotically (for large $n$) the IPR scales as $-\ln P_2 \simeq n \ln 2$ and the level statistics takes the WD form (see also Ref.~\cite{monteiro2020quantum} for a recent related discussion). Further, the localized phase is characterized by $P_2 \sim 1$ and the Poisson level statistics. Finally, the critical point is expected to be of localized character. It is worth emphasizing that the above properties of both phases are formally defined in the large-$n$ limit at fixed value of the ratio $g / g_c(n)$. 

\subsubsection{Numerical studies}

We briefly discuss now some of existing numerical results on the quantum-dot model.  The problem of Fock--space localization on a quantum dot was studied numerically in a number of papers approximately two decades ago \cite{jacquod1997emergence, georgeot1997breit, leyronas2000scaling, shepelyansky2001quantum,  PhysRevE.62.R7575, jacquod2001duality, rivas2002numerical} and very recently in \cite{monteiro2020minimal}.
In Refs.~\cite{jacquod1997emergence, georgeot1997breit, shepelyansky2001quantum} the scaling of the localization threshold $n_c$ with the quantum dot conductance $g$ was found to be consistent with the power law $n_c \sim g^{1/3}$, as in Eqs.~\eqref{qdot-rrg-gc} and \eqref{qdot-modified}. It should be noted, however, that a  logarithmic factor that was not  included in the numerical analysis in Refs.~\cite{jacquod1997emergence, georgeot1997breit, shepelyansky2001quantum} may play a quite substantial role for relatively small systems amenable for ED. In addition, we know that finite-size corrections are substantial in the RRG model, and they can be expected to be equally important in the quantum-dot problem. It is thus not surprising that an attempt of scaling analysis in Ref.~\cite{PhysRevE.62.R7575} produced a rather broad window for a possible of the exponent characterizing the scaling of $n_c$ with $g$. The paper  \cite{leyronas2000scaling} came to the conclusion that 
$n_c\sim g^{1/2}$ but it used a Hamiltonian that differs from Eq.~\eqref{qdot} by omission of diagonal interaction terms, which suppresses the spectral diffusion and thus favors localization. Summarizing, additional computational work appears to be needed to verify convincingly the analytically predicted scaling of the MBL transition, Eq.~\eqref{qdot-modified}. In particular, it would be very interesting to find out whether the exponent $\mu$ in the subleading logarithmic factor is consistent with its RRG value, $\mu=1$. 

As expected, the numerical works found ergodic behavior when the system is well on the delocalized phase of the transition. The ergodicity manifests itself in the WD form of the level statistics and in the expected scaling of the IPR that corresponds to spreading of the eigenstate over all basis many-body states within the width given by the Fermi golden rule, with the spectral function of Breit-Wigner form \cite{georgeot1997breit, rivas2002numerical,monteiro2020minimal}.

\subsubsection{Spin quantum dot}

Spin models are very popular  for investigation of MBL transitions in systems that are characterized by localization in real space in the absence of interaction, see Sec.~\ref{sec:long_range} and \ref{sec:short} below. It is thus useful to consider also a spin analog of the quantum dot model. Such a ``spin quantum dot'' model was introduced and analyzed in Ref.~\cite{gornyi2017spectral}. It is defined by the Hamiltonian
\be
H = \sum_{i=1}^n  \epsilon_i \hat S_i^z + \sum_{i,j=1}^n \sum_{\alpha, \beta \in \{x,y,z\}} v_{ij}^{\alpha\beta} \hat S_i^\alpha \hat S_j^\beta \,.
\label{H-spin-quantum-dot}
\ee
Here $\hat S_i^\alpha$ with $i=1, \ldots, n$ and $\alpha = x,y,z$ are spin-$\frac{1}{2}$ operators, $\epsilon_i$ are random fields with a box distribution on $[-W, W]$, and the interactions $v_{ij}^{\alpha\beta}$ and independent random variables with zero mean and with root-mean-square value 
\be
\langle \left(v_{ij}^{\alpha\beta}\right)^2 \rangle^{1/2} = V \,. 
\ee
The model is thus characterized by two dimensionless parameters, $n$ and $W/V$, and, in full analogy with the fermion quantum dot, there is an MBL transition line in the plane spanned by these two parameters. It is straightforward to generalize the above analysis performed for the fermionic model on the spin model. For the magnitude  $W$ of the random field and the characteristic value $V$ of the interaction matrix element (that becomes the hopping in the random-graph interpretation), we already used the same notations as in the RRG formula \eqref{rrg-large-m}. The coordination number is $m \sim n^2$, since the interaction operator affects two spins. Thus, the analog of the RRG-like formula \eqref{rrg-large-m} for the critical disorder $W_c$ reads
\be
W_c \sim V n^2 \ln n \,.
\label{spin-qd-Wc}
\ee
As in the case of a fermionic quantum dot, the applicability of  Eq.~\eqref{spin-qd-Wc} to the spin quantum dot model is not obvious, in view of correlations between many-body states that are discarded by the RRG model. Again, the spectral diffusion strongly reduces the effect of these correlations, restoring Eq.~\eqref{spin-qd-Wc}, possibly up to a different power of the logarithmic factor, $\ln n \mapsto \ln^\mu n$, as in Eq.~\eqref{qdot-modified} \cite{gornyi2017spectral}.

\subsection{Systems with power-law interaction}
\label{sec:long_range}

In this section, we discuss the MBL transition in a system with  interaction that decays according to a power law with distance. A system of fermions with random potential, spatially localized single-particle states, and a power-law interaction can be mapped onto a spin-$\frac{1}{2}$ model in a  random magnetic field with a power-law interaction \cite{burin2006energy,Demler14,gutman2015energy}. We thus start form the model in spin formulation (which also emerges as an effective model of coupled two-level systems in amorphous materials). The exposition in this section is largely based on Ref.~\cite{tikhonov18}
(see also Refs. ~\cite{burin2015many,gutman2015energy}). We consider a spin-$\frac{1}{2}$ system described by the following lattice Hamiltonian in $d$-dimensional space (we assume lattice constant to be equal to unity):
\begin{equation}
\label{long_spin}
\hat H = \sum_i\epsilon_i \hat S_i^z+4t\sum_{ij}\frac{u_{ij}\hat S_i^z \hat S_j^z+v_{ij}(\hat S_i^+ S_j^-+\hat S_i^- \hat S_j^+)}{r_{ij}^{\alpha}} \,.
\end{equation}
Here $\hat S_i^z$, $\hat S_i^+$, and  $\hat S_i^-$ are spin $1/2$ operators, the fields $\epsilon_i$ are random with the box distribution on $\left[-W,W\right]$, the interaction parameters
$u_{ij}, \; v_{ij}$ are independent random variables taking the values $u_{ij}, \; v_{ij}=\pm 1$, and $r_{ij}$ is the distance between the sites $i$ and $j$.
For simplicity, we focus on the limit of infinite temperature (which essentially means $T \gg W$) and measure energy in units of $t$, taking $t=1$ in what follows.

The exponent $\alpha$ characterizing the decay of interaction with distance can in principle take any non-negative value. The limiting case $\alpha=0$ corresponds to the quantum-dot model considered in Sec.~\ref{sec:qdot}, while the limit $\alpha=\infty$ corresponds to the case of a short-range interaction, see Sec.~\ref{sec:short} below. 
  Here, we focus on an intermediate range, $d<\alpha<2d$. As explained below, the mechanism of many-body delocalization in this range is rather peculiar and permits a quite intricate connection with the RRG model.  We will briefly discuss the whole range of $\alpha$ in Sec.~\ref{sec:summary}. 
  
  \subsubsection{Analytical considerations: Scaling of MBL transition}
  \label{sec:long-range-transition-scaling}

We consider the regime of strong disorder. The starting point  is a basis in the many-body Hilbert space formed by eigenstates of the non-interacting part of the Hamiltonian [first term in Eq.~\eqref{long_spin}].
Each basis state is characterized by definite $z$ components of all spins, $S_i^z = \pm 1/2$. 
When the $\hat S_i^z \hat S_j^z$ part of the interaction is taken into account,
the energy of each spin  $\bar{\epsilon_i}$ gets renormalized by interactions with other spins:
\be
\bar{\epsilon_i}=\epsilon_i+4\sum_k r^{-\alpha}_{ik}u_{ik}S_k^z.
\label{bar-epsilon}
\ee
The $\hat S_i^+ \hat S_j^-$ part of the interaction generates matrix elements between basis states. If energies of two spins $i$ and $j$ are sufficiently close to each other,
\be
\left|\bar{\epsilon_i}-\bar{\epsilon_j}\right|\lesssim\frac{1}{r_{ij}^{\alpha}},
\ee 
they form a resonance pair. Two levels of this pair that have zero total $z$ projection of spin get strongly hybridized by interaction. This emergent two-level system is called ``pseudospin'', with the distance the sites $i$ and $j$ being the pseudospin size. The density of pseudospins of size $\sim R$ reads
\be 
\label{rho-ps}
\rho_{\rm PS}(R)\sim \frac{R^{d-\alpha}}{W} \,.
\ee
If the interaction decays very slowly, $\alpha<d$, the density $\rho_{\rm PS}(R)$ increases with $R$. This means that every spin finds a divergent number of resonance partners, which implies delocalization. This is the ``simple''  mechanism of delocalization discussed in Ref.~\cite{anderson58}. We will focus on the case of interaction that decays faster, $\alpha > d$. In this case, the pseudospin density $\rho_{\rm PS}(R)$ decreases when $R$ increases. This means that a typical spin does not find any resonant partner at all (since the disorder is strong) and that the majority of pseudospins is of the size of the order of lattice constant. Nevertheless, pseudospins at $\alpha > d$ may drive the many-body delocalization, as we are going to discuss.

According to Eq.~\eqref{rho-ps}, the number of pseudospins of size $\sim L$ in a system of linear size $\sim L$ is
\be
\label{NR}
N_2(L)\sim  L^d \rho_{\rm PS}(L) \sim \frac{1}{W}L^{2d-\alpha}.
\ee
For $\alpha<2d$, the number $N_2(L)$  increases with $L$. In a sufficiently small system, one has $N_2(L) \ll 1$, i.e., there is essentially no pseudospins of size $\sim L$. Pseudospins present in the system are of much smaller size and disconnected from each other, so that the system is in the localized regime. On the other hand, for a sufficiently large system, we get $N_2(L) > 1$, i.e., there are multiple pseudospins of size $\sim L$. These pseudospins are coupled to each other, and processes of their flips lead to many-body delocalization very similar to delocalization of a particle on RRG, as we explain below. 

The system size at which large (size-$L$) pseudospins appear is determined by the condition  $N_2(L) \sim 1$, yielding the disorder-dependent scale \cite{burin2015many,gutman2015energy}
\be
L_{c1}(W) \sim W^{\frac{1}{2d-\alpha}}.
\label{LW}
\ee
or, equivalently, expressing $W$ through $L$, 
\be
W_{c1}(L)  \sim L^{2d-\alpha}.
\label{WL}
\ee
Consider a system of size $L$ a few times larger than $L_{c1}(W)$. A typical basis state of such system has a few pseudospins of size $\sim L$, i.e., it is resonantly coupled to a few other basis states by spin-flip interaction matrix elements. Flipping any of these pseudospins leads to another basis state well coupled to the original one. This new state will in turn possess several pseudospins, so that the process can be repeated. Crucially, by virtue of spectral diffusion, new resonances are created in the process of exploring the Fock space via these flips. Indeed, after $p$ steps of spin flips, a typical distance from a given 
site $i$ to the closest flipped spin can be estimated as $\sim Lp^{-1/d}$. Therefore, the spin will experience a shift of
the energy $\bar{\epsilon_i}$ of the order of \cite{gutman2015energy}
\be 
\Delta^{(p)}\bar{\epsilon_i} \sim L^{-\alpha}p^{\alpha/d}.
\label{Delta-epsilon}
\ee
In the considered case $\alpha > d$, the spectral diffusion thus has in fact a ``superballistic'' character.
This fast increase of $\Delta^{(p)}\bar{\epsilon_i}$ with $p$  guarantees that  resonances are efficiently reshuffled (i.e., new resonances are created at every step), so that
a tree-like network of many-body states coupled by resonances emerges \cite{gutman2015energy}. This establishes a connection of the original many-body problem and the RRG model,
leading to the conclusion that systems of sizes $ L \gtrsim L_{c1}(W)$ should be ergodic. In fact, in analogy with the RRG model with large coordination number, the delocalization is further enhanced due to
higher-order resonant processes that go via intermediate non-resonant states. In full analogy with Eq.~\eqref{rrg-large-m}, this yields an additional logarithmic factor in $W_c$ in comparison with Eq.~\eqref{WL}, thus resulting in the following prediction for the position of the MBL transition  \cite{tikhonov18}:
\begin{equation}
W_c(L)\sim L^{2d-\alpha} \ln L^d,
\label{scaling2}
\end{equation}
Equation \eqref{scaling2} determines a line of the MBL transition in the $W$--$L$ plane. 
It is seen that the critical disorder $W_c$ has a power-law dependence on $L$ (with a logarithmic correction), thus diverging in the limit $L \to \infty$. Therefore, in order to study properties of the MBL transition, one should, with increasing $L$, simultaneously rescale disorder, i.e., to consider physical observables as functions of $W/W_c(L)$. If one fixes $W/W_c(L)$ at considers the limit $L\to \infty$, the system will be in the delocalized phase for $W/W_c(L) < 1$, at criticality for $W/W_c(L) = 1$, and in the MBL phase $W/W_c(L) > 1$. This procedure was termed a ``non--standard thermodynamic limit'' in Ref.~\cite{gopalakrishnan2019instability}.

   \subsubsection{Analytical considerations: Inverse participation ratio}
   \label{sec:long-range-IPR}
   
We discuss now the scaling of the average inverse participation ratio $P_2$ of many-body eigenstates (when considered in the basis of eigenstates of the non-interacting Hamiltonian characterized by definite $z$ components of all spins).  

Let us start with the localized phase.   As was explained in Sec.~\ref{sec:long-range-transition-scaling}, there are in general many small-size pseudospins (pairs of resonant spins) in the MBL phase, $L < L_c(W)$. Specifically, according to Eq.~\eqref{rho-ps}, the majority of pseudospins has a size of order unity, and their number is 
$N_{\rm PS}\sim L^d/W$. While these resonances are not sufficient to globally delocalize the system, they affect the IPR of many-body wave functions. 
Each resonance yields a factor of $\sim 1/2$ to the IPR, which leads to
\be
- \ln P_2\sim N_{\rm PS} \sim \frac{L^d}{W}.
\label{p2-scaling}
\ee
The resulting IPR scaling has a form of fractality of eigenstates in the MBL phase. Indeed, the volume ${\cal N}$ of the Hilbert space of the many-body problem is 
\be
{\cal N} = 2^{L^d} \,.
\label{hilbert-space-volume}
\ee
Therefore, we can rewrite Eq.~(\ref{p2-scaling}) as 
\be
P_2 \sim {\cal N}^{-\tau(W)}\,, \qquad \tau(W) \sim \frac{1}{W} \,. 
\label{long-range-loc-ipr}
\ee
It is clear from this derivation that such fractal scaling of the IPR with $\cal N$ equally applies to the MBL phase of a system with short-range interaction, see also the corresponding discussion in Sec. \ref{sec:short}.

We turn now to the delocalized phase, $L > L_c(W)$.  In view of the connection to the RRG model, it is expected that the system becomes ergodic  in the large-$L$ limit in the delocalized phase---i.e., in the limit $L \to \infty$ taken at a fixed value of $W/W_c(L) < 1$. This corresponds to the IPR proportional to the inverse volume of the Hilbert space  $1/{\cal N}$, i.e.,
\be 
\label{p2-deloc}
- \ln P_2 \simeq L^d \ln 2. 
\ee

Finally, we consider the transition point, $W=W_c(L)$. In view of the relation to localization transition on RRG, it is expected that the critical point has localized character, i.e., its properties are obtained continuously from the localized phase, $W \to W_c(L) + 0$ . This means, in particular, the IPR scaling \eqref{long-range-loc-ipr}, $P_2 \sim {\cal N}^{-\tau(W_c)}$. 

Implications of the relation to RRG for the level statistics are straightforward. The level statistics is expected to be of WD form in the delocalized phase, reflecting its ergodicity, and of Poisson form in the localized phase and at criticality.

All the above analytical predictions for the model with long-range interaction are supported by numerical simulations, as we are going to discuss. 

   \subsubsection{Numerical results}
   
   %%%%%%%%%%%%%%%%%%%%%%%%%%%%%%%%%%%%%%%%%%%%%%%%%%%%%%%%%%%%%%%%%%%%%%%%%%%%%%%%%%%%%%%%%%%%%%%%%%%%%%%%%%%%%%%%%%%%%%%%%%%%%%%%%%%%%%%%%%%%%%%%%%%%%%%%%%%%%%%%%%%%%%%%%%%%%%%%%%%%%%%%%%%%%%%%%%%%%%%%%%%%%%%%%
\begin{figure}[tbp]
\centering
\includegraphics[width=0.48\textwidth]{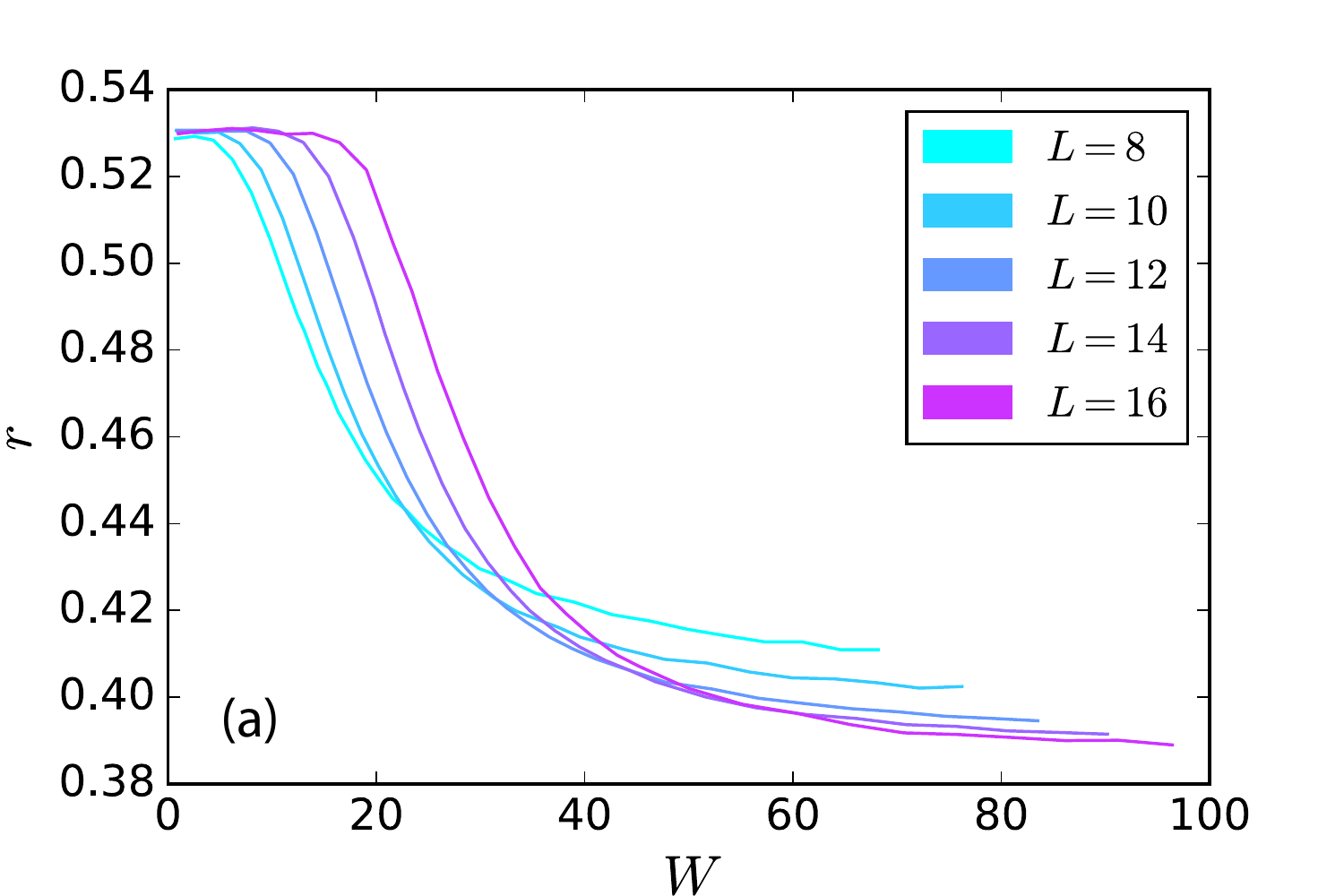}\quad
\includegraphics[width=0.48\textwidth]{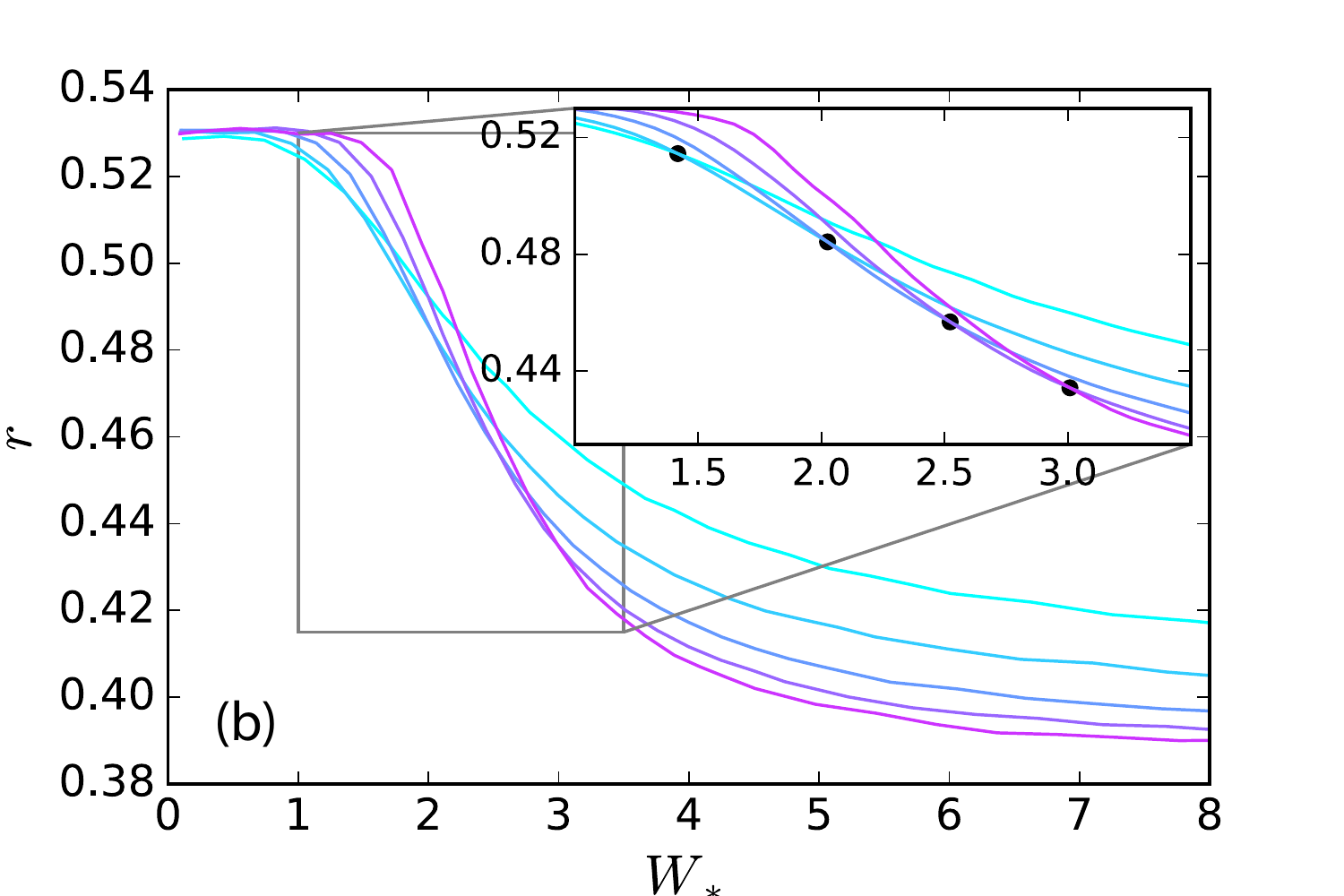}
\caption{Mean adjacent gap ratio $r$, Eq.~\eqref{agr}, characterizing spectral correlations in a spin chain (\ref{long_spin}) with long-range interaction, $\alpha=3/2$, as a function of disorder $W$ for various system lengths $L$.
(a)  $r(W)$ demonstrates ergodicity, $r \to r_{\rm WD}$, at fixed $W$ in the limit $L\to\infty$;
(b)  $r(W_*)$ plotted as a function of rescaled disorder $W^*$, Eq.~(\ref{W-star}). The drifting crossing point is expected to converge, at $L \to \infty$, to a critical value $W_{*c}$. In the limit $L \to \infty$, the system is expected to be ergodic for $W_* < W_{*c}$ and localized for $W_* > W_{*c}$.  An extrapolation (together with the data of Fig.~\ref{fig:ipr_mbl}) yields an estimate $W_{*c}\approx 4.3$. From Ref.\cite{tikhonov18}.
\label{fig:r}
}
\end{figure}
%%%%%%%%%%%%%%%%%%%%%%%%%%%%%%%%%%%%%%%%%%%%%%%%%%%%%%%%%%%%%%%%%%%%%%%%%%%%%%%%%%%%%%%%%%%%%%%%%%%%%%%%%%%%%%%%%%%%%%%%%%%%%%%%%%%%%%%%%%%%%%%%%%%%%%%%%%%%%%%%%%%%%%%%%%%%%%%%%%%%%%%%%%%%%%%%%%%%%%%%%%%%%%%%%
   
The MBL transition in the model  \eqref{long_spin} was studied by numerical simulations (ED) in Refs.~\cite{burin2015many,tikhonov18}.  We focus on results of Ref.~\cite{tikhonov18} where a detailed analysis of the level and eigenfunction statistics around the MBL transition was performed. 

\begin{figure}[tbp]
\centering
\includegraphics[width=0.48\textwidth]{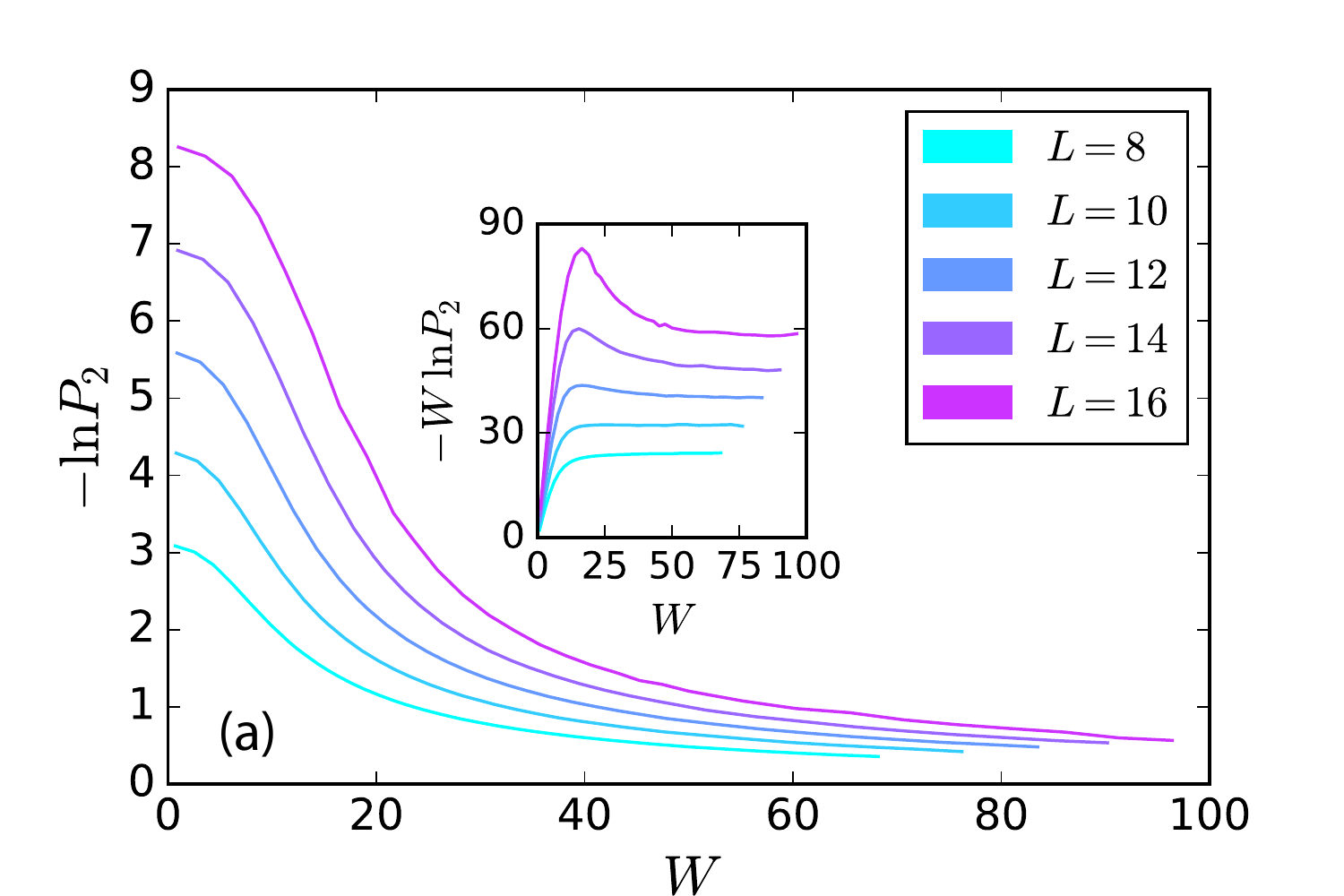}\quad
\includegraphics[width=0.48\textwidth]{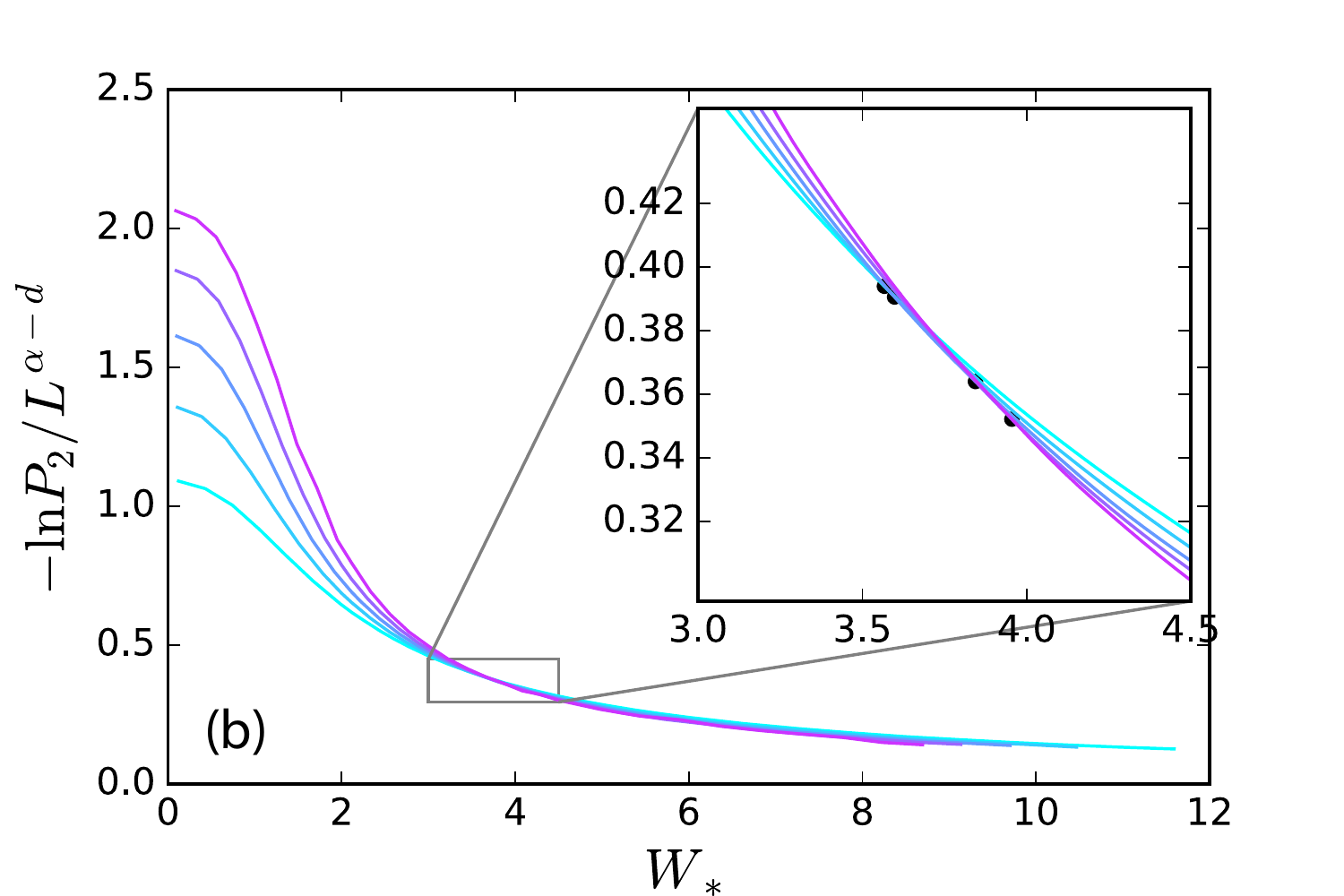}
\caption{Average IPR $P_2$  of many-body eigenstates of the 1D spin chain (\ref{long_spin}) with long-range-interaction, $\alpha=3/2$, for various system sizes $L$.
(a) $- \ln P_2$ as a function of disorder $W$ without rescaling. If $W$ is not too large, the system reaches ergodicity for these values of $L$. For large $W$, the system is still in the localized phase for these system sizes. Inset:  $- \ln P_2$ as a function of $W$ for various $L$.  The fractal scaling (\ref{p2-scaling}) is clearly seen at large $W$. (b)  $- \ln P_2 / L^{\alpha-d}$ as a function of the rescaled disorder $W_*$, Eq.~(\ref{W-star}). A crossing point slowly drifting towards larger values of $W_*$ is observed. It is expected that this drift converges to the critical value $W_{*c}$ in the limit $L \to \infty$; an extrapolation yields an estimate $W_{*c}\approx 4.3$.  From Ref. \cite{tikhonov18}.
}
\label{fig:ipr_mbl}
\end{figure}

A 1D spin chain with the interaction exponent $\alpha=3/2$ is considered.
The results for the level statistics---characterized by the mean ratio $r$ of two consecutive spacings, Eq. (\ref{agr})---are shown in Fig. \ref{fig:r}a. With increasing system size $L$, the curve $r(W)$ rapidly moves to the right, so that for a fixed $W$ the gap ratio $r$ tends to its ergodic (WD) value $r_{\rm WD} = 0.530$ at $L\to\infty$. This confirm the analytical expectation that for a fixed disorder $W$ the system is in the ergodic phase in the limit of large $L$.  In order to check the predicted scaling (\ref{scaling2}) of the critical disorder $W_c(L)$, results for $r$ are replotted in Fig. \ref{fig:r}b as a function of rescaled disorder,
\be 
W_* = \frac{W}{L^{2d-\alpha} \ln L}.
\label{W-star}
\ee
The data exhibit now a behavior similar to the one found for the RRG model, see left panel of Fig.~\ref{rrg:adjgap}. The curves become steeper with increasing $L$ and show a crossing point (between curves with consecutive values of $L$). As is seen in the inset, this crossing point drifts towards larger values of $W_*$, with the drift slowing down when $L$ increases. (The $\ln L$ factor in the denominator of Eq.~(\ref{W-star}) is important; if it is discarded, the drift accelerates, signalling a divergence at $L \to \infty$.) Thus, the level-statistics data corroborate the analytical prediction (\ref{scaling2}) for the scaling of the MBL transition point: there exists a critical value $W_{*c}$ separating the delocalized, ergodic phase at $W_* < W_{*c}$ from the localized phase at $W_* > W_{*c}$.
The finite drift of the crossing point (after a rescaling of disorder to $W_*$) is analogous to its drift in the RRG model and is related to the localized nature of the critical point, see Sec. \ref{sec:gr}. An extrapolation towards $L \to \infty$ of the data for the level statistics, together with those for the eigenfunction statistics (see text below and Fig.~\ref{fig:ipr_mbl}b), yields \cite{tikhonov18} an estimate $W_{*c}\approx 4.3$ for the critical point. 

%%%%%%%%%%%%%%%%%%%%%%%%%%%%%%%%%%%%%%%%%%%%%%%%%%%%%%%%%%%%%%%%%%%%%%%%%%%%%%%%%%%%%%%%%%%%%%%%%%%%%%%%%%%%%%%%%%%%%%%%%%%%%%%%%%%%%%%%%%%%%%%%%%%%%%%%%%%%%%%%%%%%%%%%%%%%%%%%%%%%%%%%%%%%%%%%%%%%%%%%%%%%%%%%%
%%%%%%%%%%%%%%%%%%%%%%%%%%%%%%%%%%%%%%%%%%%%%%%%%%%%%%%%%%%%%%%%%%%%%%%%%%%%%%%%%%%%%%%%%%%%%%%%%%%%%%%%%%%%%%%%%%%%%%%%%%%%%%%%%%%%%%%%%%%%%%%%%%%%%%%%%%%%%%%%%%%%%%%%%%%%%%%%%%%%%%%%%%%%%%%%%%%%%%%%%%%%%%%%%
%%%%%%%%%%%%%%%%%%%%%%%%%%%%%%%%%%%%%%%%%%%%%%%%%%%%%%%%%%%%%%%%%%%%%%%%%%%%%%%%%%%%%%%%%%%%%%%%%%%%%%%%%%%%%%%%%%%%%%%%%%%%%%%%%%%%%%%%%%%%%%%%%%%%%%%%%%%%%%%%%%%%%%%%%%%%%%%%%%%%%%%%%%%%%%%%%%%%%%%%%%%%%%%%%

We turn now to the statistics of many-body eigenfunctions in the same 1D model with $\alpha = 3/2$.  Numerical results for the dependence of IPR  $P_2$ on disorder $W$ 
are shown in Fig.~\ref{fig:ipr_mbl}a. For not too strong $W$, the ergodic behavior  (\ref{p2-deloc}) is reached already for relatively small system sizes $L$ accessible to ED.
For strong disorder $W$, the system is still on the localized side of the transition for these values of $L$. As shown in the inset of Fig.~\ref{fig:ipr_mbl}a, the numerical data fully confirm the behavior predicted analytically for the localized phase,  Eq.~(\ref{p2-scaling}), i.e., the fractal scaling \eqref{long-range-loc-ipr}. 
To determine the position of the transition, in Fig.~\ref{fig:ipr_mbl}b the rescaled logarithm of the IPR, $- \ln P_2 / L^{\alpha-d}$, is plotted as a function of the rescaled disorder $W_*$, Eq.~(\ref{W-star}).  The rescaling of the vertical axis is such that the plotted quantity increases with $L$ on the delocalized side of the transition ($W_* < W_{*c}$) and decreases in the localized phase  and at criticality ($W_* \ge W_{*c}$) according to the analytical results in Sec.~\ref{sec:long-range-IPR}. Thus, according to the analytical theory  and in analogy with Fig.~\ref{fig:r}b for the levels statistics, there should be a crossing point drifting towards larger $W_*$ and converging to $W_{*c}$ at $L\to \infty$.  The data in Fig.~\ref{fig:ipr_mbl}b fully confirm this expectation. As pointed out above, they were used, together with those from Fig. \ref{fig:r}b, to estimate the critical value of the rescaled disorder, $W_{*c}\approx 4.3$. 

Summarizing, the numerical results for the model with power-law interaction confirm the analytical predictions---which exploit a relation to the RRG model---for the scaling of the transition and properties of both phases. It is worth emphasizing that a relation to RRG here is of rather intricate character, since only large-scale resonances (pseudospins) drive the many-body delocalization. At the same time,  there are many short-scale resonances in the localized phase, which are responsible for a ``local'' spreading of an eigenstate that manifests itself in the fractal scaling of IPR, Eq.~\eqref{long-range-loc-ipr}, at variance with $P_2 \sim 1$ in the localized phase on RRG.

%%%%%%%%%%%%%%%%%%%%%%%%%%%%%%%%%%%%%%%%%%%%%%%%%%%%%%%%%%%%%%%%%%%%%%%%%%%%%%%%%%%%%%%%%%%%%%%%%%%%%%%%%%%%%%%%%%%%%%%%%%%%%%%%%%%%%%%%%%%%%%%%%%%%%%%%%%%%%%%%%%%%%%%%%%%%%%%%%%%%%%%%%%%%%%%%%%%%%%%%%%%%%%%%%

\subsection{Systems with short-range interaction}
\label{sec:short}

%%%%%%%%%%%%%%%%%%%%%%%%%%%%%%%%%%%%%%%%%%%%%%%%%%%%%%%%%%%%%%%%%%%%%%%%%%%%%%%%%%%%%%%%%%%%%%%%%%%%%%%%%%%%%%%%%%%%%%%%%%%%%%%%%%%%%%%%%%%%%%%%%%%%%%%%%%%%%%%%%%%%%%%%%%%%%%%%%%%%%%%%%%%%%%%%%%%%%%%%%%%%%%%%%
%%%%%%%%%%%%%%%%%%%%%%%%%%%%%%%%%%%%%%%%%%%%%%%%%%%%%%%%%%%%%%%%%%%%%%%%%%%%%%%%%%%%%%%%%%%%%%%%%%%%%%%%%%%%%%%%%%%%%%%%%%%%%%%%%%%%%%%%%%%%%%%%%%%%%%%%%%%%%%%%%%%%%%%%%%%%%%%%%%%%%%%%%%%%%%%%%%%%%%%%%%%%%%%%%

In this section, we discuss the MBL transition in systems with spatial localization (in the absence of interaction) and short-range interaction. 
While originally the corresponding models were formulated in terms of fermions \cite{gornyi2005interacting,basko2006metal}, the same physics can be addressed in terms of spin models.
In particular, the $S=\frac{1}{2}$ Heisenberg chain  in a random magnetic field has become a paradigmatic models for the investigation of the  MBL physics \cite{oganesyan2007localization,PhysRevB.82.174411}.  In its simplest version, the model is equivalent to the fermionic model with random potential and nearest-neighbor interaction. 
The model is governed by the Hamiltonian (${\bf  S}_{L+1} \equiv {\bf  S}_{1}$ for periodic boundary conditions)
\be 
H=\sum_{i=1}^L {\bf S}_i \cdot {\bf  S}_{i+1} -\epsilon_iS_i^z, 
\label{eq:H}
\ee
with random fields $\epsilon_i$ drawn from a uniform distribution $[-W,W]$.

\subsubsection{MBL transition: Analytical considerations}
\label{sec:short-range-analytical}

A brief review of approaches to the MBL transition in a model of the type \eqref{eq:H} was given in Sec.~\ref{sec:introduction}; we focus here on relations to localization on RRG. Connections of the many-body perturbation series for this class of models with the Anderson localization on tree-like graphs was pointed out, or is implicit, in a number of papers, see Refs.~\cite{gornyi2005interacting,basko2006metal,Ros2015420,gopalakrishnan2015low,gornyi2017spectral,tikhonov2020eigenstate}. We only briefly sketch the key point. Let $\ket{j}$ be a typical basis many-body state, i.e., an  eigenstate of all operators $S_i^z$. The spin-flipping part of interaction, $\sum_i \frac{1}{2}  (S_i^+ S_{i+1}^- + S_i^- S_{i+1}^+ )$, directly connects this basis state to $\sim L$ basis states with energies within the window $\sim W$ around the energy of the state $\ket{j}$. Each of these states is obtained from $\ket{j}$ by flipping a pair of adjacent spins. Proceeding to the second order, one finds, at first sight, $L^2$ states. However, almost all of them are formed by two remote pairs of flipped spins. Combining amplitudes for flipping these remote pairs of spins in different orders, one finds that they largely cancel, so that such processes do not contribute to many-body delocalization, see Ref. \cite{gornyi2017spectral} and references therein. The spectral diffusion is not operative in this case, since remote pairs of spins ``do not talk to each other''. As a result, such second-oder processes in fact decouple into two independent pieces (first-order processes).  This is fully analogous to short-scale resonances (``pseudospins'') in the MBL phase of models with power-law interaction, see discussion in Sec.~\ref{sec:long_range}. Therefore, to potentially yield a second-order resonance, two spin pairs should form a connected cluster. 
Consequently, the number of processes  that can lead to second-order resonances is $\sim m L$, where $m$ does not depend on $L$. For the spin chain \eqref{eq:H} one has $m \sim 1$;  a model with parametrically large (but still $L$-independent)  $m$ can be constructed by considering a chain of coupled ``spin quantum dots''  \eqref{H-spin-quantum-dot}, see Ref.~\cite{gornyi2017spectral}. 
Extending this argument to higher orders, one finds that the number of relevant processes in the order $k$ is
\be
N_{k;L}\sim L m^k \,.
\label{mbl-potential-resonances}
\ee
In Ref.~\cite{gopalakrishnan2015low}, the parameter $m$ in this formula was denoted by $e^s$, where $s$ was termed ``configuration entropy per flipped spin of the possibly resonant clusters''. 
This is the same behavior as on RRG with coordination number $m+1$, up to the overall factor $L$. Thus, for $m\sim 1$ [as for the model \eqref{eq:H}], 
we conclude that the critical disorder is 
\be
W_c \sim 1 \,,
\ee
 independent of $L$, as on RRG with $m \sim 1$.  This argument does not include, in fact, a potential effect of exponentially rare regions
with atypically strong or atypically weak disorder \cite{Agarwal2016a,Thiery2017a}. However, the role of such rare regions turns out to be less important in 1D geometry, and the conclusion about $W_c \sim 1$ remains unaffected. (In higher-dimensional systems, the avalanche instability due to rare regions lead to a slow (slower than any power law) increase of $W_c$  with system size.) In 1D, the rare regions and the resulting avalanche instability are essential ingredients of phenomenological renormalization-group approaches to the scaling at the MBL transition \cite{Dumitrescu2019a,morningstar2020a}. 
The fact that the asymptotic critical behavior at the MBL transition in systems with spatial localization and short-range interaction is different from that on RRG is clear already from the fact that the RRG value of the exponent $\nu_{\rm del} = 1/2$ [see Eq.~\eqref{xi}] would be in contradiction with the Harris criterion, $\nu \ge 2/d$. 

\subsubsection{MBL transition: Numerical results}

According to the analytical arguments discussed in Sec.~\ref{sec:short-range-analytical}, the model \eqref{eq:H} undergoes an MBL transition at a critical disorder $W_c$ that stays finite in the thermodynamic limit $L\to \infty$.
 ED studies in systems up to $L=24$  yield estimates $W_c=3.7$\:--\:$4.2$ for the critical disorder in the middle of the many--body spectrum \cite{luitz2015many,mace19multifractal,PhysRevResearch.2.042033}. However, an accurate determination of the critical point from ED data is complicated due to a rather slow convergence towards thermodynamic limit, see Ref.~\cite{abanin2019distinguishing} for a recent discussion. Specifically, the apparent transition point drifts towards larger values of $W$ with increasing system size.
 
 This behavior is similar to that seen for the RRG model. Indeed, the experience with RRG teaches us that the actual (thermodynamic-limit) critical disorder is considerably larger, due to finite-size effects, than the value suggested by the ED. Specifically, for RRG the ED would suggest $W_c \approx 15$ [e.g., as the position of the crossing point in spectral statistics, Fig.~\ref{rrg:adjgap}  (left panel), or based on maximum correlations of adjacent eigenstates, see Fig.~\ref{alpha_rrg}  (right panel)], while the actual value is $W_c = 18.17$, see analysis in Sec. \ref{sec:loctr}. In genuine interacting MBL models (such as the random-field spin chain), finite-size effects are expected to be further enhanced due to rare-region physics.  This expectation has been supported by an analysis based on the time-dependent variational principle with  matrix product states that was carried out in Ref.~\cite{Doggen2018a}, where the dynamics (imbalance relaxation) was explored for systems of much larger size, up to $L=100$. A strong drift of apparent (size-dependent) $W_c$ with system size $L$ was found, which has, however, essentially saturated at $L=100$. This yielded an estimate for the critical value $W_c\approx 5$\:--\:$5.5$ of the MBL transition in the thermodynamic limit ($L\to \infty$).  Interestingly, an advanced ED procedure supplemented by an extrapolation to $L\to \infty$
in recent Ref.~\cite{PhysRevLett.125.156601} resulted in a very similar estimate $W_c \approx 5.4$. 

Several works \cite{kjall2014many,luitz2015many,mace19multifractal} have used the ED data to carry out a finite-size scaling analysis of the MBL transition in the model \eqref{eq:H}. The results for the critical exponent $\nu$ are in the range $0.5-1$, in conflict with the Harris criterion that requires $\nu \ge 2$. This shows that the system sizes that can be treated via ED are too small to access the asymptotic critical behavior. Estimates in Refs.~\cite{Khemani2017a,PhysRevX.7.021013,Panda2020a}
indicate that spin chains of length $L \gtrsim 50$ -- 100 are needed to access the ultimate critical scaling. Interestingly, the values of the index $\nu$ obtained on the basis of ED data are quite close to the value $\nu_{\rm del}=1/2$ on RRG, especially if one takes into account that finite-size effects in small systems make the apparent $\nu_{\rm del}$ larger, see Sec.~\ref{sec:loctr}.
Thus, systems of intermediate sizes (relevant to experiments) may exhibit the critical behavior similar to that in the RRG model.

\subsubsection{Eigenfunction statistics: Scaling of IPR }
\label{sec:short-IPR}

We begin the discussion of properties of exact many-body eigenstates $\psi_k$ by the analysis of their most conventional characteristics: the inverse participation ratio (IPR) $P_2$ (defined as before in the basis of eigenstates $j$ of all $S^z_i$ operators), $P_2 = \left \langle \sum_j |\psi_k(j)|^4 \right \rangle$.

The volume of the many-body space is $\mathcal{N} \sim 2^L$.  More accurately, if we take into account that the total spin is conserved and consider the sector with total $S_z=0$, the volume of the many-body Hilbert space is found to be
$\mathcal{N} = L!/ [(L/2)!]^2 \simeq 2^L\sqrt{2/\pi L}$. The subleading power-law factor does not, play, however, any essential role.

At strong disorder $W \gg 1$ (i.e., deeply in the MBL phase), the number of resonant spin pairs is $\sim N_{1;L} / W \sim L/W$.  The number of second-order resonances is $\sim N_{2;L} / W^2 \sim L/W^2$, and higher-order resonances are suppressed by still higher powers of $W$, so that they all can be neglected. Exactly like in the case of a long-range interaction,  Sec.~\ref{sec:long-range-IPR}, each of first-order resonances brings a factor $\sim 1/2$ to IPR $P_2$. This results in a fractal scaling of the IPR \cite{gornyi2017spectral,tikhonov18} that has the same form as Eq.~\eqref{long-range-loc-ipr}:
\be
P_2 \sim \mathcal{N}^{-\tau(W)}\,, \qquad \tau(W) \sim \frac{1}{W} \,.
\label{short-range-loc-ipr}
\ee
This fractal scaling of IPR was indeed found numerically \cite{Luitz2016a,mace19multifractal} in the MBL phase of the spin-chain model \eqref{eq:H}. Furthermore, the numerical analysis in Ref.~\cite{mace19multifractal} led to the conclusion that this behavior holds up to the critical point, with $\tau(W_c) = \lim_{W \to W_c +0} \tau(W)$. This means that the critical point of the MBL problem has properties of the localized phase, in analogy with the RRG model. Also, it was found in Ref.~\cite{mace19multifractal} that the scaling $\tau(W) \sim 1/W$ that is derived analytically for $W \gg 1$ in fact holds with a good accuracy up to the transition point $W_c$, with $\tau(W_c) \approx 0.2$. 

In the delocalized phase, an analogy with the RRG model, see Sec. \ref{sec:IPR}, suggests  the ergodic behavior (also supported by general expectations)
\be
\label{mbl:erg}
P_2 \propto \frac{1}{\mathcal{N}}\,.
\ee 
Indeed, the ergodic scaling  \eqref{mbl:erg} of the IPR was found in numerical simulations (ED) in Ref.~\cite{mace19multifractal}.

\subsubsection{Eigenfunction statistics:  Dynamical correlations}

Let us now consider correlations of different many-body eigenstates. For the RRG model, the corresponding dynamical correlation function $\beta(\omega)$ was
analyzed in Sec.~\ref{sec:eigenfunc_corr_dynamical} and \ref{sec:eigenfunc_corre_loc}. We recall that $\beta(\omega)$ is the average overlap of two eigenstates separated in energy by $\omega$, 
\be
\beta(\omega) = \left \langle \sum_j |\psi_k(j) \psi_l(j)|^2 \right \rangle \,, \qquad \omega = E_k - E_l\,.
\ee
It was found in Ref. \cite{tikhonov2020eigenstate} that properties of the correlation function $\beta(\omega)$ for the MBL problem are remarkably similar (although not identical) to those for RRG. We briefly review below the corresponding analytical and numerical results.

We begin with analytical consideration. On the delocalized side of the transition, the ergodic behavior is expected, which should have the same form as the first line of Eq.~\eqref{betascres} derived for the RRG model:
\be
\mathcal{N}^2\beta(\omega) \sim \mathcal{N}_\xi \,, \qquad \omega<\omega_{\xi} \,.
\label{beta-omega-mbl-ergodic}
\ee
Here  $\mathcal{N}_\xi$, which  has a meaning of Hilbert-space correlation volume, depends on disorder $W$ only (i.e., does not depend on $\mathcal{N}$ and on $\omega$). For higher frequencies, $\omega > \omega_\xi$, a critical behavior is expected, in analogy with the second line of Eq.~\eqref{betascres}. As we discuss below, it is characterized by a power-law dependence of $\beta(\omega)$ on $\omega$, as on RRG. 
 
 We turn now to the analysis of eigenstates correlations in the MBL phase. It largely parallels the corresponding calculation for the RRG model,  Sec.~\ref{sec:eigenfunc_corre_loc}, with
 the correlation function $\beta(\omega)$ being governed by Mott--type resonances. The result reads \cite{tikhonov2020eigenstate} \be
\mathcal{N}^2\beta(\omega)\sim \omega^{-\mu(W)}\frac{(\log_2 \mathcal{N})^{3/2}}{\mathcal{N}^{\tau(W)}}.
\label{mbl-final-scaling}
\ee
Equation \eqref{mbl-final-scaling} has largely the same form as Eq.~(\ref{betafit}) for RRG; the key factor is the power-law dependence on frequency, $\omega^{-\mu(W)}$, with a disorder-dependent exponent $\mu(W)$.  Specifically,  $\mu(W)$ decreases logarthmically at large $W$ as in the RRG model, Eq.~\eqref{mu-strong-disorder}.
The additional factor $\mathcal{N}^{-\tau(W)}$ in Eq.~\eqref{mbl-final-scaling} originates from the suppression of resonant overlap, which is of order unity in the RRG model, see Eq. (\ref{resonance-overlap}) and becomes
  \be
  \sum_j |\psi_k(j) \psi_l(j)|^2 \sim \mathcal{N}^{-\tau(W)}
  \label{P-res}
  \ee
for the genuine many-body problem (spin chain), for the same reason as the IPR, Eq. \eqref{short-range-loc-ipr}.  The exponent $\tau(W)$ is parametrically small ($\sim 1/W$) in the MBL phase and remains quite small numerically at the critical point [$\tau(W_c) \approx 0.2$ according to Ref.~\cite{mace19multifractal}, see Sec.~\ref{sec:short-IPR}].  Thus, the difference between the results for the RRG and spin-chain models is not so significant.
 
%%%%%%%%%%%%%%%%%
\begin{figure}[tbp]
\centerline{\includegraphics[width=0.5\textwidth]{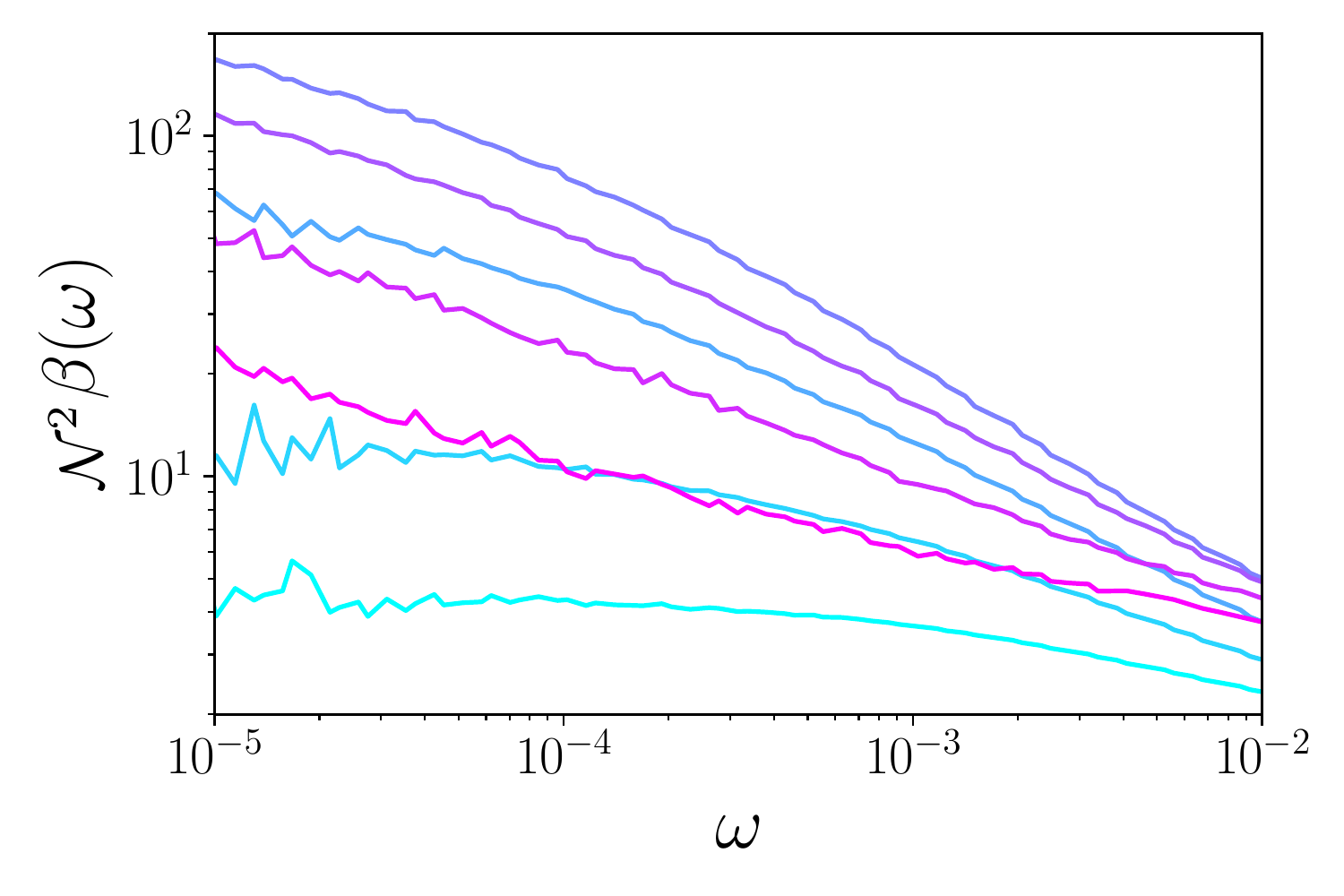}}
\caption{Dynamical eigenstate correlation function $\beta(\omega)$ for spin chain of size $L=16$ and disorder $W = 1.5, 1.7, 2, 3, 4, 6, 10$ (from cyan to magenta).
The three smallest values of $W$ are well in the ergodic phase,  the next two are also on the delocalized side but correspond to the critical regime for system sizes accessible to ED, and the two largest values are in the MBL phase. The figure is a counterpart of Fig.~\ref{beta_rrg_full} for the RRG model. From Ref. \cite{tikhonov2020eigenstate}.}
\label{beta_spins_full}
\end{figure}
%%%%%%%%%%%%%%%%%%

We discuss now  ED results for the dynamical eigenstate correlations in the model (\ref{eq:H}), which corroborate analytical expectations.  In Fig. \ref{beta_spins_full}, the correlation function $\beta(\omega)$ is shown for disorder strength from $W=1.5$ to $W=10$, i.e., across the MBL transition. The behavior of $\beta(\omega)$ is very similar to that for the RRG model, Fig.~\ref{beta_rrg_full}. For relatively weak disorder, $W=1.5$, 1.7, 2 (i.e., deeply in the delocalized phase), $\beta(\omega)$ exhibits a power-law critical behavior at higher frequencies and saturates at lower frequencies, 
in agreement with the expectation \eqref{beta-omega-mbl-ergodic}. The saturation value is fully consistent with the $\mathcal{N}$-independence of $\mathcal{N}_\xi$ in Eq.~\eqref{beta-omega-mbl-ergodic}, thus confirming the ergodicity of the delocalized phase, as discussed below. For $W=3$ and 4, the correlation function $\beta(\omega)$ shows a clear bending towards saturation at small $\omega$ but the saturation is not reached. This implies that these values of $W$ are on the delocalized side of the transition (in the limit $L \to \infty$) but the accessible system sizes are insufficient to observe ergodic behavior: for these values of $L$ the system is still in the critical regime.   At strong disorder, $W=6$ and 10, the function $\beta(\omega)$ shows a power-law scaling in the full range of $\omega$, as predicted for the MBL phase, Eq.~\eqref{mbl-final-scaling}.

%%%%%%%%%%%%%%%%%%%%%%%%%%%%%%%%%%%%%%%%%%%%%%%%%%%%%%%%%%%%%%%%%%%%%%%%%%%%%%%%%%%%%%%%%%%%%%%%%%%%%%%%%%%%%%%%%%%%%%%%%%%%%%%%%%%%%%%%%%%%%%%%%%%%%%%%%%%%%%%%%%%%%%%%%%%%%%%%%%%%%%%%%%%%%%%%%%%%%%%%%%%%%%%%%
\begin{figure*}[tbp]
\minipage{0.5\textwidth}\includegraphics[width=\textwidth]{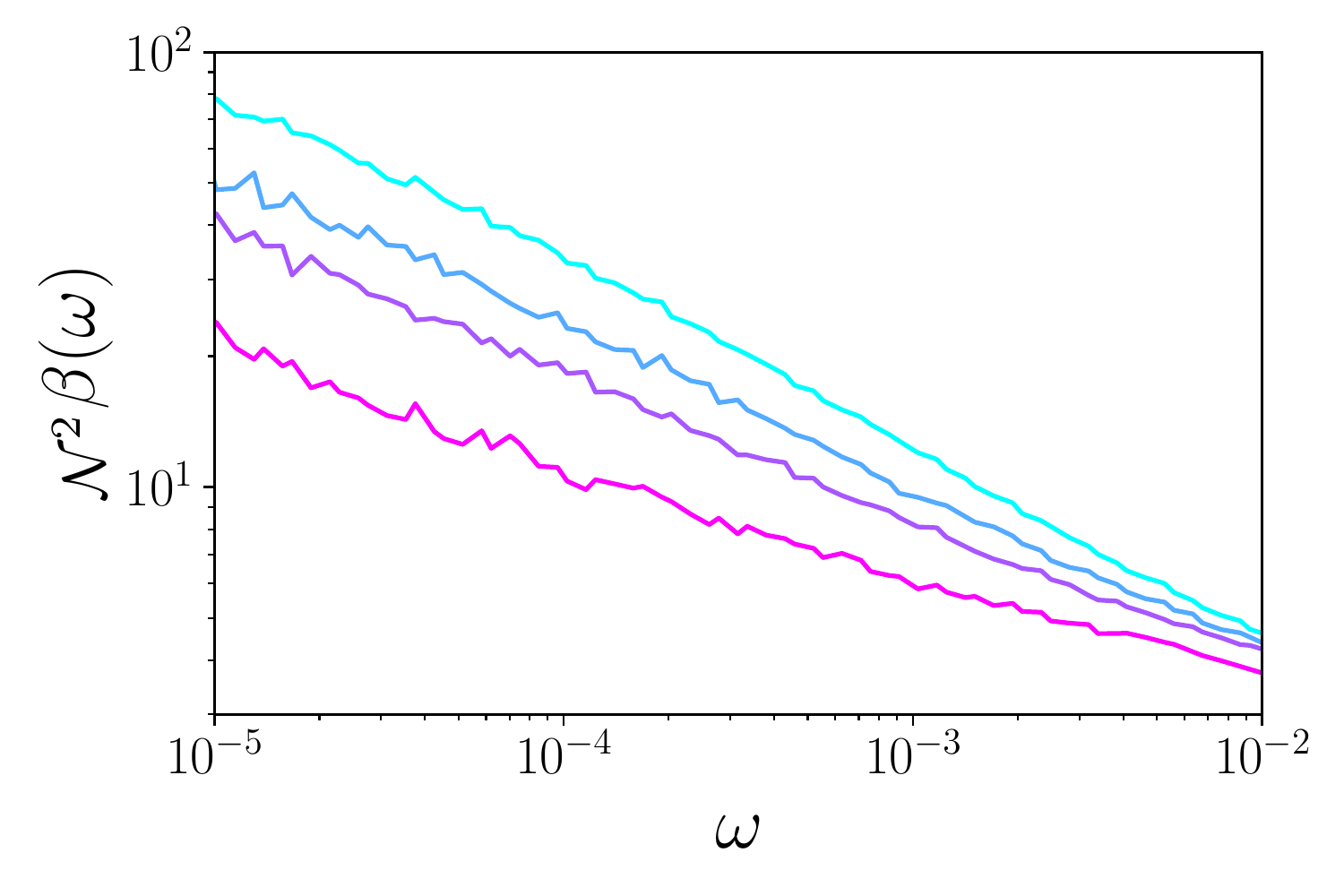}\endminipage
\minipage{0.5\textwidth}\includegraphics[width=\textwidth]{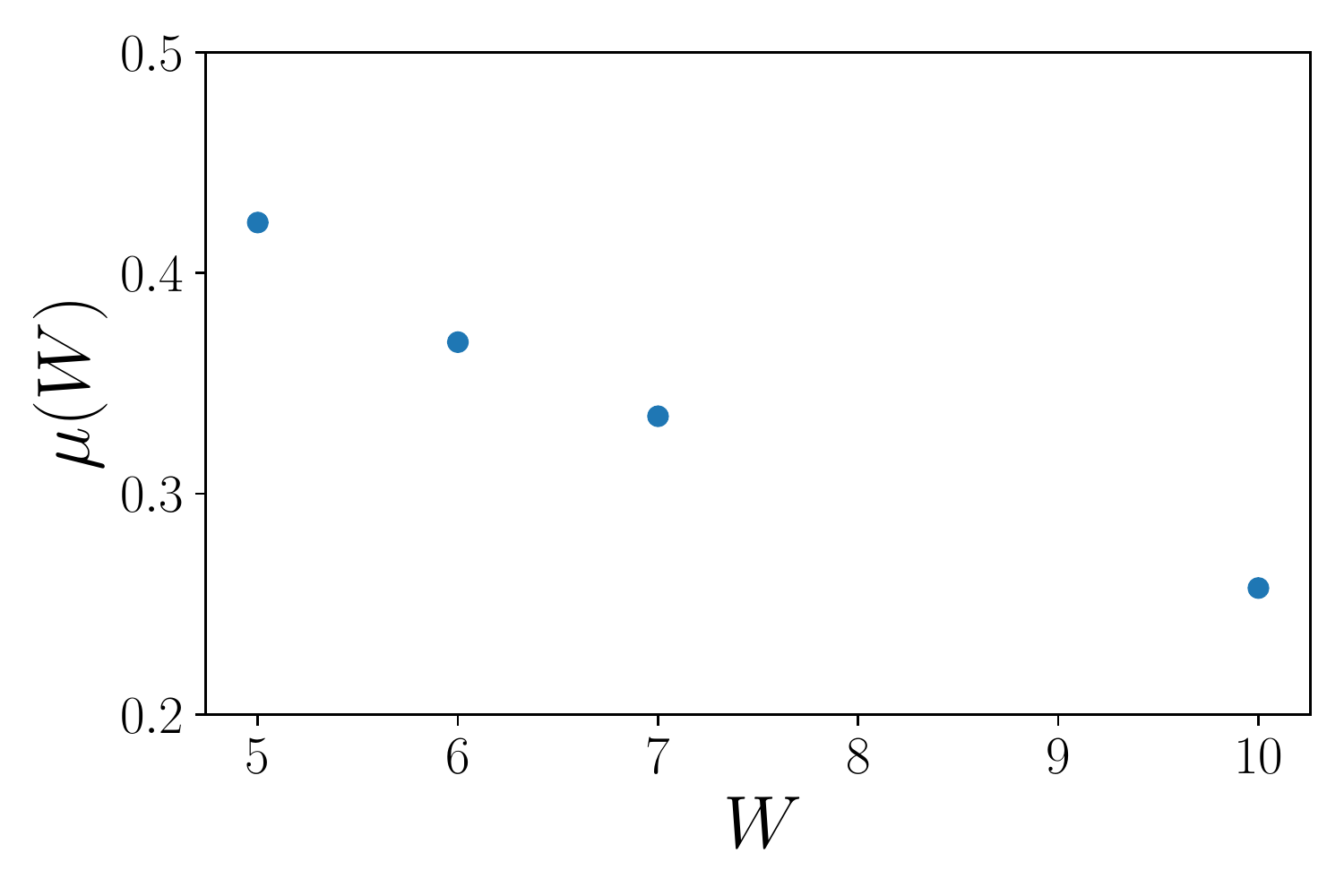}\endminipage
\caption{Dynamical eigenstate correlations for spin chain of size $L=16$. {\it Left:}  Correlation function $\beta(\omega)$ for disorder strengths $W = 5, 6, 7, 10$ (from cyan to magenta). {\it Right:}  Disorder dependence of the exponent $\mu(W)$ governing the power-law scaling $\beta(\omega) \propto \omega^{-\mu(W)}$, see Eq.~\eqref{mbl-final-scaling}. The figure is a counterpart of Fig.~\ref{beta_rrg} for the RRG model. From Ref. \cite{tikhonov2020eigenstate}.
}
\label{beta_spins}
\end{figure*}
%%%%%%%%%%%%%%%%%%%%%%%%%%%%%%%%%%%%%%%%%%%%%%%%%%%%%%%%%%%%%%%%%%%%%%%%%%%%%%%%%%%%%%%%%%%%%%%%%%%%%%%%%%%%%%%%%%%%%%%%%%%%%%%%%%%%%%%%%%%%%%%%%%%%%%%%%%%%%%%%%%%%%%%%%%%%%%%%%%%%%%%%%%%%%%%%%%%%%%%%%%%%%%%%%

Figure \ref{beta_spins} displays the data for stronger disorder, from $W=5$ (close to the MBL transition point $W_c$) to $W=10$ (well in the MBL phase). It is very similar to its RRG counterpart, Fig.~\ref{beta_rrg}. Data in the left panel clearly demonstrate the power-law frequency scaling $\beta(\omega) \propto \omega^{-\mu(W)}$, in agreement with Eq.~\eqref{mbl-final-scaling}. The disorder dependence of the exponent $\mu(W)$ is shown in the right panel.

\subsubsection{Adjacent eigenstate correlations}

\begin{figure*}[tbp]
\minipage{0.5\textwidth}\includegraphics[width=\textwidth]{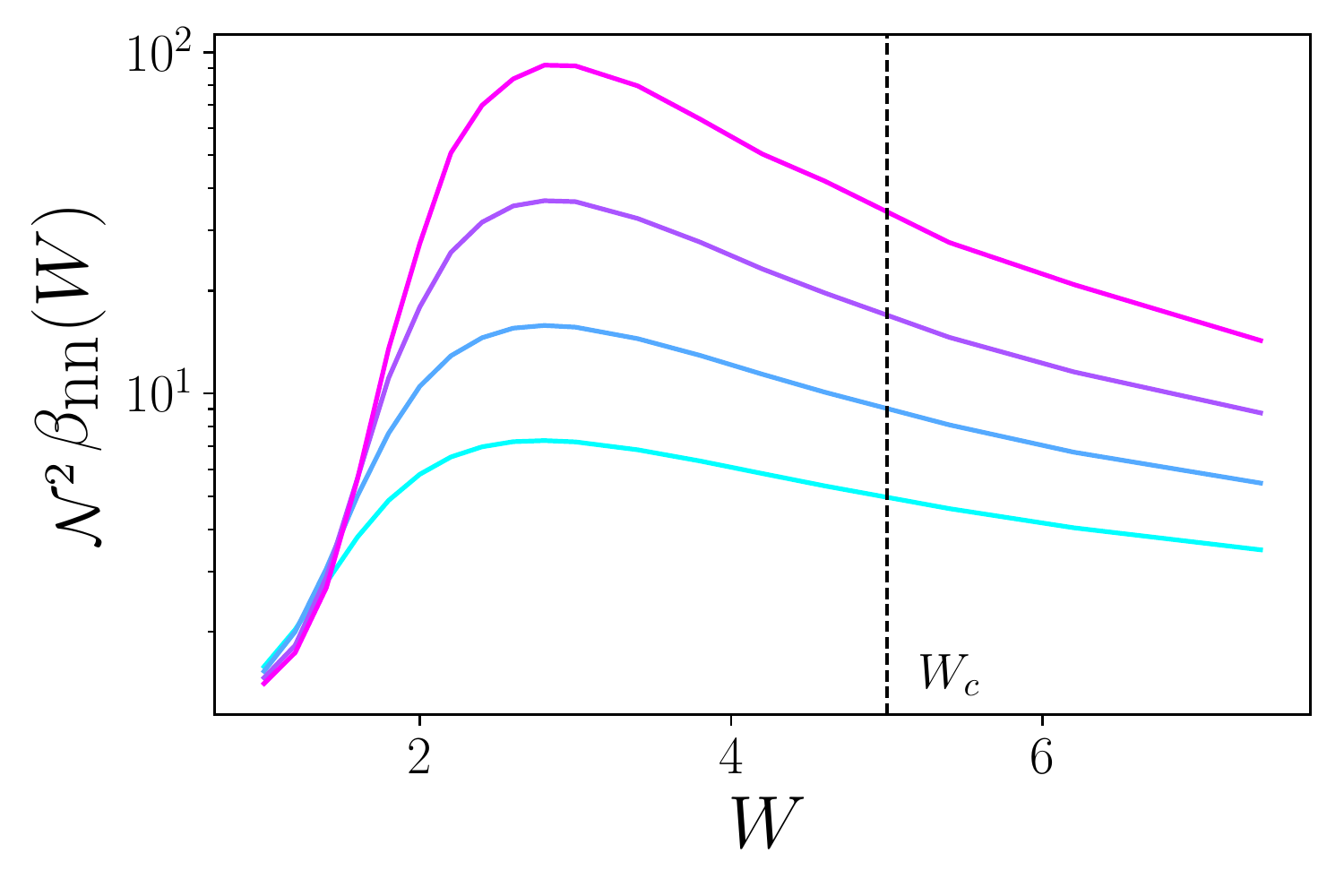}\endminipage
\minipage{0.5\textwidth}\includegraphics[width=\textwidth]{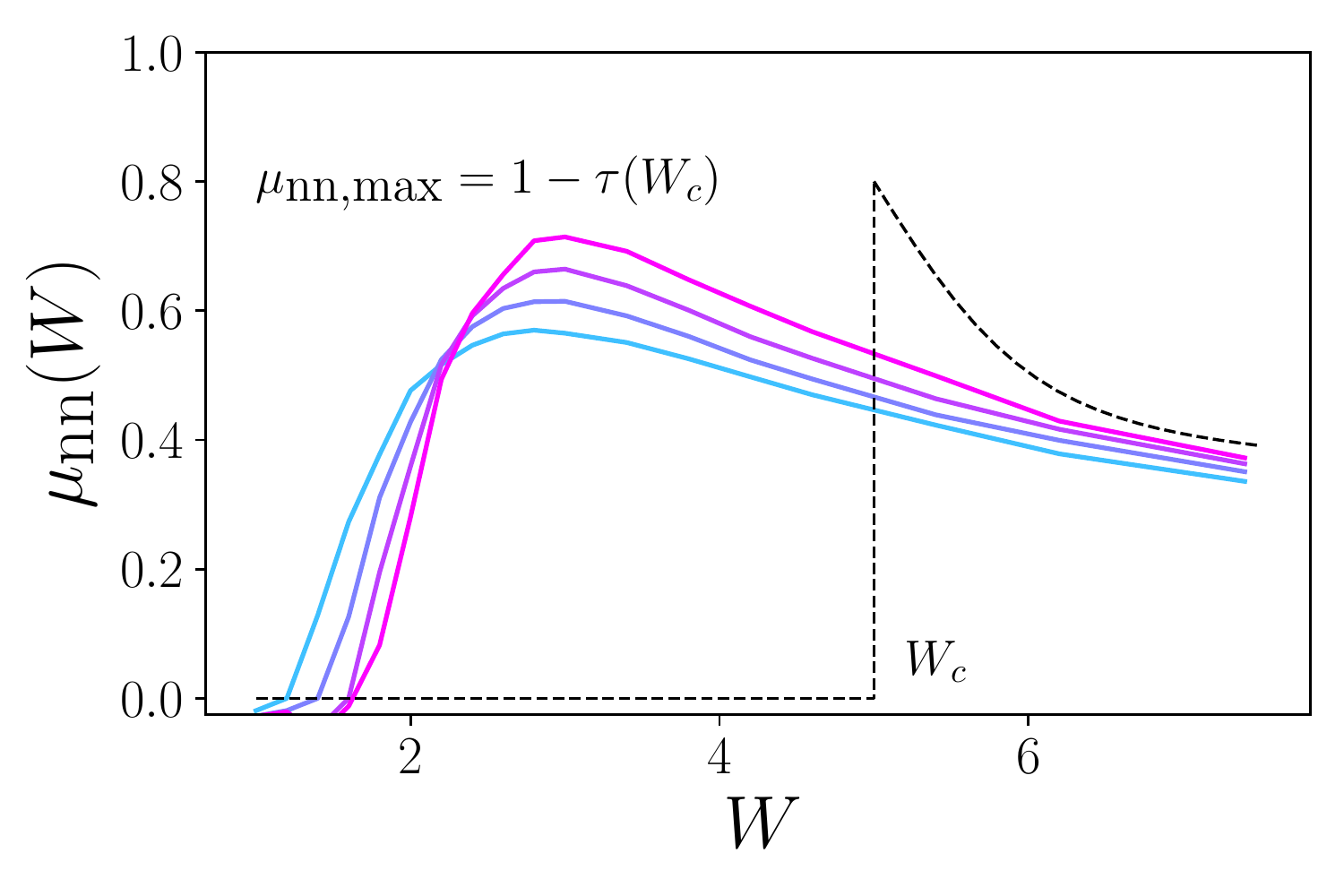}\endminipage
\caption{Correlation of adjacent eigenstates for spin chain with system sizes $L=12, 14, 16, 18$ (from cyan to magenta).
{\it Left:}  Correlation function $\beta_{\textrm{nn}}(W)$. Dashed vertical line marks an estimated value of the MBL transition in the thermodynamic limit,  $W_c \simeq 5$, as obtained by an approach based on matrix-product states for large chains \cite{Doggen2018a}.
{\it Right:}  Exponent $\mu_{\textrm{nn}}$  that characterizes the $\mathcal{N}$ scaling of adjacent-state correlation function, see Eq. (\ref{alphafit}) (with the replacement $N\rightarrow\mathcal{N}$).  Dashed line shows expected $\mathcal{N}\to\infty$ behavior, see Eq.~(\ref{betann-del}) for the ergodic phase and Eq.~(\ref{betann}) for the MBL phase (this section of the line is schematic). 
The figure is a counterpart of Fig.~\ref{alpha_rrg} for the RRG model. From Ref. \cite{tikhonov2020eigenstate}.
}
\label{alpha_spins}
\end{figure*}

As for the RRG model [see Sec.~\ref{sec:RRG-adjacent-state-corr}], it is useful to consider the correlation function of adjacent-in-energy eigenstates, Eq.~(\ref{beta-nn}). The corresponding data are presented in Fig.~\ref{alpha_spins}, which is very similar to its RRG analog, Fig.~\ref{alpha_rrg}. In the left panel of Fig.~\ref{alpha_spins},
the function $\beta_{\textrm{nn}}(W)$ is shown in a broad interval of disorder $W$ across the MBL transition, for several values of $L$.  For $W < W_c$, the data approach, with increasing $L$, a limiting curve  [$\mathcal{N}_\xi(W)$ in Eq.~\eqref{beta-omega-mbl-ergodic}], which is a manifestation of the ergodic character of the delocalized phase. For values of $L$ accessible to ED, this large-$L$ ergodic behavior is reached for disorder strengths $W \lesssim 2$.  As for the RRG model, the $\beta_{\textrm{nn}}(W)$ curves have a maximum near $W \approx 3$. It serves as a finite-size estimate for the transition point, drifting slowly towards the true ($L \to \infty$) value of $W_c$.

%%%%%%%%%%%%%%%%%%%%%%%%%%%%%%%%%%%%%%%%%%%%%%%%%%%%%%%%%%%%%%%%%%%%%%%%%%%%%%%%%%%%%%%%%%%%%%%%%%%%%%%%%%%%%%%%%%%%%%%%%%%%%%%%%%%%%%%%%%%%%%%%%%%%%%%%%%%%%%%%%%%%%%%%%%%%%%%%%%%%%%%%%%%%%%%%%%%%%%%%%%%%%%%%%
%%%%%%%%%%%%%%%%%%%%%%%%%%%%%%%%%%%%%%%%%%%%%%%%%%%%%%%%%%%%%%%%%%%%%%%%%%%%%%%%%%%%%%%%%%%%%%%%%%%%%%%%%%%%%%%%%%%%%%%%%%%%%%%%%%%%%%%%%%%%%%%%%%%%%%%%%%%%%%%%%%%%%%%%%%%%%%%%%%%%%%%%%%%%%%%%%%%%%%%%%%%%%%%%%

The right panel  of Fig.~\ref{alpha_spins} presents the  flowing exponent $\mu_{\textrm{nn}}$ defined by Eq.~(\ref{alphafit}) (with the replacement $N\rightarrow\mathcal{N}$).  
This figure is also similar to its RRG counterpart, right panel of Fig.~\ref{alpha_rrg}. There is, however, a difference: while for the RRG model the maximum value of $\mu_{\textrm{nn}}$ for the largest $L$ is unity with a good accuracy (as expected analytically), for the spin-chain model the maximum value is $\approx 0.75$. This is partly due to finite-size effects (which are stronger for the spin chain), but there is also a deeper reason. In full analogy with the IPR scaling  at the transition point, $P_2 \sim \mathcal{N}^{-\tau(W_c)}$ [see Eq.~\eqref{short-range-loc-ipr}], we expect the overlap of two adjacent states at criticality to exhibit the same scaling, $\mathcal{N}^2 \beta_{\rm nn} \sim \mathcal{N}^{-\tau(W_c)}$   [cf. Eq.~\eqref{P-res}], so that
\be
\mu_{\textrm{nn}}(W_c)=1-\tau(W_c) \,.
\ee
Thus, in the limit $\mathcal{N}\to \infty$,  we have $\mu_{\textrm{nn}}(W_c) \approx 0.8$, which is the maximum of $\mu_{\textrm{nn}}(W)$. 
 This result can be also obtained from Eq.~\eqref{mbl-final-scaling}  by extending it from the MBL phase to the transition point and setting $\mu(W_c)= 1$ (as in the RRG model).  More generally, setting $\omega \sim 1/\mathcal{N}$ in Eq.~\eqref{mbl-final-scaling}, one gets a relation between the exponents in the MBL phase,
\be
\label{exponent-relation}
\mu_{\textrm{nn}}(W)= \mu(W) - \tau(W) \,.
\ee
At strong disorder, the exponent $\tau(W)$ is small (it decays with increasing disorder as $\sim 1/W$, while $\mu(W)$ decays only logarithmically), so that $\tau(W) \ll \mu(W)$ and thus
\be
\label{exponent-relation-approx}
\mu_{\textrm{nn}}(W) \approx \mu(W) \,.
\ee
Numerically, this remains valid with reasonable accuracy up to the transition point, since $\tau(W_c)$ is rather small.  It should be also mentioned that logarithmic corrections to scaling, such as the logarithmic factor in Eq.~\eqref{mbl-final-scaling}, as well as further finite-size effects, significantly affect numerical values of MBL-phase exponents as obtained by ED.

The expected behavior of $\mu_{\textrm{nn}}(W)$ in the limit $L\to\infty$ is shown by a dashed line in the right panel of Fig. \ref{alpha_spins}. 
In analogy with the $\beta_{\textrm{nn}}$ peak, the position of the maximum of $\mu_{\textrm{nn}}$ yields a finite-size estimate for the critical point and drifts, with increasing $L$, towards the true (thermodynamic-limit)  transition point $W_c$. The drift is approximately linear with system size $L$;  the accessible system sizes are way too small to allow for a reliable estimate of the $L\to \infty$ critical disorder $W_c$. In similarity with the RRG model, a substantial part of the delocalized phase belongs to a broad critical regime, $2.5 \lesssim W \lesssim 5$,  for sizes $L$ accessible to ED.

\subsubsection{Dynamical correlations: Real-space observables}

Finally, we briefly discuss connections between the eigenstate correlation function $\beta(\omega)$ and other dynamical observables. In particular, Ref.~\cite{serbyn2017thouless} 
(see also a recent paper \cite{sels2020dynamical}) studied matrix elements of local (in real space) operator $S_z^i$,
\be
F(\omega) = \left \langle \left | \left(S^z_i \right)_{kl} \right |^2 \right \rangle    \equiv 
\left \langle | \langle \psi_k | S^z_i |\psi_l \rangle |^2 \right \rangle \,,
\label{local-spin-corr}
\ee
as a function of the frequency $\omega = E_k - E_l$.   The $\omega \to t$ Fourier transform of Eq.~(\ref{local-spin-corr}) has a meaning of the return probability in real space. At the same time, the Fourier transform of $\beta(\omega)$ yields the return probability in the many-body space $p(t)$ [cf. Sec.~\ref{sec:return_prob}] , which is clearly very different in general. Indeed, in the ergodic phase the behavior of $\beta(\omega)$ and $F(\omega)$ is very different: while $\beta(\omega)$ is a constant at not too large frequency, $F(\omega)$ has a power-law behavior reflecting diffusive or subdiffusive transport. Let us note that, contrary to $\beta(\omega)$, the function $F(\omega)$ does not have a direct counterpart in the RRG model, since the latter mimics the many-body space but not the real space. 
Nevertheless, there is a remarkable similarity in the behavior of $\beta(\omega)$ and $F(\omega)$ in the MBL phase (and at criticality): they both exhibit a power-law dependence on $\omega$, with a continuously varying exponent. In Ref.~\cite{gopalakrishnan2015low}, a related power-law frequency scaling of the conductivity (which is another real-space-related observable) was found, $\sigma(\omega) \sim \omega^\alpha$, with the exponent $\alpha$ varying in the range $1 < \alpha < 2$ in the MBL phase. It appears that the correlation functions defined in many-body space (like $\beta(\omega)$) and in the real space (like $F(\omega)$ or  $\sigma(\omega)$) are closely related in the MBL phase, since both these classes of correlation functions are governed by Mott-type resonances. More work is needed to better understand  these relations. 

 %%%%%%%%%%%%%%%%%%%%%%%%%%%%%%%%%%%%%%%%%%%%%%%%%%%%%%%%%%%%%%%%%%%%%%%%%%%%%%%%%%%%%%%%%%%%%%%%%%%%%%%%%%%%%%%%%%%%%%%%%%%%%%%%%%%%%%%%%%%%%%%%%%%%%%%%%%%%%%%%%%%%%%%%%%%%%%%%%%%%%%%%%%%%%%%%%%%%%%%%%%%%%%%%%

\section{Summary}
\label{sec:summary}

%%%%%%%%%%%%%%%%%%%%%%%%%%%%%%%%%%%%%%%%%%%%%%%%%%%%%%%%%%%%%%%%%%%%%%%%%%%%%%%%%%%%%%%%%%%%%%%%%%%%%%%%%%%%%%%%%%%%%%%%%%%%%%%%%%%%%%%%%%%%%%%%%%%%%%%%%%%%%%%%%%%%%%%%%%%%%%%%%%%%%%%%%%%%%%%%%%%%%%%%%%%%%%%%%
\begin{figure}[tbp]
\centerline{\includegraphics[width=0.75\textwidth]{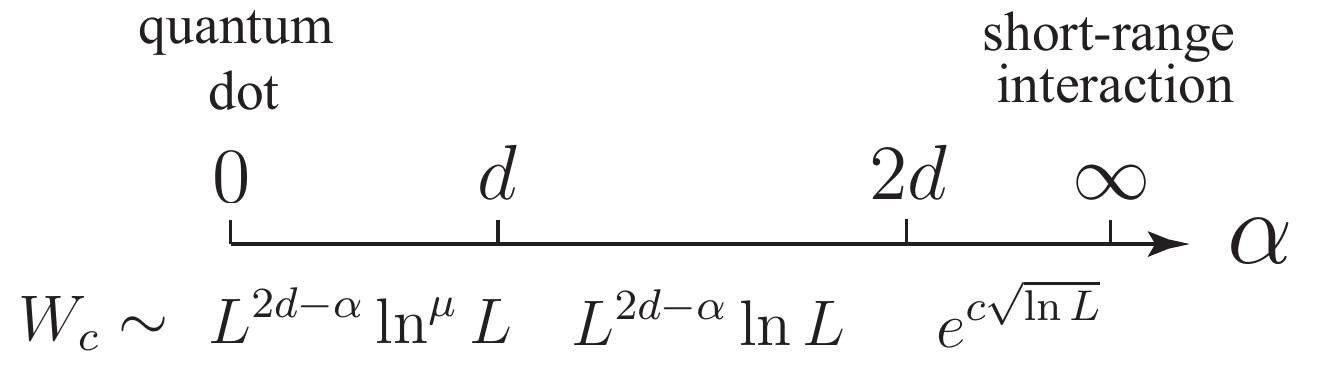}}
\caption{Evolution of $L$-scaling of the critical disorder $W_c$ of the MBL transition in the spin model
\eqref{long_spin}
with the power-law-interaction exponent $\alpha$. 
The range $d \le \alpha < 2d$ is considered in Sec.~\ref{sec:long_range}, and the corresponding critical disorder is given by Eq.~(\ref{scaling2}). Extreme cases are the limits of infinite-range interaction ($\alpha=0$, quantum dot, Sec.~\ref{sec:qdot}) and of short-range interaction ($\alpha = \infty$, Sec.~\ref{sec:short}). In the range $0 \le \alpha < d$ the mechanism of ergodization is analogous to that in the quantum dot ($\alpha=0$) model discussed in Sec.~\ref{sec:qdot}. In the range $2d \le \alpha < \infty$, the delocalization is expected to take place due to rare ergodic spots \cite{Thiery2017a,roeck17,thiery2017microscopically}. 
Adapted from Ref. \cite{tikhonov18}.}
\label{comm}
\end{figure}
%%%%%%%%%%%%%%%%%%%%%%%%%%%%%%%%%%%%%%%%%%%%%%%%%%%%%%%%%%%%%%%%%%%%%%%%%%%%%%%%%%%%%%%%%%%%%%%%%%%%%%%%%%%%%%%%%%%%%%%%%%%%%%%%%%%%%%%%%%%%%%%%%%%%%%%%%%%%%%%%%%%%%%%%%%%%%%%%%%%%%%%%%%%%%%%%%%%%%%%%%%%%%%%%%

In this article, we have reviewed properties of Anderson localization in the RRG model and its connection to a broad class of MBL problems, i.e. to interacting many-body models that exhibit a localization transition in Fock space. For models that are in the spatially localized phase in the absence of interaction, this MBL transition manifests itself also in real-space localization properties.  

The RRG model has a great advantage that it can be treated analytically in a controllable way within a field-theoretical approach using supersymmetry. Many of the physical observables characterizing eigenstate and level statistics can be then expressed in terms of a solution of a self-consistency equation. Key properties of the solution of this equation are understood analytically. Furthermore, this equation can be very efficiently solved numerically by means of PD. Remarkably, this allows one to proceed effectively to system sizes $N$ (Hilbert-space volume) as huge as $N \sim 10^{19}$, which is many orders of magnitude larger than systems that can be studied via ED ($N \lesssim 10^6$). This permits to determine the position of the critical point with an outstanding accuracy; in particular, for the most frequently studied model with coordination number $m+1 = 3$ and box distribution of disorder it is $W_c = 18.17 \pm 0.01$. 

Statistical properties of eigenstates and energy levels in the RRG problem have been studied very systematically, as reviewed  in Sec.~\ref{sec:RRG} of this article. This has been done by using three approaches: (i) purely analytical; (ii) analytical supplemented by numerical solution of the self-consistency equation, and (iii) ED. Results of all three approaches are in perfect agreement with each other.  We summarize key properties of Anderson localization in the RRG model:

\begin{itemize}

\item The delocalized phase, $W < W_c$, is ergodic. The ergodicity implies, in particular, the $1/N$ asymptotic scaling of the IPR $P_2$ as well as the universal, WD form of the level statistics (for not too large frequencies $\omega$). In the vicinity of the transition, the ergodicity is reached for system sizes $N \gg N_\xi$, where $\xi$ is the correlation length and $N_\xi \sim m^\xi$ is the correlation volume. The correlation length diverges according to a power law,  $\xi \sim (W_c-W)^{-\nu_{\rm del}}$, with the critical index  $\nu_{\rm del} = \frac{1}{2}$, so that $N_\xi$ diverges exponentially fast. In view of this, there is a sizeable range of disorders on the ergodic side of the transition for which the condition $N \gg N_\xi$ cannot be reached with sufficient margin in ED, so that the ED data can not fully reveal the ergodic behavior. 

\item The critical point of the Anderson transition, $W=W_c$ has a localized character. The leading behavior of various observables at criticality can be obtained by an extrapolation from the localized phase to the critical point, $W \to W_c +0$. For example, the IPR in the critical point is $P_2 \sim 1$  as in the localized phase, the level statistics at criticality is of Poisson form as in the localized phase etc. This should be contrasted to Anderson transition in $d$ dimensions where the corresponding behavior is in a sense intermediate between those in localized and delocalized phases.

\item
 The localized character of the critical point leads to strong finite-size effects. A  system in the ergodic phase ($W< W_c$) but close to the transition first evolves, with increasing size $N$,  towards criticality, i.e., towards localized behavior of energy-level and eigenstate statistics. Only when $N$ exceeds $N_\xi$, the flow changes direction and the system starts evolving towards ergodicity.  This non-monotonic $N$-dependence of various observables manifests itself in a substantial drift of the apparent transition point (as obtained on the basis of ED data) towards stronger disorder with increasing system size. 
 
 \end{itemize}
 
 In Sec.~\ref{sec:MBL}, we have briefly reviewed some of main properties of the MBL transitions. We have considered three classes of models.  In Sec.~\ref{sec:qdot}, ``quantum-dot'' models were discussed, with single-particle states spread over the whole system. In these models, the MBL transition happens in the Fock space only. Sections \ref{sec:long_range} and \ref{sec:short} deal with models that are characterized by single-particle states that are localized in real space, with long-range and short-range interaction, respectively. For each of these three classes, there are closely related models formulated in terms of fermions and in terms of spins. Since our main focus is on relations between the RRG and MBL problems, we discussed those observables in MBL models that have direct counterparts in the RRG model. These include many of key properties of the MBL systems, such as the scaling of MBL transition points, level statistics, as well as fluctuations and correlations of many-body eigenstates.  
 For all of the considered MBL models, there are strong connections with the RRG problem, and in some cases approximate mapping to RRG is used to infer physical properties. 
 Main results can be summarized as follows:
 
 \begin{itemize}
 
 \item For the quantum-dot model (or, equivalently, the SYK model perturbed by a kinetic-energy term), Sec.~\ref{sec:qdot}, the Fock-space MBL transition takes place in the plane spanned by two dimensionless parameters: the number $n \gg 1$ of single-particle orbitals and the ratio $g^{-1} = V / \Delta \ll 1$ of the interaction matrix element $V$ to the single-particle level spacing $\Delta$.
 The model has approximately a structure of an RRG model with the Hilbert-space volume $\mathcal{N} \sim 2^n$, disorder strength $W \sim n\Delta$, hopping $V \sim \Delta/g$ and coordination number $m+1 \sim n^4$. The RRG position of the critical point, $W/V \sim m \ln m$,  translates, in this approximation, into the following line of the MBL transition in the $g$--$n$ plane:  $ g \sim n^3 \ln n$, Eq.~\eqref{qdot-rrg-gc}.
 In fact, the accuracy of this approximation is not fully clear at this stage. The reason is that a model with two-body interaction involves correlations between many-body states coupled by interaction that are not present in the RRG model. It turns out, however, that spectral diffusion efficiently reduces the effect of these correlations in the quantum-dot problem.
Specifically,  the analysis in Ref.~\cite{gornyi2017spectral} comes to the same result  as the RRG model suggests but with possibly smaller power of the logarithm: $ g \sim n^3 \ln^\mu n$ with $\mu \le 1$. Furthermore, Ref.~\cite{monteiro2020minimal} proposes that the RRG result even holds exactly, including the power of the logarithm. The argument is based, however, on an approximation of effective-medium type that, in our view, has to be studied more accurately. Independently of these details,  the MBL transition in the quantum-dot (perturbed-SYK) model is closely related to the localization transition on RRG. This concerns not only the scaling of the MBL transition but also properties of both phases and of the critical point.

 \item In Sec.~\ref{sec:long_range}, we have considered the MBL transition in a model with  random long-range interaction decaying with distance as $1/r^\alpha$. While  we discussed (following Ref.~\cite{tikhonov18}) a spin model with random interaction and random on-site magnetic field, interacting fermionic models can be (approximately) mapped on a spin model of this type and have similar properties. Parameters characterizing the model are the linear size $L$, the spatial dimensionality $d$, the power-law exponent $\alpha$, and the strength $W$ of the random field (normalized to the nearest-neighbor interaction strength). Depending on the value of the exponent $\alpha$, there are three distinct situations, see Fig.~\ref{comm}. For $\alpha < d$, the number of direct resonances that each spin finds increases without bound with increasing $L$. This can be viewed as a ``quantum-dot-like'' situation. The analysis can be performed analogously to the quantum-dot model, with the critical disorder of the MBL transition behaving like $W_c \sim L^{2d-\alpha} \ln^\mu L$. For $d < \alpha < 2d$, only rare spins form resonances (``pseudospins''). At the same time, the total number of such pseudospins increases with $L$, so that they eventually proliferate and start to interact.  It turns out that the system of interacting pseudospins can be approximately mapped on the RRG model, yielding the result for the critical point
 $W_c \sim L^{2d-\alpha} \ln L$, Eq.~\eqref{scaling2}.  Finally, for $ \alpha > 2d$, such mechanism of delocalization is not operative. This regime is similar to the case of short-range interaction. Delocalization at large $L$ (and fixed large $W$) is expected to happen due to rare events only (as in short-range-interaction models in $d>1$), so that $W_c$ increases with $L$  in a very slow fashion (more slowly than any power law). 
 
 We have focussed in Sec.~\ref{sec:long_range} on the case $d < \alpha < 2d$. The mapping of the MBL model with long-range interaction with such $\alpha$ onto RRG model strongly suggests that most of key properties of the RRG model hold also for the MBL problem. This includes, in particular the ergodic character of the delocalized phase and the localized nature of the critical point $W_c$ (i.e., properties of 
the critical point are obtained continuously by an extrapolation $W \to W_c +0$ from the localized side.)

At the same time, there is one important difference between the MBL models that involve localization in real space and their RRG counterparts. The scaling of the IPR in the localized phase of the MBL problems is of multifractal form, $P_2 \sim \mathcal{N}^{-\tau(W)}$, with the disorder dependent exponent $\tau \sim 1/W$, at variance with $P_2 \sim 1$ in the localized phase on RRG. This multifractality of the MBL phase results in a natural way from rare short-scale resonances that have a concentration $1/W$. While such resonances are not sufficient to establish delocalization for $W > W_c$, each of them yields a factor $\sim 1/2$ into $P_2$, thus leading to a fractal scaling. This equally applies to models with power-law interaction (with $\alpha > d$) and with short-range interaction.

The analytical predictions resulting from the mapping of a model with $d < \alpha < 2d$ on RRG have been verified by numerical simulations (for $d=1$ and $\alpha=3/2$).  The numerical results are in agreement with the analytically obtained scaling of  critical disorder $W_c(L)$. Further, numerical results for the IPR scaling confirm the ergodicity of the delocalized phase and the fractal scaling of IPR in the localized phase.

  \item  In Sec.~\ref{sec:short} we discussed the MBL problem with a short-range interaction. Specifically, we have focussed on 1D $S=\frac{1}{2}$ spin chain in random field that represents the most popular MBL model. For a short-range interaction model, the connection to RRG is somewhat more delicate than in the case of, e.g., quantum-dot model. Indeed, a number of states to which each basis many-body state (eigenstate of the non-interacting problem) is connected by interaction grows $\sim L$, where $L$ is the chain length. At first sight, one could think that the model is similar to RRG with a coordination number $\sim L$. This is not true, however in view of a combination of two reasons. First, there are strong correlations between states coupled by the interaction. Second, distant resonances ``do not talk to each other'' in a model with short-range interaction, so that the spectral diffusion does not play a role that it plays in the quantum-dot model. The number of resonances growing as $\sim L$ leads only to fractal behavior of IPR,  $P_2 \sim \mathcal{N}^{-\tau(W)}$, as in the case of power-law interaction. 
 At the same time, the number of potential resonances that can be responsible for many-body delocalization grows with the order $n$ of the perturbation theory as
 $N_{n;L}\sim L m^n$, with $m \sim 1$, Eq.~\eqref{mbl-potential-resonances}. The delocalization is thus similar to that on RRG with a coordination number $m \sim 1$, so that the critical disorder is $L$-independent, $W_c \sim 1$, in the limit $L \to \infty$. This argument does not include a possible effect of rare ergodic regions. However, in 1D geometry, rare regions do not affect the conclusion about the $L$-independence of $W_c$ at large $L$ (although they do affect the exact value of $W_c$ and probably the critical behavior). 
 
 Numerical results support the analytical conclusion that $W_c$ saturates at an $L$-independent value in the large-$L$ limit. Furthermore, they indicate that qualitative properties of the MBL transition are in many key respects analogous to those of the RRG model. Specifically, the delocalized phase is found to be ergodic, the critical point has the localized characacter, and the apparent critical disorder $W_c$ exhibits a sizeable drift towards larger values with increasing $L$. Of course, the ED is limited by systems of moderate length (typically $L \le 24$); at such lengths, a sizeable range of disorder near $W_c$ belongs to a critical region. 
 
 As an important characteristic of the many-body state, we have analyzed in Sec.~\ref{sec:short} dynamical correlations of eigenstates $\beta(\omega)$ across the MBL transition. The correlations reveal a remarkable similarity between the MBL and RRG problems. In particular, the Figs. \ref{beta_spins_full}, \ref{beta_spins}, and \ref{alpha_spins} for the spin-chain problem are very similar to their RRG counterparts, Figs.~\ref{beta_rrg_full}, \ref{beta_rrg}, and \ref{alpha_rrg}. On the delocalized side of the MBL transitions, these correlations support (along with various other observables) ergodicity of the system. On the MBL side, $W > W_c$, they yield a power-law dynamical scaling, $\beta(\omega) \propto \omega^{-\mu(W)}$, with a disorder-dependent exponent $\mu(W)$, in analogy with the RRG problem.
 
  \end{itemize}
 
Clearly, the RRG model is not equivalent to any of the MBL models discussed above. There are many observables in the MBL problems (such as properties of single-particle excitations or real-space observables) that do not have direct counterparts in the RRG  model. Indeed, the Hilbert space of the RRG problem, although mimicking important features of the Fock space of the many-body problem, is not the Fock space. For the same reason, notions of rare ergodic or localized spots that play an important role in phenomenological theories of the MBL transition do not have their counterparts in the RRG model either. The RRG critical behavior cannot be the asymptotic behavior for MBL problems since it would violate the Harris criterion.  All these differences are manifestations of the fact that the RRG model is only a toy model representing a simplistic version of the genuine MBL problem. Nevertheless, as was discussed in the present review and emphasized in this section, the RRG model captures many key physical properties of the MBL problem. In models with long-range interaction, the relation between the MBL and RRG problems can be even made quantitative. 

\section{Acknowledgments}

We acknowledge collaboration with M. Skvortsov on Ref.~\cite{tikhonov2016anderson}, which was the paper that started our investigations of the RRG model and its relations to the MBL problem. In course of these studies, we enjoyed useful discussions with many colleagues, including A. Altland, A. Burin, I. P. Castillo, F. Evers, M. V. Feigelman, A. Knowles, V. E. Kravtsov, N. Laflorencie, G. Lemari\'e, F. L. Metz, A. Scardicchio, M. Serbyn, and M. Tarzia.

\bibliography{rrg.bib}
%\medskip
%\printbibliography

\end{document}